\documentclass[11pt,a4paper]{article}
\usepackage{jheppub}
\usepackage{mathrsfs}
\usepackage{amsfonts}
\usepackage{setspace}
\usepackage{cellspace}
\usepackage{amsmath,amssymb,bm}
\usepackage[colorlinks=true,linkcolor=blue]{hyperref}
\usepackage{xcolor}
\usepackage{epsfig}
\usepackage{slashed}
\usepackage{caption}
\usepackage{hhline,multirow,tabularx}  
\usepackage{dcolumn}    
\usepackage{url}        
\usepackage{braket}     
\renewcommand{\bra}[1]{\left<#1\left|}
\renewcommand{\ket}[1]{\right|#1\right>}
\preprint{CNF-UMD-2022}
\newcommand{\Hcff}{\mathcal{H}}
\newcommand{\Ecff}{\mathcal{E}}
\newcommand{\Htcff}{\widetilde{\mathcal{H}}}
\newcommand{\Etcff}{\widetilde{\mathcal{E}}}
\newcommand{\HTcff}{\bar{\mathcal{H}}_{2T}}
\newcommand{\ETcff}{\bar{\mathcal{E}}_{2T}}
\newcommand{\HTtcff}{\bar{\widetilde{\mathcal{H}}}_{2T}}
\newcommand{\ETtcff}{\bar{\widetilde{\mathcal{E}}}_{2T}}

\title{Twist-three cross-sections in deeply virtual Compton scattering}

\author[a]{Yuxun~Guo}
\author[a,b]{, Xiangdong~Ji}
\author[b]{, Brandon~Kriesten}
\author[b]{and Kyle~Shiells}
\affiliation[a]{Maryland Center for Fundamental Physics, Department of Physics,\\ University of Maryland, College Park, MD 20742, USA}
\affiliation[b]{Center for Nuclear Femtography, SURA,\\ 1201 New York Ave. NW, Washington, DC 20005, USA}
\emailAdd{yuxunguo@umd.edu}
\emailAdd{xji@umd.edu}
\emailAdd{bkriesten@sura.org}
\emailAdd{kshiells@sura.org}
\abstract{We study the deeply virtual Compton scattering process with both twist-two and twist-three Compton form factors and present our cross-sections formulas with all polarization configurations. While the twist-three contributions are generally assumed to be negligible in the literature due to the kinematical suppression, we compare them with the twist-two ones at typical JLab 6 GeV and 12 GeV kinematics as well as EIC kinematics and show their kinematical suppression explicitly, justifying the leading-twist approximation made in the literature. In addition, we also estimate the twist-three Compton form factors using Wandzura-Wilczek relations and inputs of twist-two generalized parton distributions based on a reggeized spectator model. With those estimated Compton form factors, we analyze the kinematical behavior of twist-two and twist-three cross-sections in a wide range of kinematics, and discuss the optimal regions for separating the leading-twist effects from the higher-twist ones. }
\date{\today}
\begin{document}
\maketitle

\section{Introduction}

Deeply virtual Compton scattering (DVCS)~\cite{Ji:1996ek,Ji:1996nm}, the process where a space-like photon with large virtuality $Q^2$ collides with the nucleon while keeping it intact and creates an on-shell photon in the final state, serves as a clean probe of the generalized parton distributions (GPDs)~\cite{Muller:1994ses, Ji:1996ek} of the nucleon. Encoded in this process are the important information of the nucleon such as the mass, angular momentum and mechanical properties~\cite{Ji:1994av, Ji:1996ek, Polyakov:2002yz} and the three-dimensional structure of the nucleon~\cite{Burkardt:2000za, Ji:2003ak, Belitsky:2003nz}.  There have been numerous measurements from HERA (H1 \cite{H1:2001nez,H1:2005gdw,H1:2007vrx,H1:2009wnw}, ZEUS~\cite{ZEUS:2003pwh,ZEUS:2008hcd} and HERMES~\cite{HERMES:2012gbh,HERMES:2012idp}) and Jefferson Lab (JLab) (CLAS~\cite{CLAS:2007clm,CLAS:2008ahu,Niccolai:2012sq,CLAS:2015bqi,CLAS:2015uuo,CLAS:2018bgk,CLAS:2018ddh,CLAS:2021gwi} and Hall A~\cite{JeffersonLabHallA:2006prd,JeffersonLabHallA:2007jdm,JeffersonLabHallA:2012zwt,Georges:2017xjy,JeffersonLabHallA:2022pnx}) spanning decades, and more
programs are planned in the future such as the JLab 24 GeV, EIcC~\cite{Anderle:2021wcy} and EIC~\cite{AbdulKhalek:2021gbh}. 

The theoretical foundation of studying the DVCS process is the collinear factorization theorem proven in quantum chromodynamics (QCD) to the leading power accuracy of $Q$~\cite{Ji:1997nk,Ji:1998xh,Collins:1998be}, where the twist expansion is introduced. With the twist expansion, two types of corrections shall be considered for a systematical analysis of the higher-twist effects --- the kinematically higher-order corrections associated with the leading-twist GPDs and the effects of the higher-twist GPDs. The kinematical twist-three effects ~\cite{Anikin:2000em,Penttinen:2000dg,Belitsky:2000vx,Kivel:2000cn,Radyushkin:2000ap,Guo:2021gru} and twist-four effects such as finite $t$ and target mass corrections~\cite{Blumlein:2006ia,Blumlein:2009hit,Braun:2011zr,Braun:2011dg,Braun:2012bg,Braun:2012hq,Braun:2014sta} in the DVCS process have been studied extensively in the literature. In this work, we focus on the effects of higher-twist GPDs, specifically the twist-three ones. On the one hand, analysis of the leading-twist effects usually assumes the suppression of higher-twist effects at high $Q^2$, which needs to be justified explicitly. On the other hand, once the leading-twist effects are determined with enough precision, we will then be allowed to measure the twist-three effects at relatively low $Q^2$ and potentially constrain the twist-three GPDs, which plays an important role in the angular momentum of the nucleon\cite{Jaffe:1989jz,Lorce:2011kd,Burkardt:2012sd,Rajan:2016tlg,Rajan:2017cpx,Guo:2021aik}. 

The analysis of higher-twist effects gets complicated when the Wandzura-Wilczek (WW) relations are taken into account, which relate GPDs of different twists as required by Lorentz invariance and QCD equations of motion. Twist-three GPDs can be split into the WW parts that are expressible in terms of the leading-twist GPDs and the genuine twist-three parts that are related to the higher-twist quark-gluon-quark operators~\cite{Belitsky:2000vx,Belitsky:2005qn,Kivel:2000rb,Radyushkin:2000jy,Radyushkin:2001fc,Aslan:2018zzk}. 
However, implementation of such relations in the cross-section analysis is non-trivial, as the DVCS cross-sections are not direct measurements of the GPDs but the so-called Compton form factors (CFFs), which are the convolutions of GPDs and complex-valued Wilson coefficients~\cite{Belitsky:2001ns,Diehl:2003ny,Belitsky:2005qn}. Thus, even with explicit relations between twist-two and twist-three GPDs, it will be extremely hard, if not impossible, to find out the relations between twist-two and twist-three CFFs explicitly, which requires one to deconvolute the twist-two CFFs first to get the twist-two GPDs. Consequently, we will need extra inputs for the twist-two GPDs, such as a parameterization of GPDs, in order to compare the twist-two and twist-three CFFs.

Using the twist-three CFFs estimated with WW approximation, the twist-three cross-sections can be calculated. One of the most important test for the twist expansion is the kinematical suppression of higher-twist effects, which ensures that one can extract the leading-twist quantities without the interference of the unknown higher-twist contributions. However, due to the existence of the
Bethe-Heitler (BH) process, a photon emission process driven by quantum electrodynamics (QED), even the leading-twist effects are suppressed compared to the BH background and thus are hard to extract. Therefore, it is crucial to find the proper kinematical regions where the higher-twist effects are suppressed while the leading-twist effects are still sizable. As we will show, this is possible at higher beam energy/center of mass energy with reasonably large $Q^2$.

The organization of the paper is as follows. In section \ref{sec:general}, we present our cross-sections formulas for all polarization configurations with twist-three CFFs, following our previous works for the twist-two ones \cite{Guo:2021gru}. In section \ref{sec:t3scalar}, we study the twist-three scalar coefficients, which are the kinematical prefactors of the twist-three CFFs in the cross-section formulas. In section \ref{sec:t3cff}, we employ a GPD parameterization and use it to estimate the WW twist-three CFFs, with which the twist-three cross-sections are studied as well. In the end, we conclude in  section \ref{sec:conc}.

\section{Twist-three DVCS cross-section with polarized beam and target}
\label{sec:general}
In this section, we present the twist-three DVCS cross-section formulas, see for instance refs. \cite{Ji:1996nm,Belitsky:2001ns,Belitsky:2005qn,Kriesten:2019jep,Guo:2021gru} for more details on DVCS cross-sections. The set-up follows our previous works in ref. \cite{Guo:2021gru}, and we review part of them in this paper for consistency. Consider the electroproduction of  a photon off a proton as,
\begin{equation}
    e(k,h)+N(P,S)\to e(k',h')+N(P',S')+\gamma(q',\Lambda')\ ,
\end{equation}
where the $k,k',P,P',q'$ are the momenta of the 5 particles respectively, and $h,h',S,S',\Lambda'$ correspond to their helicities or polarization vectors. The helicities $h$ and $h'$ take the value of $\pm 1/2$, $\Lambda'$ takes $\pm 1$ while the target polarization vector $S$ satisfies $S^2=-1$ and $S\cdot P=0$ and similarly for $S'$. We also define the virtual photon momentum $q\equiv k-k'$ and its helicity $\Lambda$. 
The full amplitude is given mainly by the sum of two sub-processes, the BH process and the DVCS process, and can be written as \cite{Belitsky:2001ns,Guo:2021gru},
\begin{equation}
    \mathcal T=\mathcal T_{\rm{BH}}+\mathcal T_{\rm{DVCS}}\ .
\end{equation}
The electroproduction cross-section in the lab frame can be expressed in terms of the squared amplitude combined with some kinematical prefactors as \cite{Belitsky:2001ns,Kriesten:2019jep},
\begin{equation}
    \frac{\text{d}^5\sigma}{\text{d}x_B \text{d}Q^2\text{d}|t|\text{d}\phi \text{d}\phi_S}=\frac{\alpha_{\rm{EM}}^3 x_B y^2}{16\pi^2 Q^4\sqrt{1+\gamma^2}} \left|\mathcal T\right|^2 \ ,
\end{equation}
where we have the following definitions: fine structure constant $\alpha_{\rm{EM}}\equiv e^2/(4\pi)$ , photon virtuality $Q^2\equiv -q^2$, the Bjorken scaling variable $x_B\equiv Q^2/(2 P\cdot q)$, the electron energy loss variable $y \equiv (P\cdot q)/(P\cdot k)$ and momentum transfer square $t\equiv (P'-P)^2$. The two angles $\phi$ and $\phi_S$ are the azimuthal angle between the reaction plane and the leptonic plane, and the azimuthal angle between the target polarization vector  and the leptonic plane in the case of transversely polarized target, respectively. The notation $ \gamma\equiv 2M x_B/Q \ $ is introduced with $M$ the target mass. We also define the notation $\bar P\equiv (P+P')/2$, $\bar q\equiv (q+q')/2$ and $\Delta\equiv P'-P=q-q'$. 

The squared amplitude consists of three parts,
\begin{equation}
\label{eq:sqrtamp}
    \left|\mathcal T \right|^2 =\left|\mathcal T_{\rm{BH}}\right|^2 +\left|\mathcal T_{\rm{DVCS}}\right|^2+T_{\rm{BH}}^*\mathcal T_{\rm{DVCS}}+\mathcal T_{\rm{DVCS}}^*\mathcal T_{\rm{BH}}\ ,
\end{equation}
where the last two terms define the interference contribution,
\begin{equation}
    \mathcal I \equiv \mathcal T_{\rm{BH}}^*\mathcal T_{\rm{DVCS}}+\mathcal T_{\rm{DVCS}}^*\mathcal T_{\rm{BH}}=2 \text{Re}\left[\mathcal T_{\rm{BH}}^*\mathcal T_{\rm{DVCS}}\right]\ .
\end{equation}
Consequently, the cross-sections can be split into three contributions,
\begin{equation}
    \text{d}\sigma_{\rm{Total}} =   \text{d}\sigma_{\rm{BH}}+ \text{d}\sigma_{\rm{DVCS}}+\text{d}\sigma_{\rm{INT}}\ .
\end{equation}
As the BH amplitude can be calculated relatively easily given the well-determined Dirac and Pauli form factors from the elastic scattering processes \cite{Punjabi:2015bba}, the pure BH cross-sections will be treated as the background and will not be discussed in detail. 

\subsection{Comparison of fixed-target and collider coordinates}

\begin{figure}[t]
\centering
\begin{minipage}[b]{0.49\textwidth}
\centering
\includegraphics[width=\textwidth]{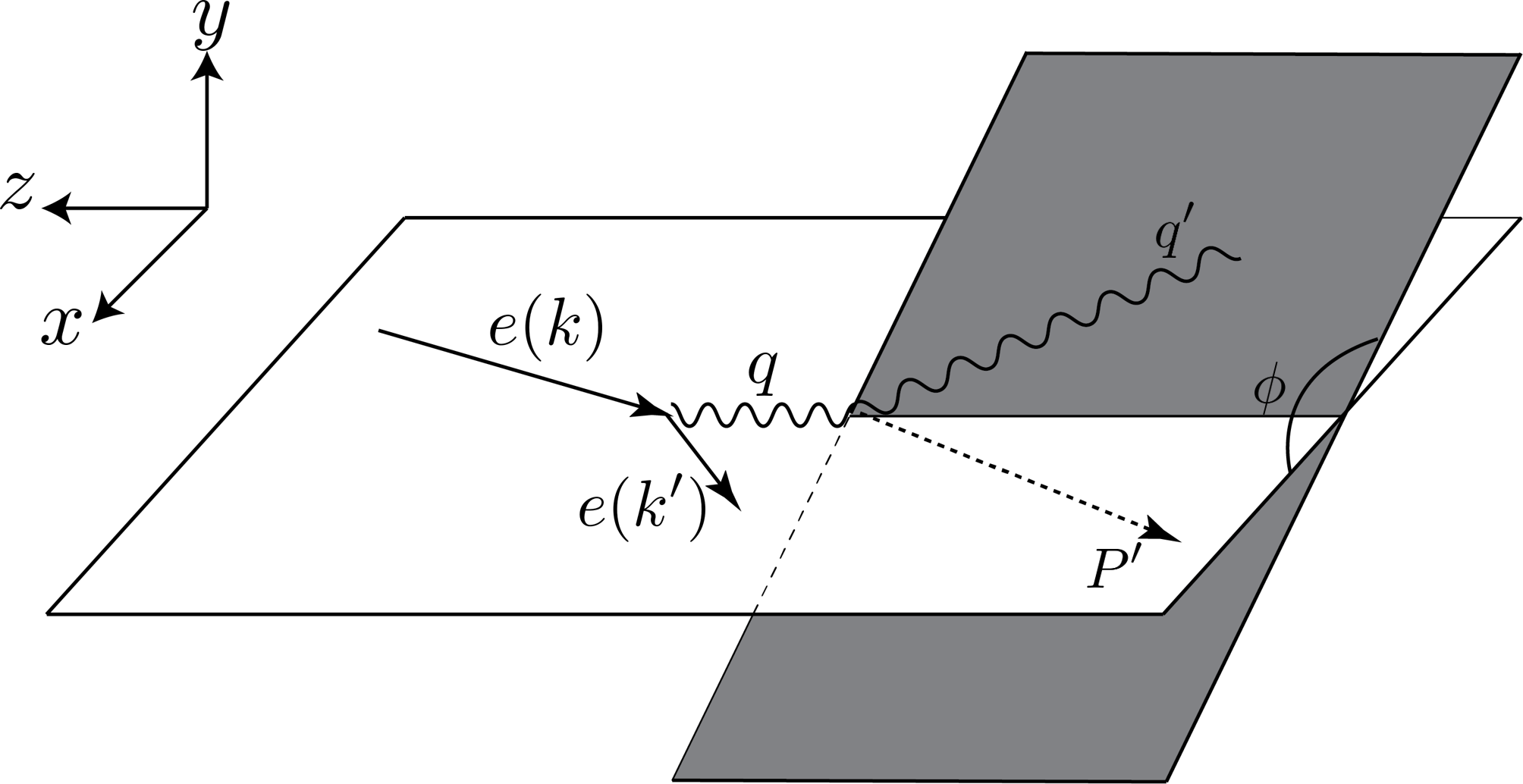}
\end{minipage}
\begin{minipage}[b]{0.49\textwidth}
\centering
\includegraphics[width=\textwidth]{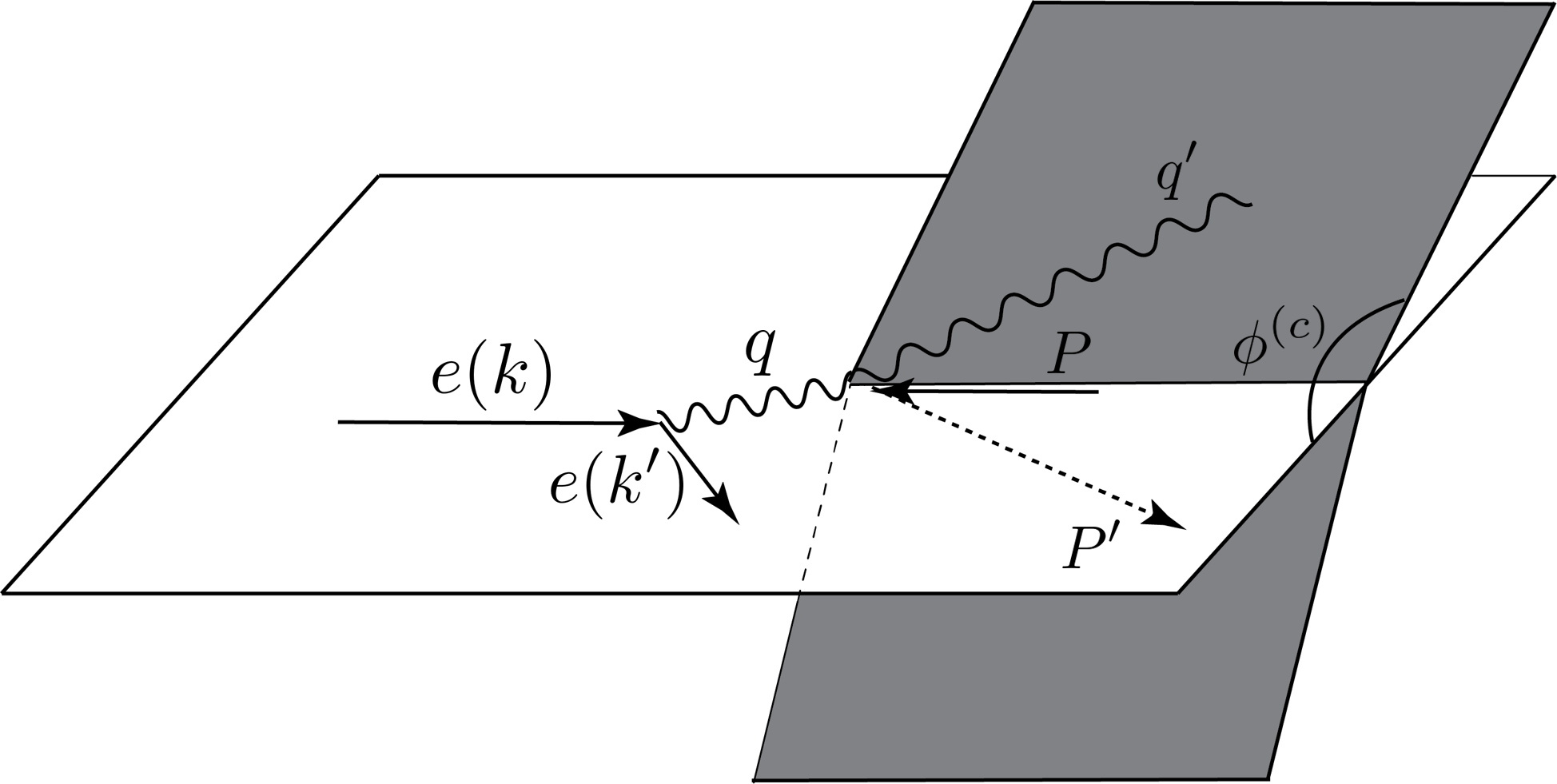}
\end{minipage}
\caption{\label{fig:coordinatesetup} The coordinate choice for fixed-target experiments (left) and the collider experiments (right), where the target momentum is not shown for the fixed-target plot. The $z$-axis in the fixed-target coordinates is chosen such that the virtual photon momentum $\vec q$ is in $-z$ direction, whereas the $z$-axis in the collider coordinates is defined by the electron beam ($-z$ direction) and proton beam ($+z$ direction).} 
\end{figure}

The above set-up, especially the azimuthal variables, is developed majorly for fixed-target experiments, where we used the conventions for coordinates in ref. \cite{Kriesten:2019jep}. This coordinate set-up needs to be modified for collider experiments such as the future EIcC~\cite{Anderle:2021wcy} and EIC~\cite{AbdulKhalek:2021gbh} program. For comparison, we show plots for the coordinate set-ups for both coordinate choices in figure \ref{fig:coordinatesetup}.
 
\begin{figure}[t]
\centering
\includegraphics[width=0.95\textwidth]{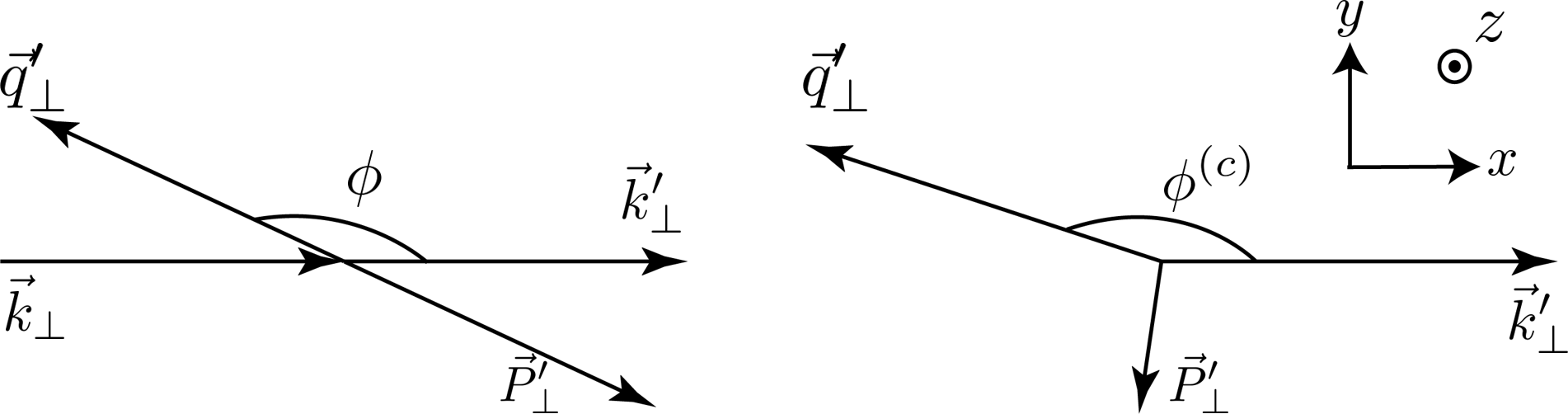}
\caption{\label{fig:coordinatesetupT} The coordinate choice in transverse $x$-$y$ plane with the $z$-axis pointing outwards for fixed-target experiments (left) and the collider experiments (right). The transverse momenta in the fixed-target coordinates satisfy $\vec q'_\perp = -\vec P'_\perp$ and $\vec k'_\perp=-\vec k_\perp$, while the transverse momenta in the collider coordinates are not align with each other though the momentum conservation ensures $\vec q'_\perp +\vec k'_\perp+\vec P'_\perp =0$.} 
\end{figure}

Note that in ref.~\cite{Kriesten:2019jep} the $z$-axis is reversely plotted, which leads to a redefinition of azimuthal angle $\phi\to -\phi$. Besides, we define the azimuthal angle $\phi$ with the real photon momentum $q'$, different from the convention in ref. \cite{Belitsky:2001ns} where the final proton momentum is used. This corresponds to a redefinition $\phi \to \phi +\pi$. As for the collider coordinates, the azimuthal variables $\phi^{(c)}$ will be related to the $\phi$ in the fixed-target coordinates in a more complicated way. Their difference is more intuitive in the transverse $x$-$y$ plane, as shown in figure \ref{fig:coordinatesetupT}. While in both coordinates the leptonic plane is chosen to be the $x$-$z$ plane, the $z$-axis is chosen differently.
For the fixed-target coordinates, the $z$-axis is chosen according to the virtual photon momentum such that $\vec q$ is in $-z$ direction, whereas the $z$-axis for the collider coordinates is defined by the electron beam ($-z$ direction) and proton beam ($+z$ direction).

Each coordinate choice has its own advantages. For the fixed-target coordinates, the reaction plane made by the final photon momentum $\vec q'$ and proton momentum $\vec P'$ contains the $z$-axis, since $\vec q'_\perp = -\vec P'_\perp$. Consequently, the two momenta have the same azimuthal angle $\phi$ and the products (or contractions with Levi-Civita symbol) of any momenta are always in the form of $A+B \cos(\phi)$ (or $A'\sin(\phi)$). Therefore, the cross-sections can always be written in terms of a finite-order polynomial of $\cos(\phi)$ or $\sin(\phi)$ except for some extra angular dependence in the denominator due to the BH propagators. This is guaranteed by the geometry of this coordinate choice and the fact that the kinematical coefficients can only be written with the contractions of all the four-vectors.

The collider coordinates, on the other hand, do not have this simplification. As we can see on the right of figure \ref{fig:coordinatesetupT}, although the transverse final proton momentum $\vec P'_\perp$ is fixed by momentum conservation $\vec P'_\perp+\vec q'_\perp+\vec k'_\perp=0$, the angle it makes with respect to the $x$-axis is non-trivially related to $\phi^{(c)}$, making the final $\phi^{(c)}$ dependence extremely cumbersome. However, a different coordinate system is needed for collider experiments for practical reasons. Since the angle $\phi$ is not invariant under boost in the electron beam direction, it can not be directly measured in collider experiments where the target is not at rest, whereas the angle $\phi^{(c)}$ and other azimuthal angles in the collider coordinates are.

With this in mind, the best way to analyze the azimuthal dependence is to measure $\phi$ and $\phi^{(c)}$ for fixed-target and collider experiments, respectively. The two coordinates can then be connected by a rotation in the target rest frame, and the transformation relations are given in the appendix. \ref{app:eicframe}. Therefore, the cross-sections measurements can be transformed according to the different coordinates choices. It is worth noting that the difference between $\phi$ and $\phi^{(c)}$ is suppressed as higher-twist effects. Since in the leading-twist picture, all the hard momenta approach the light cone such that they are collinear and the transverse plane makes no difference in different coordinates.

We also note that the polynomial behavior of the azimuthal dependence in the fixed-target coordinate is very helpful for azimuthal analysis, and lots of techniques are built assuming this property, see for instance refs. \cite{Belitsky:2001ns,Shiells:2021xqo}. Therefore, we will always present the results in terms of the fixed-target variables $\phi$, even for collider kinematics, and assume the cross-sections in the collider coordinates are always transformed to the fixed-targets ones using the relations in the appendix. \ref{app:eicframe} in order for consistency with those analysis methods.

\subsection{Twist-three Compton tensor and promoted Compton tensor coefficients}
With our explicit choice of frame and coordinates, the cross-section formulas can be calculated. The DVCS amplitude can be expressed in terms of the Compton tensor as,
\begin{equation}
\label{eq:DVCSamp}
    \mathcal T_{\rm{DVCS}}=\frac{e_l}{Q^2}\bar u(k',h') \gamma^\nu u(k,h) T^{\mu\nu} \varepsilon^*_{\mu}(q',\Lambda')\ ,
\end{equation}
with $e_l$ the lepton charge in the unit of electron charge that is positive (negative) for electron (positron), while the Compton tensor $T^{\mu\nu}$ is defined as,
\begin{equation}
   T^{\mu\nu}\equiv i \int \text{d}^4x e^{i(q+q')z/2} \bra{P',S'} \text{T}\left\{J^{\mu}\left(\frac{z}{2}\right)J^{\nu}\left(-\frac{z}{2}\right)\right\} \ket{P,S}\ ,
\end{equation}
with $J^\mu(z)$ the electromagnetic current operator and T the time order operator. With the twist expansion introduced by the collinear factorization theorem, the Compton tensor $T^{\mu\nu}$ can be expressed as,
\begin{align}
\begin{split}
   T^{\mu\nu}=T^{\mu\nu}_{(2)}+T^{\mu\nu}_{(3)}+\cdots\ ,
\end{split}
\end{align}
where the ellipsis stands for the twist-four and higher contributions that will not be considered. The leading and next-to-leading Compton tensor can be written as \cite{Belitsky:2005qn},
\begin{align}
\label{eq:comptontensor}
\begin{split}
   T^{\mu\nu}_{(2)}=&\int_{-1}^1\text{d}x \sum_{q}\left(\mathscr {T}^{\mu\nu}_{(2)} C^{q[-]}_{(0)}(x,\xi) n^\rho W^{\left[\gamma_\rho\right]}(x,\xi,t)  +\widetilde{\mathscr {T}}^{\mu\nu}_{(2)}C^{q[+]}_{(0)}(x,\xi) n^\rho W^{\left[\gamma_\rho\gamma^5\right]}(x,\xi,t)\right) \ ,\\
              T^{\mu\nu}_{(3)}=&\int_{-1}^1\text{d}x \sum_{q}\left(\mathscr {T}^{\mu\nu,\rho}_{(3)} C^{q[-]}_{(0)}(x,\xi)  W^{\left[\gamma_\rho^\perp\right]}(x,\xi,t)  +\widetilde{\mathscr {T}}^{\mu\nu,\rho}_{(3)}C^{q[+]}_{(0)}(x,\xi) W^{\left[\gamma_\rho^\perp\gamma^5\right]}(x,\xi,t)\right)\ ,
\end{split}
\end{align}
where $W^{\left[\Gamma\right]}$ are GPDs defined as
\begin{align}
\begin{split}
   W^{\left[\Gamma\right]}\equiv \int \frac{\text{d}\lambda}{2\pi} e^{i\lambda x} \bra{P',S'} \bar \psi\left(-\frac{\lambda n}{2}\right) \Gamma \psi\left(\frac{\lambda n}{2}\right)\ket{P,S}\ ,
\end{split}
\end{align}
with $\Gamma$ a certain combination of Dirac matrices: $\Gamma=\{1,\gamma^\mu,\sigma^{\mu\nu},\gamma^\mu\gamma^5,\gamma^5\}$ and $\sigma^{\mu\nu} \equiv \frac{i}{2} \left[\gamma^\mu,\gamma^\nu\right]$. The tree level Wilson coefficient functions $C^{q[\pm]}_{(0)}$ read~\cite{Ji:1996nm,Belitsky:2001ns},
\begin{align}
\begin{split}
C^{q[\pm]}_{(0)}=-Q_q^2\left(\frac{1}{x-\xi+i0}\mp \frac{1}{x+\xi-i0}\right) \ ,
\end{split}
\end{align}
with $Q_q$ the charge of quarks in the unit of proton charge. Then the Compton tensor can be expressed in terms of those Compton tensor coefficients $\mathscr {T}^{\mu\nu}$s, which are perturbatively calculable, and the result reads~\cite{Belitsky:2005qn},
\begin{align}
\label{eq:LFcompton}
\begin{split}
\mathscr {T}^{\mu\nu}_{(2)} &=-\frac{1}{2}\left[g^{\mu\nu}_\perp- \frac{1}{p \cdot \bar q}\left(p^\mu {q'}^\nu_\perp+q^\mu_\perp p^\nu\right) \right]\ ,\\
\widetilde{\mathscr {T}}^{\mu\nu}_{(2)} &=\frac{i}{2}\left[\epsilon^{\mu\nu}_\perp- \frac{1}{p \cdot \bar q}\left(-p^\mu \epsilon^{\nu\rho}_\perp {q'}_\rho^\perp+\epsilon^{\nu\rho}_\perp q_\rho^\perp p^\nu\right) \right]\ ,\\
\mathscr {T}^{\mu\nu,\rho}_{(3)} &=\frac{1}{2 p \cdot \bar q}\left[{q'}^\mu g_\perp^{\nu\rho}+g_\perp^{\mu\rho}\left(q^\nu+4\xi p^\nu\right)\right]\ ,\\
\widetilde{\mathscr {T}}^{\mu\nu,\rho}_{(3)} &=\frac{i}{2 p \cdot \bar q}\left[\epsilon^{\mu\nu\rho\sigma}\bar q_\sigma+\xi \left(p^\mu \epsilon_\perp^{\rho\nu}+p^\nu \epsilon_\perp^{\rho\mu}\right)\right]\ .
\end{split}
\end{align}
Note that in order to perform the twist expansion, we introduced the light-cone vector $n$ (mass dimension $-1$) and its conjugate light-cone vector $p$ (mass dimension +1) that satisfies $n^2=0,\ p^2=0$ and $n\cdot p=1$, such that all four-vector $V^\mu$ can be written as
\begin{equation}
    V^\mu = V^+ p^\mu+V^- n^\mu +V_\perp^\mu\ ,
\end{equation}
where $V^+\equiv (V\cdot n)$, $V^-\equiv (V\cdot p)$ and $V_\perp^\mu$ are the remaining transverse components. The skewness parameter $\xi$ is defined as $\xi\equiv -\Delta \cdot n/(2\bar P \cdot n)$. In addition, two transverse tensors can be defined in terms of the light-cone vectors,
\begin{align}
\label{eq:gperpdef}
\begin{split}
   g^{\mu\nu}_\perp\equiv g^{\mu\nu}-p^\mu n^\nu -n^\mu p^\nu\ , \qquad  \epsilon^{\mu\nu}_\perp\equiv \epsilon^{\mu\nu\rho\sigma} p_\rho n_\sigma \ .
\end{split}
\end{align}
and the above Compton tensor coefficients $\mathscr {T}^{\mu\nu}$s are written with those light vectors with twist-three accuracy.

However, the expressions of the above Compton tensor coefficients $\mathscr {T}^{\mu\nu}$s are not unique. Since the Compton tensor is evaluated from the handbag diagrams expanded to twist-three accuracy, all expressions with the same twist-three behavior are equally acceptable. This ambiguity of the Compton tensor leads to the light-cone dependence of the Compton tensor, as was discussed in our previous work \cite{Guo:2021gru}. Motivated by our observation and similar arguments in refs. \cite{Radyushkin:2000ap,Braun:2011dg,Braun:2011zr,Braun:2012bg,Braun:2012hq,Braun:2014sta}, we find the so-called covariant Compton tensor coefficients most suitable for our analysis, as they take parts of the kinematical corrections into account and satisfy the current conservation relation manifestly, see more details in ref. \cite{Guo:2021gru} for how we get those covariant Compton tensor coefficients and ref. \cite{Braun:2014sta} where the same choice of twist-two Compton tensor coefficients is made. Therefore, we promote the above light-cone Compton tensor coefficients to the covariant Compton tensor coefficients and define
\begin{align}
\begin{split}
\label{eq:covtensor}
\widetilde{g}^{\mu\nu}\equiv g^{\mu\nu}-\frac{q^\mu q'^\nu+q^\nu q'^\mu}{q\cdot q'}+\frac{q'^\mu q'^\nu q^2}{(q\cdot q')} \ ,
\widetilde{\epsilon}^{\mu\nu} &=\frac{1}{(q\cdot q')}\epsilon^{\mu\nu q q'}\ ,
\end{split}
\end{align}
as well as the covariant Compton tensor coefficients,
\begin{align}
\begin{split}
\label{eq:covCompton}
\mathscr {T}^{\mu\nu}_{(2),\rm{C}} =-\frac{1}{2}\widetilde{g}^{\mu\nu} \ ,&\qquad
\widetilde{\mathscr {T}}^{\mu\nu}_{(2),\rm{C}} =\frac{i}{2}\widetilde{\epsilon}^{\mu\nu}\ ,\\
\mathscr {T}^{\mu\nu,\rho}_{(3),\rm{C}} =-\frac{\left(q^\nu-\frac{q^2}{q\cdot q'} q'^\nu\right)}{2\bar P\cdot q} \widetilde{g}^{\mu\rho}\ ,&\qquad
\widetilde{\mathscr {T}}^{\mu\nu,\rho}_{(3),\rm{C}} =i\frac{\left(q^\nu-\frac{q^2}{q\cdot q'} q'^\nu\right)}{2\bar P\cdot q} \widetilde{\epsilon}^{\mu\rho}\ .
\end{split}
\end{align}
We emphasize again the covariant Compton tensor coefficients are equal to the light-cone ones with twist-three accuracy, and they are chosen based on the observations of higher-order kinematical effects. We also mention another two approximate relations,
\begin{align}
\label{eq:epsapprox}
\begin{split}
\widetilde{\epsilon}^{\mu\nu} \approx -\frac{1}{n\cdot \Delta}\epsilon^{\mu\nu n\Delta} + \mathcal{O}\left(\frac{M^2}{Q^2}\right)\ ,
\end{split}
\end{align}
and
\begin{align}
\label{eq:PTapprox}
\begin{split}
\widetilde{g}^{\mu\nu}\bar P_\mu \bar P_\nu \approx -\frac{-4M^2\xi^2+(\xi^2-1)t}{4\xi^2}\ ,
\end{split}
\end{align}
which become exact if one defines the light cone with the two photon momenta. These relations lead to important simplifications to the cross-section formula, and the differences resulting from them will be twist-five suppressed for the analysis of twist-three CFFs.

Besides the Compton tensor coefficients, explicit definitions of GPDs are needed to express the Compton tensor. For twist-two GPDs, we use \cite{Ji:1996nm,Meissner:2009ww},
\begin{align}
    W^{\left[\gamma^+\right]}&=\bar u(P',S')\left[
   \gamma^+ H(x,\xi,t)+\frac{i\sigma^{+\nu} \Delta_\nu}{2M} E(x,\xi,t)\right] u(P,S) \ ,\\
   W^{\left[\gamma^+\gamma^5\right]}&=\bar u(P',S')\left[
   \gamma^+ \gamma^5 \widetilde H(x,\xi,t)+\frac{ \Delta^+ \gamma^5 }{2M} \widetilde E(x,\xi,t)\right] u(P,S)\ ,
\end{align}
and for the twist-three GPDs, we use \cite{Meissner:2009ww}
\begin{equation}
\label{eq:Meiztw3v}
\begin{split}
    W^{\left[\gamma^j\right]} &= \frac{M}{\bar{P}^+} \bar u(P',S')\Bigg{[} i\sigma^{+j}H_{2T}(x,\xi,t) + \frac{\gamma^+\Delta^j-\Delta^+\gamma^j}{2M}E_{2T}(x,\xi,t) \\
   & \qquad\qquad\quad+\frac{\bar{P}^+\Delta^j-\Delta^+\bar{P}^j}{M^2}\widetilde{H}_{2T}(x,\xi,t) + \frac{\gamma^+\bar{P}^j-\bar{P}^+\gamma^j}{M}\widetilde{E}_{2T}(x,\xi,t) \Bigg{]}u(P,S)\ ,
\end{split}
\end{equation}
and
\begin{equation}
\label{eq:Meiztw3a}
\begin{split}
    W^{\left[\gamma_j\gamma^5\right]} &= \frac{i \widetilde{\epsilon}_{j k}M}{\bar{P}^+} \bar u(P',S')\Bigg{[} i\sigma^{+k}H'_{2T}(x,\xi,t) + \frac{\gamma^+\Delta^k-\Delta^+\gamma^k}{2M}E'_{2T}(x,\xi,t) \\
   & \qquad\qquad\quad+\frac{\bar{P}^+\Delta^k-\Delta^+\bar{P}^k}{M^2}\widetilde{H}'_{2T}(x,\xi,t) + \frac{\gamma^+\bar{P}^k-\bar{P}^+\gamma^k}{M}\widetilde{E}'_{2T}(x,\xi,t) \Bigg{]}u(P,S)\ .
\end{split}
\end{equation}
It is worth noting that our definition of the matrix elements of $W^{\left[\gamma_j\gamma^5\right]}$ are slightly different from the original ones in ref. \cite{Meissner:2009ww} due to the usage of $\widetilde{\epsilon}^{\mu\nu}$ instead of the ${\epsilon}^{\mu\nu}_{\perp}$ there. The two definitions can be matched if one chooses the light cone according to the two photon momenta. The corresponding CFFs can then be defined with those GPDs as,
\begin{align}
   \mathcal F (\xi,t) \equiv \int_{-1}^1 \text{d}x C^{q[-]}_{(0)}(x,\xi) F (x,\xi,t)\ ,\\
  \widetilde{ \mathcal F} (\xi,t) \equiv \int_{-1}^1 \text{d}x C^{q[+]}_{(0)}(x,\xi) \widetilde F (x,\xi,t)\ ,
\end{align}
with $F=\{H$, $E$, $H_{2T}$, $E_{2T}$, $\widetilde H_{2T}$, $\widetilde E_{2T}\}$ and $\widetilde F=\{\widetilde H$, $\widetilde E$, $H'_{2T}$, $E'_{2T}$, $\widetilde H'_{2T}$, $\widetilde E'_{2T}\}$ which are GPDs of different parities, and $\mathcal F=\{\mathcal H$, $ \mathcal E$, $\mathcal{H}_{2T}$, $\mathcal{E}_{2T}$, $\widetilde{ \mathcal{H}}_{2T}$, $\widetilde{\mathcal {E}}_{2T}\}$ and $\widetilde {\mathcal{F}}=\{\widetilde{\mathcal{ H}}$, $\widetilde{ \mathcal{E}}$, $\mathcal{H}'_{2T}$, $\mathcal{E}'_{2T}$, $\widetilde {\mathcal{H}}'_{2T}$, $\widetilde {\mathcal{E}}'_{2T}\}$ which are their corresponding CFFs. It is worth noting that eq. (\ref{eq:covCompton}) indicates that the two sets of twist-three GPDs always enter the cross-section in the form of 
\begin{align}
\label{eq:t3GPDcomb}
 \widetilde{ g}^{\mu\rho}\int_{-1}^1 \text{d}x C^{q[-]}_{(0)}(x,\xi) W^{\left[\gamma_\rho\right]}(x,\xi,t)-i\widetilde{\epsilon}^{\mu\rho}\int_{-1}^1 \text{d}x C^{q[+]}_{(0)}(x,\xi) W^{\left[\gamma_\rho\gamma^5\right]}(x,\xi,t)\ .
\end{align}
Together with the GPDs defined in eq. (\ref{eq:Meiztw3v}) and eq. (\ref{eq:Meiztw3a}), one immediately find that the twist-three CFFs can only enter the cross-section formulas in the form of 
\begin{align}
\label{eq:CFFbar}
  \bar{ \mathcal {H}}_{2T}(\xi,t)\equiv  \mathcal {H}_{2T}(\xi,t)-\mathcal {H}'_{2T}(\xi,t)
\end{align}
and similarly for the other three combinations of twist-three CFFs, which are noticed in refs. \cite{Belitsky:2005qn,Aslan:2018zzk}. The degeneracy shows up at the level of Compton tensor, making it impossible to identify all the 8 twist-three CFFs associated with the 8 twist-three GPDs in the twist-three DVCS cross-section at leading order of $\alpha_S$. Instead, only 4 combinations of these 8 CFFs enter the DVCS amplitude and consequently the cross-section formulas.  On the other hand, this degeneracy simplifies the cross-section formulas significantly, as we could drop all the $W^{\left[\gamma^j\gamma^5\right]}(x,\xi,t)$ terms in the Compton tensor and substitute $\mathcal {H}_{2T}(\xi,t)$ with $\bar{ \mathcal {H}}_{2T}(\xi,t)$ defined above, and similarly for the other three combinations.

\subsection{Twist-three pure DVCS cross-section}

\label{subsec:puredvcs}
Using the above Compton tensor coefficients and CFFs, the twist-three cross-sections can be calculated. The squared DVCS amplitude $\left| {\mathcal T}_{\rm{DVCS}}\right|^2$  in eq. (\ref{eq:sqrtamp}) can be split into the leptonic and hadronic parts as,
\begin{align}
    \left| {\mathcal T}_{\rm{DVCS}}\right|^2= \frac{1}{Q^4}L_{\rm{DVCS}}^{\rho\sigma} H^{\rm{DVCS}}_{\rho\sigma} \ ,
\end{align}
where we have
\begin{align}
       L_{\rm{DVCS}}^{\rho\sigma}\equiv&\sum_{h'}\bar u(k,h)\gamma^\rho u(k',h')  \bar u(k',h') \gamma^\sigma u(k,h) \ ,\\
   \label{eq:hdvcs}
   H_{\rm{DVCS}}^{\rho\sigma}\equiv & \sum_{S',\Lambda'} T^{* \mu\rho} T^{ \nu\sigma} \varepsilon_\mu(q',\Lambda')\varepsilon^*_{\nu}(q',\Lambda') \ .
\end{align}
The leptonic tensor $L_{\rm{DVCS}}^{\rho\sigma}$ can always be expressed in terms of the unpolarized part and the polarized part where the $h$-dependence is made explicit as,
\begin{align}
\label{eq:ldvcspol}
\begin{split}
   L_{\rm{DVCS}}^{\rho\sigma}= L_{\rm{DVCS,U}}^{\rho\sigma}+ i2h L_{\rm{DVCS,L}}^{\rho\sigma}\ ,
\end{split}
\end{align}
where each term is \cite{Ji:1996nm},
\begin{align}
   L_{\rm{DVCS,U}}^{\rho\sigma}&= 2\left(k^\rho k'^\sigma+k'^\rho k^\sigma-g^{\rho\sigma}k\cdot k' \right)\ ,\\
   L_{\rm{DVCS,L}}^{\rho\sigma}&= 2 \epsilon^{\rho\sigma\alpha\beta} k_\alpha k'_\beta \ ,
\end{align}
whereas hadronic matrix element of the pure DVCS cross-section can be written with the polarization vectors introduced in ref. \cite{Guo:2021gru},
\begin{equation}
\label{eq:hdvcspol}
    \begin{aligned}
         H_{\rm{DVCS}}^{\rho\sigma}= &H_{\rm{DVCS,U}}^{\rho\sigma} + (2 \Lambda_L)   H_{\rm{DVCS,L}}^{\rho\sigma}\\
         &+2\Lambda_T \left[H_{\rm{DVCS,T,in}}^{\rho\sigma}\cos\left(\phi_S-\phi\right) + H_{\rm{DVCS,T,out}}^{\rho\sigma}\sin\left(\phi_S-\phi\right)\right]\ ,
    \end{aligned}
\end{equation}
where we defined
\begin{align}
 H_{\rm{DVCS,L}}^{\rho\sigma}&\equiv \frac{1}{2}\Big[ H_{\rm{DVCS}}^{\rho\sigma}(S_{\rm{L}})-H_{\rm{DVCS}}^{\rho\sigma}(-S_{\rm{L}})\Big]\ ,\\
  H_{\rm{DVCS,T,in}}^{\rho\sigma}&\equiv \frac{1}{2}\Big[ H_{\rm{DVCS}}^{\rho\sigma}(S_{\rm{T,in}})-H_{\rm{DVCS}}^{\rho\sigma}(-S_{\rm{T,in}})\Big]\ ,\\
  H_{\rm{DVCS,T,out}}^{\rho\sigma}&\equiv \frac{1}{2}\Big[ H_{\rm{DVCS}}^{\rho\sigma}(S_{\rm{T,out}})-H_{\rm{DVCS}}^{\rho\sigma}(-S_{\rm{T,out}})\Big]\ .
\end{align}
 The hadronic matrix element includes both twist-two and twist-three CFFs, and thus it can be split into three parts as
\begin{align}
\label{eq:twistdecdvcs}
 H_{\rm{DVCS}}^{\rho\sigma} =H_{\rm{DVCS},(2)}^{\rho\sigma}+H_{\rm{DVCS},(3)}^{\rho\sigma}+H_{\rm{DVCS},(4)}^{\rho\sigma}
\end{align}
such that they consist of twist-two CFFs only, both twist-two and twist-three CFFs, and twist-three CFFs only, respectively. The same twist decomposition applies to each unpolarized/polarized hadronic tensor on the right-hand side of eq. (\ref{eq:hdvcspol}). As the twist-two pieces are already presented in the previous work in ref. \cite{Guo:2021gru}, here we will focus on the twist-three and twist-four hadronic tensor of each polarization, see eqs. (\ref{eq:dvcsut3}) -- (\ref{eq:dvcsTout4}) in appendix \ref{app:dvcsstructurefunc}. The following tensor structures that emerge in the hadronic matrix element can be defined,
\begin{equation}
\label{eq:h3ampdef}
    \begin{aligned}
   {\mathscr H}^{\rho\sigma}_{(3)}&\equiv  -g_{\mu\nu} \mathscr{T}^{\mu\rho}_{(2)} \mathscr{T}^{\nu\sigma,\gamma}_{(3)} \left(2\xi \bar P_\gamma\right) \ ,&  \widetilde{{\mathscr H}}^{\rho\sigma}_{(3)}&\equiv i g_{\mu\nu} \widetilde{\mathscr{T}}^{\mu\rho}_{(2)} \mathscr{T}^{\nu\sigma,\gamma}_{(3)} \epsilon_{\gamma n \bar P \Delta} \ ,\\
   {\mathscr H'}^{\rho\sigma}_{(3)}&\equiv  ig_{\mu\nu} \widetilde{\mathscr{T}}^{\mu\rho}_{(2)} \mathscr{T}^{\nu\sigma,\gamma}_{(3)} \left(2\xi \bar P_\gamma\right) \ ,&  \widetilde{{\mathscr H'}}^{\rho\sigma}_{(3)}&\equiv g_{\mu\nu} \mathscr{T}^{\mu\rho}_{(2)} \mathscr{T}^{\nu\sigma,\gamma}_{(3)} \epsilon_{\gamma n\bar P \Delta} \ ,\\
   \mathscr{H}^{\rho\sigma}_{(4)}&\equiv \frac{M^2}{\left(2\bar P\cdot q\right)^2}\left(q^\rho-\frac{q^2}{q\cdot q'} q'^\rho\right)\left(q^\sigma-\frac{q^2}{q\cdot q'} q'^\sigma\right) \ , 
\end{aligned}
\end{equation}
The above tensors can be further simplified with the help of eq. (\ref{eq:epsapprox}), and get
\begin{equation}
\label{eq:h3ampsim}
\begin{aligned}
    &{\mathscr H}^{\rho\sigma}_{(3)}= \widetilde{{\mathscr H}}^{\rho\sigma}_{(3)}= -\frac{\xi \widetilde{g}^{\rho\nu} \bar P_\nu}{2\bar P\cdot q}\left(q^\sigma-\frac{q^2}{q\cdot q'} q'^\sigma\right) \\
    &{\mathscr H'}^{\rho\sigma}_{(3)}= \widetilde{{\mathscr H'}}^{\rho\sigma}_{(3)}= -\frac{\xi \widetilde{\epsilon}^{\rho\nu} \bar P_\nu}{2\bar P\cdot q}\left(q^\sigma-\frac{q^2}{q\cdot q'} q'^\sigma\right)
\end{aligned}
\end{equation}
The last step is to contract our hadronic part in eq. (\ref{eq:hdvcspol}) with the leptonic part in eq. (\ref{eq:ldvcspol}), and we have
\begin{equation}
\label{eq:amp2dvcs}
    \begin{aligned}
         \left|\mathcal T_{\rm{DVCS}}\right|^2=&\frac{1}{Q^4} \Bigg\{F_{\rm{UU}}+(2\Lambda_L) F_{\rm{UL}}+(2\Lambda_T)\left(\cos\left(\phi_S-\phi\right) F_{\rm{U T,in}}+\sin\left(\phi_S-\phi\right)F_{\rm{UT,out}}\right)\\
         +&(2h)\Big[F_{\rm{LU}}+(2\Lambda_L) F_{\rm{LL}}+(2\Lambda_T) \left(\cos\left(\phi_S-\phi\right) F_{\rm{LT,in}}+\sin\left(\phi_S-\phi\right)F_{\rm{LT,out}}\right)\Big] \Bigg\}\ ,
    \end{aligned}
\end{equation}
where we define
\begin{equation}
\label{eq:dvcsstruct}
    \begin{aligned}
        F_{\rm{UU}} &\equiv  L^{\rm{DVCS,U}}_{\rho\sigma}H_{\rm{DVCS,U}}^{\rho\sigma}\ ,&\qquad  F_{\rm{LU}} &\equiv  iL^{\rm{DVCS,L}}_{\rho\sigma}H_{\rm{DVCS,U}}^{\rho\sigma} \ ,\\
        F_{\rm{UL}} &\equiv  L^{\rm{DVCS,U}}_{\rho\sigma} H_{\rm{DVCS,L}}^{\rho\sigma}\ ,&\qquad F_{\rm{LL}} &\equiv  i L^{\rm{DVCS,L}}_{\rho\sigma} H_{\rm{DVCS,L}}^{\rho\sigma}\ ,\\
         F_{\rm{UT,in}} &\equiv   L^{\rm{DVCS,U}}_{\rho\sigma}H_{\rm{DVCS,T,in}}^{\rho\sigma}\ ,&F_{\rm{LT,in}} &\equiv  i L^{\rm{DVCS,L}}_{\rho\sigma}H_{\rm{DVCS,T,in}}^{\rho\sigma}\ ,\\
         F_{\rm{UT,out}} &\equiv  L^{\rm{DVCS,U}}_{\rho\sigma} H_{\rm{DVCS,T,out}}^{\rho\sigma}\ ,&\qquad  F_{\rm{LT,out}} &\equiv  i L^{\rm{DVCS,L}}_{\rho\sigma} H_{\rm{DVCS,T,out}}^{\rho\sigma}\ ,
    \end{aligned}
\end{equation}
analogous to those in refs. \cite{Kriesten:2019jep, Guo:2021gru}. Note that with eq. (\ref{eq:twistdecdvcs}), the structure functions $F_{\rm{P}}$s can be written into three parts as well,
\begin{align}
\label{eq:twistdecdvcsstruct}
 F_{\rm{P}} =F_{\rm{P}}^{(2)}+F_{\rm{P}}^{(3)}+F_{\rm{P}}^{(4)}\ ,
\end{align}
with $\rm{P} = \{\rm{UU},\rm{LU},\rm{UL},\rm{LL},\rm{UT,in},\rm{LT,in},\rm{UT,out},\rm{LT,out}\}$ different polarization configurations. Again, as the twist-two parts $F_{\rm{P}}^{(2)}$s are given in the previous work \cite{Guo:2021gru}, we focus on the structure functions related to higher-twist CFFs, which are given in eqs. (\ref{eq:fuu3}) -- (\ref{eq:flu4}), and the following scalar coefficients are defined by contracting the hadronic tensor with the leptonic tensor,
\begin{equation}
\label{eq:dvcsscalaramp}
    \begin{aligned}
   h^{\rm{U}}_{(3)}\equiv L^{\rm{DVCS,U}}_{\rho\sigma} {\mathscr H}^{\rho\sigma}_{(3)} \  ,&\qquad h^{\rm{L}}_{(3)}\equiv L^{\rm{DVCS,L}}_{\rho\sigma} {\mathscr H}^{\rho\sigma}_{(3)} \\
    h'^{\rm{U}}_{(3)}\equiv L^{\rm{DVCS,U}}_{\rho\sigma} {\mathscr H'}^{\rho\sigma}_{(3)} \  ,&\qquad h'^{\rm{L}}_{(3)}\equiv L^{\rm{DVCS,L}}_{\rho\sigma} {\mathscr H'}^{\rho\sigma}_{(3)}\\
     h^{\rm{U}}_{(4)}\equiv L^{\rm{DVCS,U}}_{\rho\sigma} {\mathscr H}^{\rho\sigma}_{(4)}\ ,& \qquad h^{\rm{L}}_{(4)}=0\ .
\end{aligned}
\end{equation}
Then the structure functions can be written in terms of those scalar coefficients and the CFFs, which are also explicitly presented in appendix \ref{app:dvcsstructurefunc}.

\subsection{Twist-three interference cross-section}
\label{subsec:intxsection}

Similarly, we write the interference squared amplitude as a product of their leptonic and hadronic parts as
\begin{equation}
     \mathcal I  =-\frac{e_l}{Q^2 t} L_{\rm{INT}}^{\mu\rho\sigma} H^{\rm{INT}}_{\mu\rho\sigma} +\text{c.c.}\ .
\end{equation}
where
\begin{align}
\label{eq:lint}
L_{\rm{INT}}^{\mu\rho\sigma}\equiv & \sum_{h'} \bar u(k,h)\gamma^\rho u(k',h') l_{\rm{BH}}^{\mu\sigma}\ ,\\
       \label{eq:hint}
   H_{\rm{INT}}^{\mu\rho\sigma }\equiv&\sum_{S',\Lambda'}  T^{*\nu\rho}\varepsilon_\nu(q',\Lambda')\varepsilon^*_{\mu}(q',\Lambda') \bar u(P',S')\left[(F_1+F_2)\gamma^\sigma-\frac{\bar P^\sigma}{M}F_2\right]u(P,S)  \ ,
\end{align}
and we split the leptonic term into the polarized and unpolarized part as, 
\begin{equation}
    L^{\rm{INT}}_{\mu\rho\sigma}=L^{\rm{INT,U}}_{\mu\rho\sigma}+i 2 h L^{\rm{INT,L}}_{\mu\rho\sigma}\ .
\end{equation}
Following the same polarization decomposition in the last subsection, we can write the hadronic tensor in terms of four different terms as
\begin{equation}
\label{eq:hintpol}
    \begin{aligned}
         H_{\rm{INT}}^{\mu\rho\sigma}= &H_{\rm{INT,U}}^{\mu\rho\sigma} + (2\Lambda_L)   H_{\rm{INT,L}}^{\mu\rho\sigma}+2\Lambda_T  \left[H_{\rm{INT,T,in}}^{\mu\rho\sigma}\cos\left(\phi_S-\phi\right) + H_{\rm{INT,T,out}}^{\mu\rho\sigma}\sin\left(\phi_S-\phi\right)\right]\ ,
    \end{aligned}
\end{equation}
see ref. \cite{Guo:2021gru} for how those polarizations are defined. For the interference cross-section, the hadronic tensor consists of only two parts,
\begin{equation}
\label{eq:hintdecom}
    \begin{aligned}
         H_{\rm{INT}}^{\mu\rho\sigma}=H_{\rm{INT},(2)}^{\mu\rho\sigma}+H_{\rm{INT},(3)}^{\mu\rho\sigma}\ ,
    \end{aligned}
\end{equation}
which involve twist-two CFFs and twist-three CFFs, respectively. As the twist-two pieces are presented in the previous work, here we present the twist-three part, more details are given in appendix \ref{app:structurefunc}. Again, we have eight different polarization configurations as,
\begin{align}
\begin{split}
    \mathcal I=\frac{-e_l}{Q^2 t}\Bigg\{&F^{{\rm{I}}}_{\rm{UU}}+2\Lambda_L F^{{\rm{I}}}_{\rm{UL}}+2\Lambda_T \left(F^{{\rm{I}}}_{\rm{UT,in}}\cos(\phi_S-\phi)+F^{{\rm{I}}}_{\rm{UT,out}}\sin(\phi_S-\phi)\right)\\
    &+2h \Big[F^{{\rm{I}}}_{\rm{LU}}+2\Lambda_L F^{{\rm{I}}}_{\rm{LL}}+2\Lambda_T \left(F^{{\rm{I}}}_{\rm{LT,in}}\cos(\phi_S-\phi)+F^{{\rm{I}}}_{\rm{LT,out}}\sin(\phi_S-\phi)\right)\Big]\Bigg\}\ ,
\end{split} 
\end{align}
where those eight polarized cross-sections can be expressed as,
\begin{equation}
\label{eq:intxsecdef}
    \begin{aligned}
        F^{\rm{I}}_{\rm{UU}} &\equiv  L^{\rm{INT,U}}_{\mu\rho\sigma}H_{\rm{INT,U}}^{\mu\rho\sigma}+\text{c.c.}\ ,& F^{\rm{I}}_{\rm{LU}} &\equiv  i L^{\rm{INT,L}}_{\mu\rho\sigma}H_{\rm{INT,U}}^{\mu\rho\sigma}+ \text{c.c.}\\
        F^{\rm{I}}_{\rm{UL}} &\equiv  L^{\rm{INT,U}}_{\mu\rho\sigma} H_{\rm{INT,L}}^{\mu\rho\sigma}+\text{c.c.}\ ,&\qquad F^{\rm{I}}_{\rm{LL}} &\equiv  i L^{\rm{INT,L}}_{\mu\rho\sigma} H_{\rm{INT,L}}^{\mu\rho\sigma}+\text{c.c.}\ ,\\
        F^{\rm{I}}_{\rm{UT,in}} &\equiv   L^{\rm{INT,U}}_{\mu\rho\sigma}H_{\rm{INT,T,in}}^{\mu\rho\sigma}+\text{c.c.}\ ,&\qquad F^{\rm{I}}_{\rm{LT,in}} &\equiv  i L^{\rm{INT,L}}_{\mu\rho\sigma}H_{\rm{INT,T,in}}^{\mu\rho\sigma}+\text{c.c.}\ ,\\
       F^{\rm{I}}_{\rm{UT,out}} &\equiv  L^{\rm{INT,U}}_{\mu\rho\sigma} H_{\rm{INT,T,out}}^{\mu\rho\sigma}+\text{c.c.}\ ,& \qquad F^{\rm{I}}_{\rm{LT,out}} &\equiv  i L^{\rm{INT,L}}_{\mu\rho\sigma} H_{\rm{INT,T,out}}^{\mu\rho\sigma}+\text{c.c.}\ .
    \end{aligned}
\end{equation}
analogous to those in refs. \cite{Kriesten:2019jep,Guo:2021gru}. Similar to the case of pure DVCS cross-sections, with the help of eq. (\ref{eq:hintdecom}), we can split the structure functions into two parts as
\begin{equation}
\label{eq:inttwistsep}
    \begin{aligned}
         F^{\rm{I}}_{\rm{P}}= F^{\rm{I}}_{\rm{P},(2)}+ F^{\rm{I}}_{\rm{P},(3)}\ ,
    \end{aligned}
\end{equation}
with $\rm{P}$ different polarization configuration and $F^{\rm{I}}_{\rm{P},(2)}$s are given in our previous work \cite{Guo:2021gru}. The structure functions $F^{\rm{I}}_{\rm{P},(3)}$s that are related to the twist-three CFFs are given in appendix \ref{app:structurefunc}, where we define 6 coefficients $A^{\rm{I}}_{(3)}$, $B^{\rm{I}}_{(3)}$,$C^{\rm{I}}_{(3)}$, $\widetilde A^{\rm{I}}_{(3)}$, $\widetilde B^{\rm{I}}_{(3)}$ and $\widetilde C^{\rm{I}}_{(3)}$ by contracting the hadronic tensor with the leptonic tensor as,
\begin{equation}
\label{eq:intstructfunc}
\begin{aligned}
    A^{\rm{I}}_{(3)}& \equiv8 \xi\bar P^\sigma \bar P_\gamma \mathscr{T}^{\mu\rho,\gamma}_{(3)} L^{\rm{INT}}_{\mu\rho\sigma} \ ,&\qquad \widetilde{A}^{\rm{I}}_{(3)}& \equiv8 i \xi \bar P^\sigma \bar P_\gamma \widetilde{\mathscr{T}}^{\mu\rho,\gamma}_{(3)}  L^{\rm{INT}}_{\mu\rho\sigma} \ ,\\
    B^{\rm{I}}_{(3)} &\equiv 2 t n^\sigma \bar P_\gamma \mathscr{T}^{\mu\rho,\gamma}_{(3)} L^{\rm{INT}}_{\mu\rho\sigma}\ ,& \qquad \widetilde{B}^{\rm{I}}_{(3)} &\equiv 2i t n^\sigma \bar P_\gamma \widetilde{\mathscr{T}}^{\mu\rho,\gamma}_{(3)}    L^{\rm{INT}}_{\mu\rho\sigma}\ ,\\
    C^{\rm{I}}_{(3)}& \equiv 8 M^2 \mathscr{T}^{\mu\rho,\sigma}_{(3)} L^{\rm{INT}}_{\mu\rho\sigma} \ ,&\qquad \widetilde{C}^{\rm{I}}_{(3)}& \equiv 8i M^2 \widetilde{\mathscr{T}}^{\mu\rho,\sigma}_{(3)} L^{\rm{INT}}_{\mu\rho\sigma}  L^{\rm{INT}}_{\mu\rho\sigma}\ .
\end{aligned}
\end{equation}
Each of them can be written as the sum of their unpolarized and polarized parts,
\begin{align}
\begin{split}
    \mathcal A^{\rm{I}}\equiv \mathcal A^{\rm{I,U}} +i 2 h \mathcal A^{\rm{I,L}}\ ,
\end{split}
\end{align}
with $\mathcal A= \{A_{(3)},B_{(3)},C_{(3)},\widetilde A_{(3)},\widetilde B_{(3)},\widetilde C_{(3)}\}$ such that each $\mathcal A^{\rm{I,U/L}}$ is given by the same definition in eq. (\ref{eq:intstructfunc}) but with $L^{\rm{INT}}$ replaced by $ L^{\rm{INT,U}}/ L^{\rm{INT,L}}$ respectively. The cross-section formulas of all polarization configurations can then be expressed as combinations of scalar coefficients and CFFs, where we assume the knowledge of the electromagnetic form factors.

\subsection{Comments on the cross-section formulas}

In the end of the section, we comment on the other cross-section formulas in the literature. We denote one of the most popular formulas as the Belitsky-M\"{u}ller-Kirchner (BMK01) one \cite{Belitsky:2001ns}, which receives higher-order kinematical corrections \cite{Belitsky:2010jw}(BMK10) and then gets refreshed in the Belitsky-M\"{u}ller-Ji (BMJ) formulas \cite{Belitsky:2012ch} to include the transversely polarized target. The latest version of this series used in the fitting of the recent JLab Hall A DVCS measurement \cite{JeffersonLabHallA:2022pnx} is the Braun-Manashov-M\"{u}ller-Pirnay (BMMP) formulas \cite{Braun:2014sta}, where kinematical twist-four effects such as target-mass and finite-$t$ corrections are considered. Another recent work in the literature, denoted VA, is given in ref. \cite{Kriesten:2019jep} which utilizes helicity amplitudes to separate out the different contributions to the cross-section at both leading and sub-leading twist. 

Several differences between our formulas and the BMK01, BMK10, BMJ and BMMP formulas should be noted here. In both the BMK01 and BMK10 formulas, the twist-three parts are given in terms of the so-called effective CFFs which are mixtures of twist-two and twist-three CFFs. In the later BMJ formulas \cite{Belitsky:2012ch}, the mixing of CFFs gets more complicated due to the introduction of the helicity-dependent CFFs, where each of them might involve CFFs of different twist. On the other hand, our results have them separated manifestly. The importance of twist separation has been stressed in our previous work \cite{Guo:2021gru}, where we showed that the effective CFFs defined in refs. \cite{Belitsky:2001ns,Belitsky:2010jw,Belitsky:2012ch} contain twist-two contributions that should be combined with the twist-two CFFs in order to cancel the kinematically twist-three effects. In the BMMP formulas \cite{Braun:2014sta}, such mixing is resolved by choosing a different set of photon polarization vectors, which rotates the helicity-dependent CFFs and disentangles the mixing. This agrees with our argument in ref. \cite{Guo:2021gru} that with the BMMP coordinate choice, such kinematical twist-three effects vanish.

Another difference worth noting is the choice of definitions of twist-three GPDs. Commonly in the literature (as well as in this work), the definitions in refs. \cite{Diehl:2001pm,Meissner:2009ww} are used, where the GPDs are defined according to their different Dirac structures that do not mix the Dirac tensor (without $\gamma^5$) and pseudo-tensor (with $\gamma^5$ and Levi-Civita symbol $\epsilon$ to preserve the parity). With this choice, the physical interpretation of GPDs gets clearer. For instance, the GPD $E_{2T}(x,\xi,t)$ is closely related to the orbital angular momentum of partons \cite{Penttinen:2000dg,Kiptily:2002nx,Aslan:2018zzk,Guo:2021aik}. On the other hand, in the BMK01, BMK10, BMJ and BMMP formulas, another set of twist-three GPDs $H^{3}_{\pm},E^{3}_{\pm},\tilde{H}^{3}_{\pm},\tilde{E}^{3}_{\pm}$ (or ones with equivalent definitions) are defined, resembling the definition of $H,E,\tilde{H},\tilde{E}$. This choice makes the calculation simpler, as it will be similar to the twist-two one, while they make the physical interpretation of GPDs, such as the orbital angular momentum for $E_{2T}$, less obvious.

There are several reasons we perform another individual calculation here. It has been noted that the VA formulas differ from the BMK10 one in ref. \cite{Kriesten:2020wcx}. Our formulas \cite{Guo:2021gru} serves an independent check to resolve the discrepancy, and we find that the difference is mainly caused by an extra $\cos(\phi)$ phase factor in the interference cross-section. Our previous work \cite{Guo:2021gru}, without the extra factor, agrees with the BMK10 results with great precision.

Besides the extra factor, other higher-order differences are associated to the kinematical higher-order effects, including choices of light-cone vectors and choices of gauge fixing condition for the final photon. For instance, the definitions of GPDs (and the corresponding CFFs) in the BMK01, BMK10, BMJ and BMMP formulas are tied to the physical vectors, especially the photon momenta $q,q'$, which do not exist for general applications of GPDs. Instead, the light-cone vectors $n$ and $p$, which could have different definitions from theirs, are used in the universal definitions of GPDs. Therefore, the light-cone dependence must be explicitly studied to avoid the ambiguity, which is done with our more general set-up in the previous work \cite{Guo:2021gru}. This work, including also the effects of twist-three CFFs, then completes the twist-three analysis.

At last, we comment on the quantitative comparison of our formulas to the other formulas, which can be done by choosing specific light-cone vectors according to the choices made in the other formulas. We focus on the comparison to the BMK10 and BMJ formulas for the following reasons. The BMK01 formulas lack kinematical higher order effects and the VA formulas differs with an extra $\cos(\phi)$ phase factor in the interference cross-section. The connections between the BMMP and BMJ formulas are discussed with details in ref. \cite{Braun:2014sta}. Besides the kinematical twist-four corrections added in the BMMP formula, the major difference between them is the redefinition of light-cone vectors in the BMMP formulas, while the cross-section formulas are in essence the same in the BMJ and BMMP formulas.

We also note that the BMJ formulas is closely related to the BMK01 and BMK10 formulas --- the scalar coefficients in the appendices of BMJ are identical to those of the BMK10, and there are no new scalar coefficients in the BMJ formula. Besides, the structures of CFFs in the BMJ formulas reduce to those of the BMK01 up to kinematical twist-four terms. Therefore, the comparison to the BMJ formulas is essentially represented by the comparison to the BMK01 and BMK10 ones, which is done in ref. \cite{Guo:2021gru} in details. Here we collect some of the statements there, and address the comparison to the BMJ formulas as well.

Due to the complexity of converting the different definitions of twist-three GPDs as stated above, it is impractical to put in the same twist-three CFFs and directly compare the outcome twist-three cross-sections for all the formulas. Therefore, the comparison is  separated into two parts: comparing the scalar coefficients numerically and comparing the combinations of CFFs analytically. And we have the following comparisons done.
\begin{itemize}
    \item In ref. \cite{Guo:2021gru}, we compare our twist-two scale coefficients  ${A}^{\rm{I,U}},{B}^{\rm{I,U}},{C}^{\rm{I,U}},\tilde{A}^{\rm{I,U}},\tilde{B}^{\rm{I,U}}$ and $\tilde{C}^{\rm{I,U}}$ for unpolarized and longitudinally-polarized targets with the corresponding BMK10 scalar coefficients numerically, and the results agree perfectly.
    \item In addition, the scalar coefficients associated with the effective CFFs are compared and shown the same. 
    \item Since the BMJ scalar coefficients are identical to the BMK10 ones, our scalar coefficients also agree with the BMJ ones.
    \item At last, we compare our twist-two structure functions $F$s with the corresponding $\mathcal C^{\rm{VCS}}$ in the BMJ formula, and confirm that they have the same combination of CFFs up to twist-four terms.
\end{itemize}
The above comparisons show the consistency of the twist-two cross-sections as well as the twist-three scalar coefficients between ours and both the BMK10 and BMJ formulas, whereas the twist-three CFFs are not directly comparable due to the different definitions of GPDs.

\section{Numerical studies of twist-three scalar coefficients }

\label{sec:t3scalar}
In the previous section, we show that the twist-three cross-section formulas can be expressed in terms of the twist-three scalar coefficients and twist-three CFFs. Thus, comparisons of both twist-three scalar coefficients and CFFs are needed to compare the twist-three cross-sections. Due to the complexity of evaluating the twist-three CFFs, we focus on comparing the scalar coefficients in this section, whereas the twist-three CFFs at different kinematics are considered to be purely numeric inputs, assuming their weak $Q^2$ dependence resulting from radiative corrections.
When the twist-three CFFs are comparable to the twist-two ones, the comparisons of the scalars coefficients are equivalent to the comparisons of the cross-sections. This assumption will be tested in section \ref{sec:t3cff} with WW approximation and inputs of a GPD model.

\subsection{Twist-three pure DVCS scalar coefficients}

\begin{figure}[t]
\centering
\begin{minipage}[b]{\textwidth}
\includegraphics[width=0.49\textwidth]{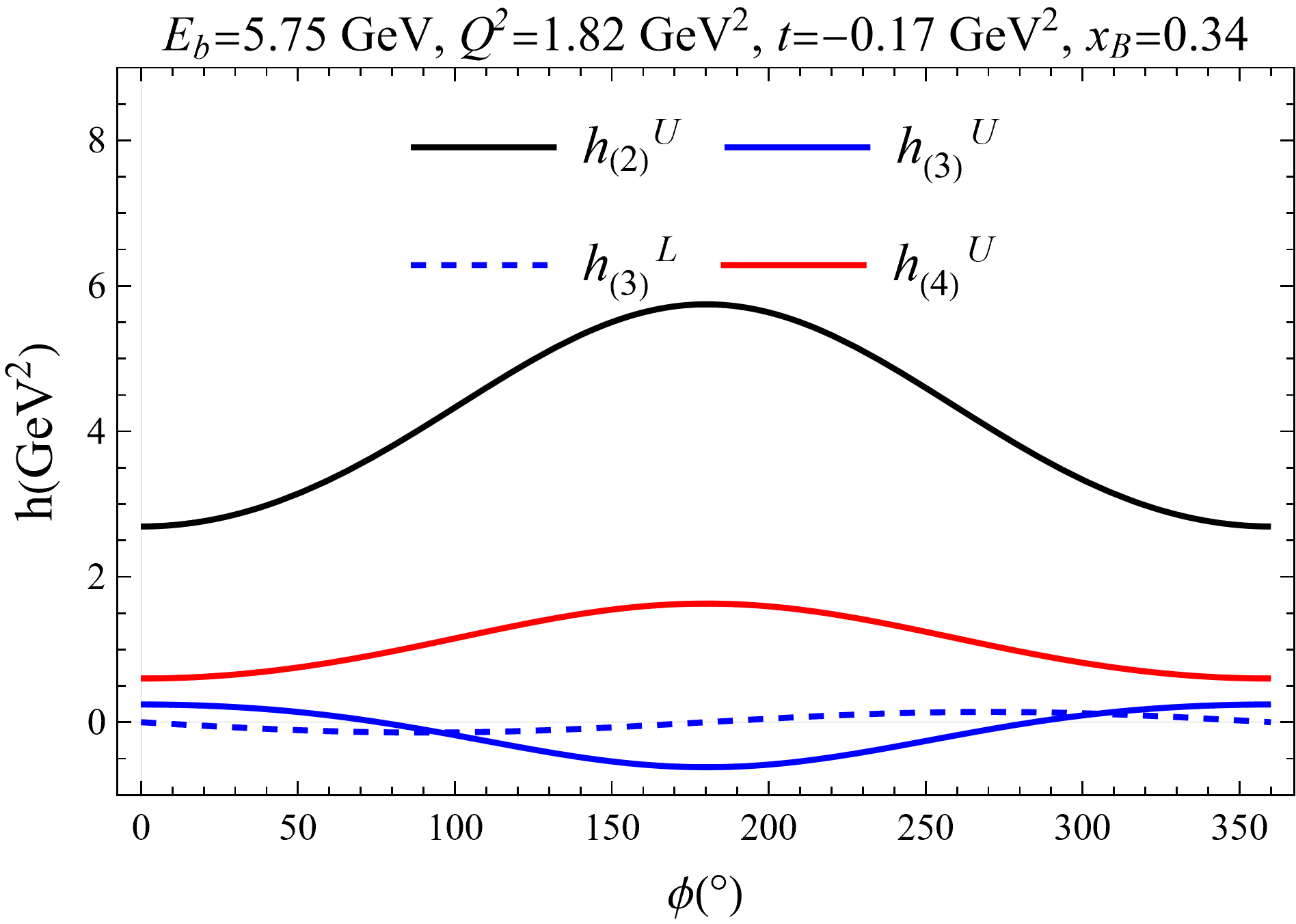}
\includegraphics[width=0.49\textwidth]{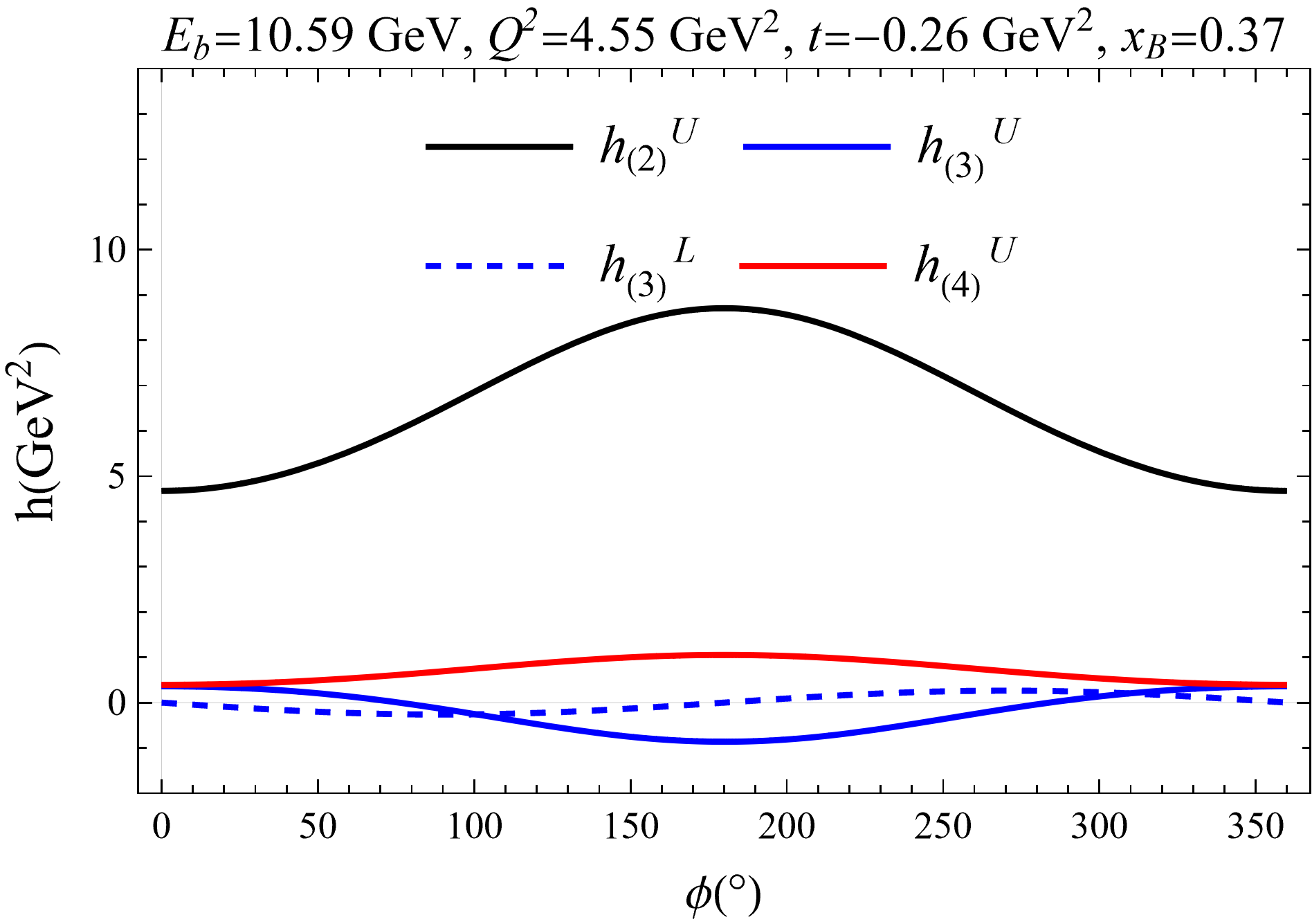}
\end{minipage}
\begin{minipage}[b]{\textwidth}
\includegraphics[width=0.49\textwidth]{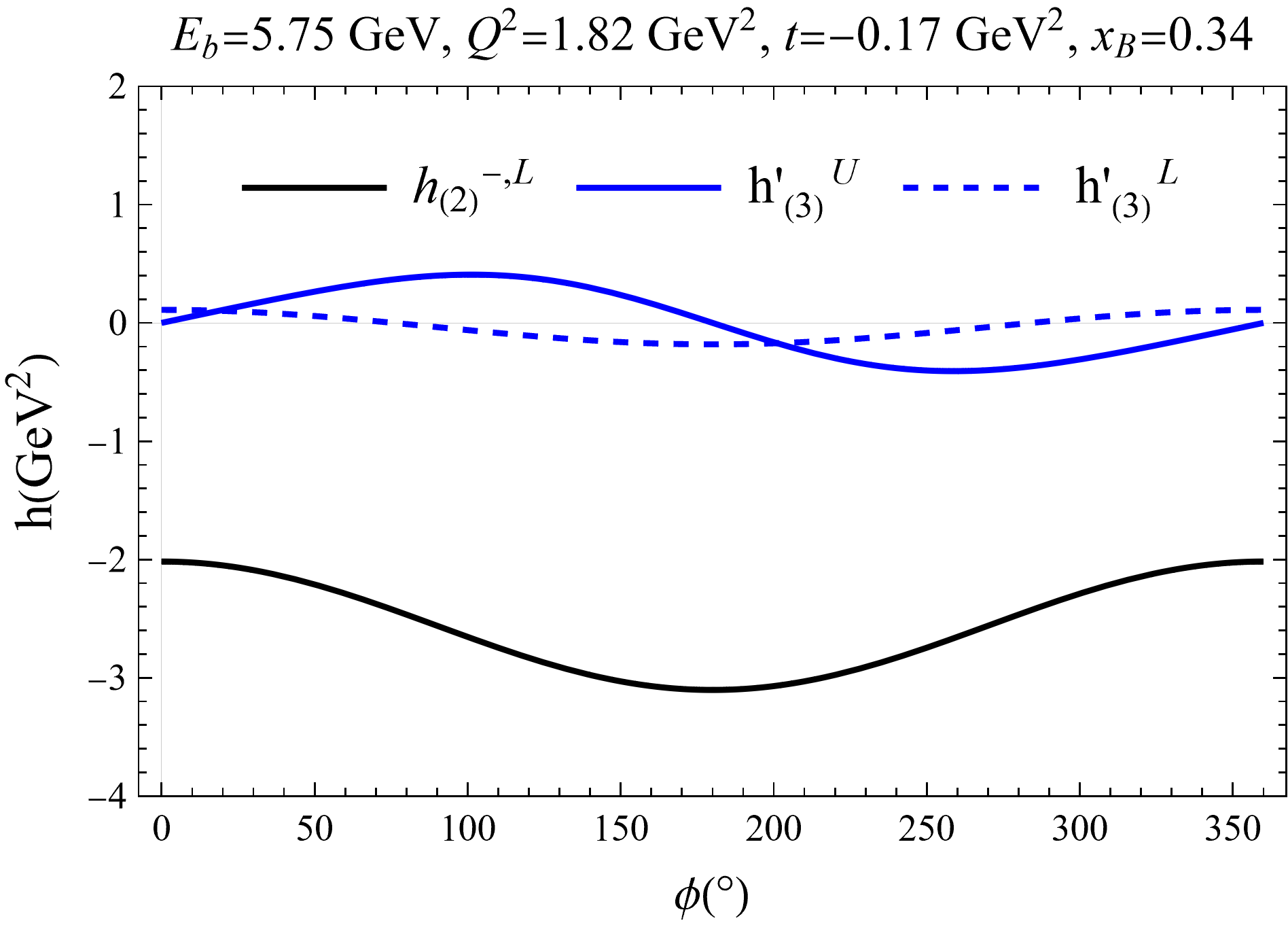}
\includegraphics[width=0.49\textwidth]{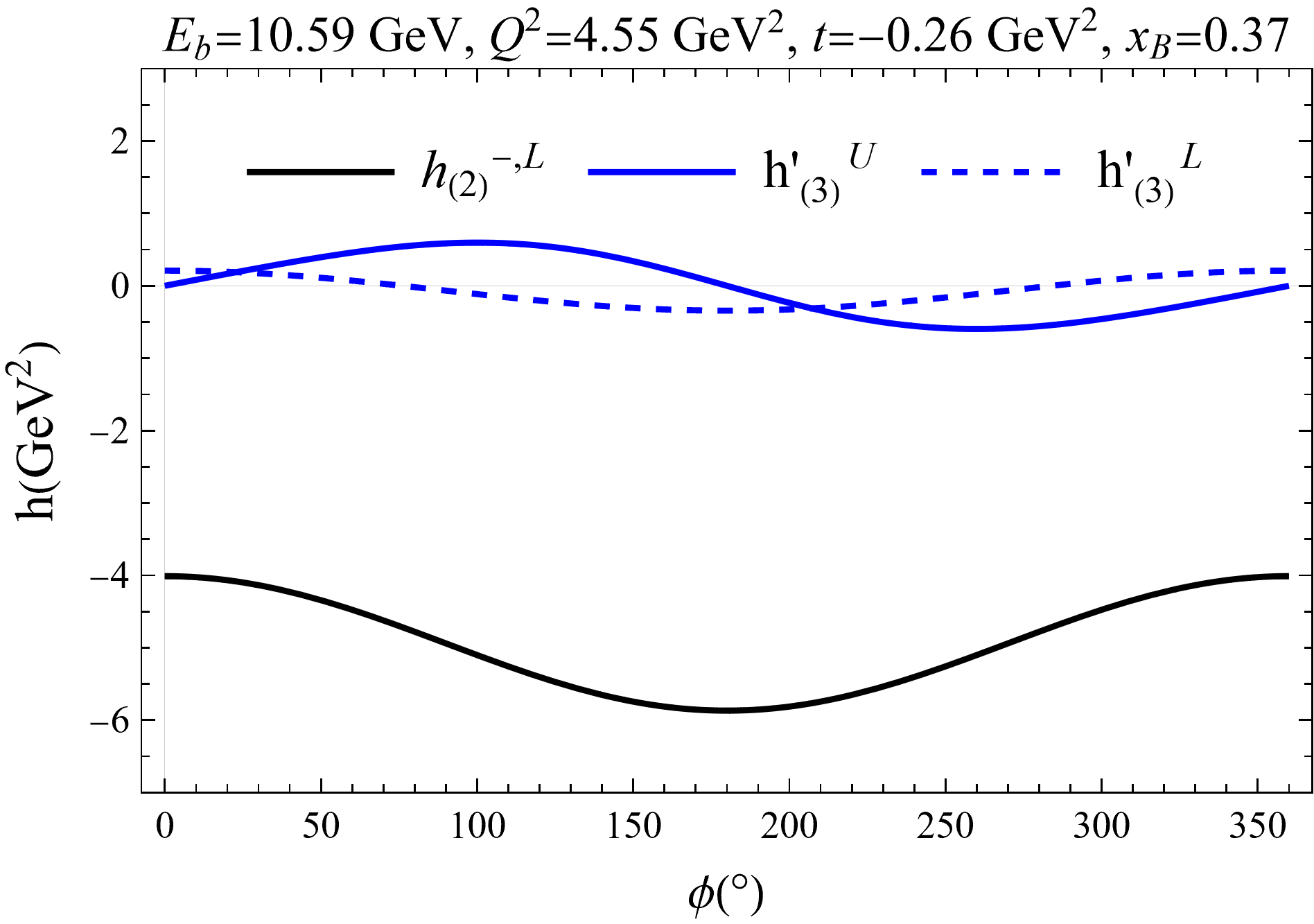}
\end{minipage}
\caption{\label{fig:dvcsscalar} Comparison of the coefficients $h^{\rm{U}}_{(2)}$, $h^{\rm{U/L}}_{(3)}$ and $h^{\rm{U}}_{(4)}$, and $h^{-,\rm{L}}_{(2)}$ and $h'^{\rm{U/L}}_{(3)}$ at typical JLab 6 GeV (left) and 12 GeV (right) kinematics \cite{Georges:2018kyi}. We plot the twist-two coefficients $h^{\rm{U}}_{(2)}$ and $h^{-,\rm{L}}_{(2)}$ as the background for comparison of the higher-twist scalar coefficients. In all the plots, the twist-two scalar coefficients (black lines) dominates the other higher-twist scalar coefficients. }
\end{figure}

In eq. (\ref{eq:dvcsscalaramp}), we define 4 twist-three scalar coefficients and 1 twist-four scalar coefficient for pure DVCS cross-sections, whereas we have 2 independent twist-two scalar coefficients $h^{\rm{U}}$  and $h^{-,\rm{L}}$ (which we rename into $h^{\rm{U}}_{(2)}$ and $h^{-,\rm{L}}_{(2)}$ hereafter for consistency) as defined in ref. \cite{Guo:2021gru}. Each of those scalar coefficients enters the DVCS cross-section formulas with a different combination of the CFFs depending on the polarization configuration. Specifically, the scalar coefficients $h^{\rm{U}}_{(2)}$ , $h^{\rm{U/L}}_{(3)}$ and $h^{\rm{U}}_{(4)}$ contribute to unpolarized target and also out-of-plane transversely polarized target while $h^{-,\rm{L}}_{(2)}$ and $h'^{\rm{U/L}}_{(3)}$ contribute to the longitudinally polarized target and in-plane transversely polarized target. Therefore, we will compare these two sets of scalar coefficients, which are shown in figure \ref{fig:dvcsscalar} with different kinematics chosen respectively.

From the plots in figure \ref{fig:dvcsscalar}, the twist-three scalar coefficients are actually comparable to the twist-four coefficients at the given kinematics, which seems to indicate some extra suppression of twist-three parameters besides the kinematical twist suppression. This results from the fact that the twist-three parameters are associated with an extra factor of $\xi$ which are quite small when $x_{B}$ are small. Consequently, although $\xi$ itself is considered to be an order $\mathcal{O}(1)$ parameter in terms of twist counting, it can suppress the twist-three scalar coefficients and make them comparable to or even smaller than the twist-four scalar coefficients at the given kinematics when $Q^2$ is not sufficiently large while $x_B$ is small. 
As shown by the numerical calculations, for relatively low $Q^2$ of $ 1.82 \text{ GeV}^2$ in the left plots, the twist-four scalar coefficient $h_{(4)}^{\rm{U}}$ is can be more significant than the twist-three scalar coefficients $h_{(4)}^{\rm{U}/\rm{L}}$, whereas the twist suppression gets more relevant as $Q^2$ increases as shown in the right plots with $Q^2= 4.55 \text{ GeV}^2$. The same arguments apply to the interference scalar coefficients, as we will see in the next subsection.
However, this $x_B$ suppression could be compensated by the small $x_B$ behavior of CFFs, as the CFFs might be divergent in the $x_B \to 0$ limit. Therefore, the $x_B$ suppression of scalar coefficients does not necessary indicates the suppression of the cross-sections.

\begin{figure}[t]
\centering
\begin{minipage}[b]{\textwidth}
\includegraphics[width=0.49\textwidth]{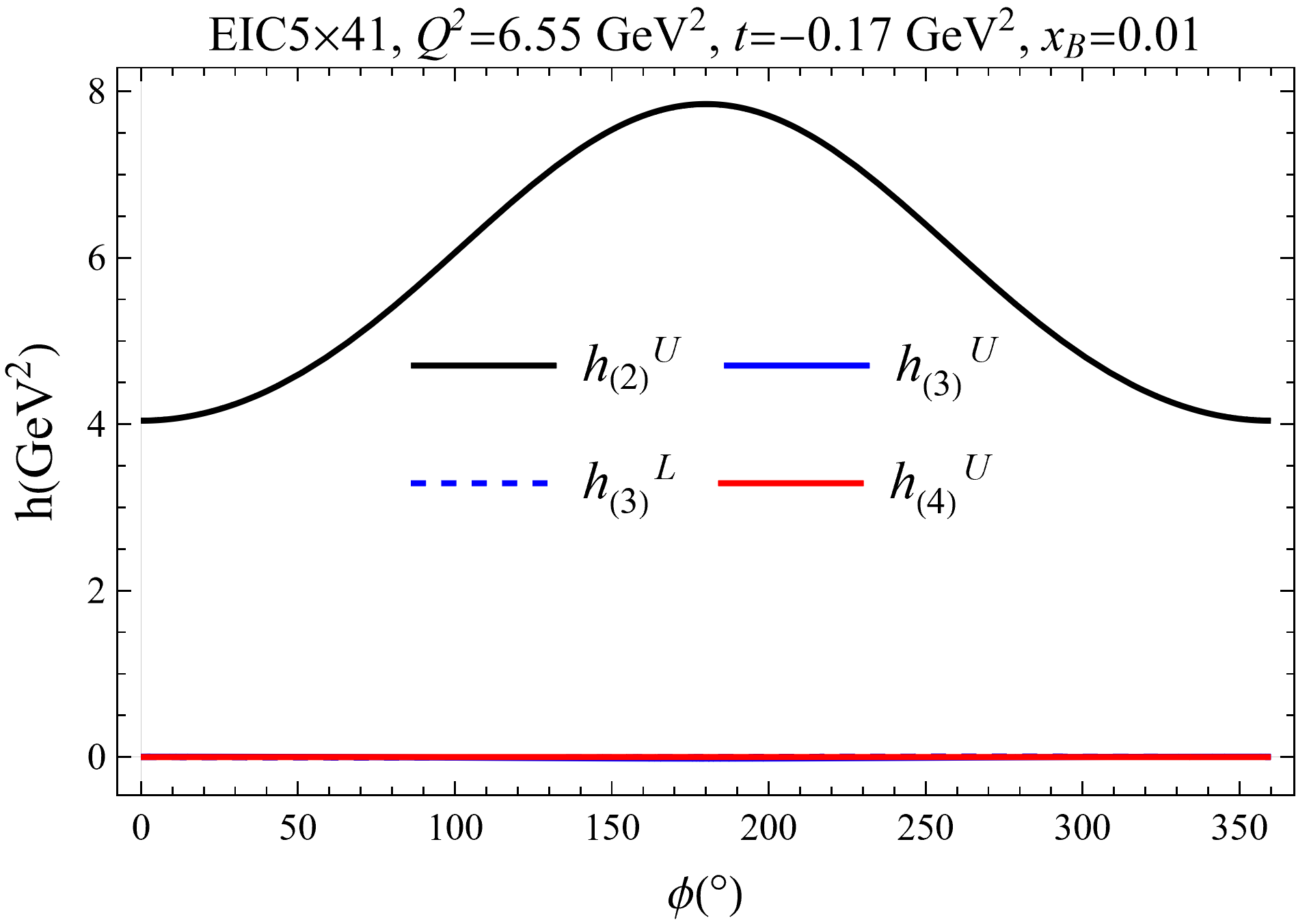}
\includegraphics[width=0.49\textwidth]{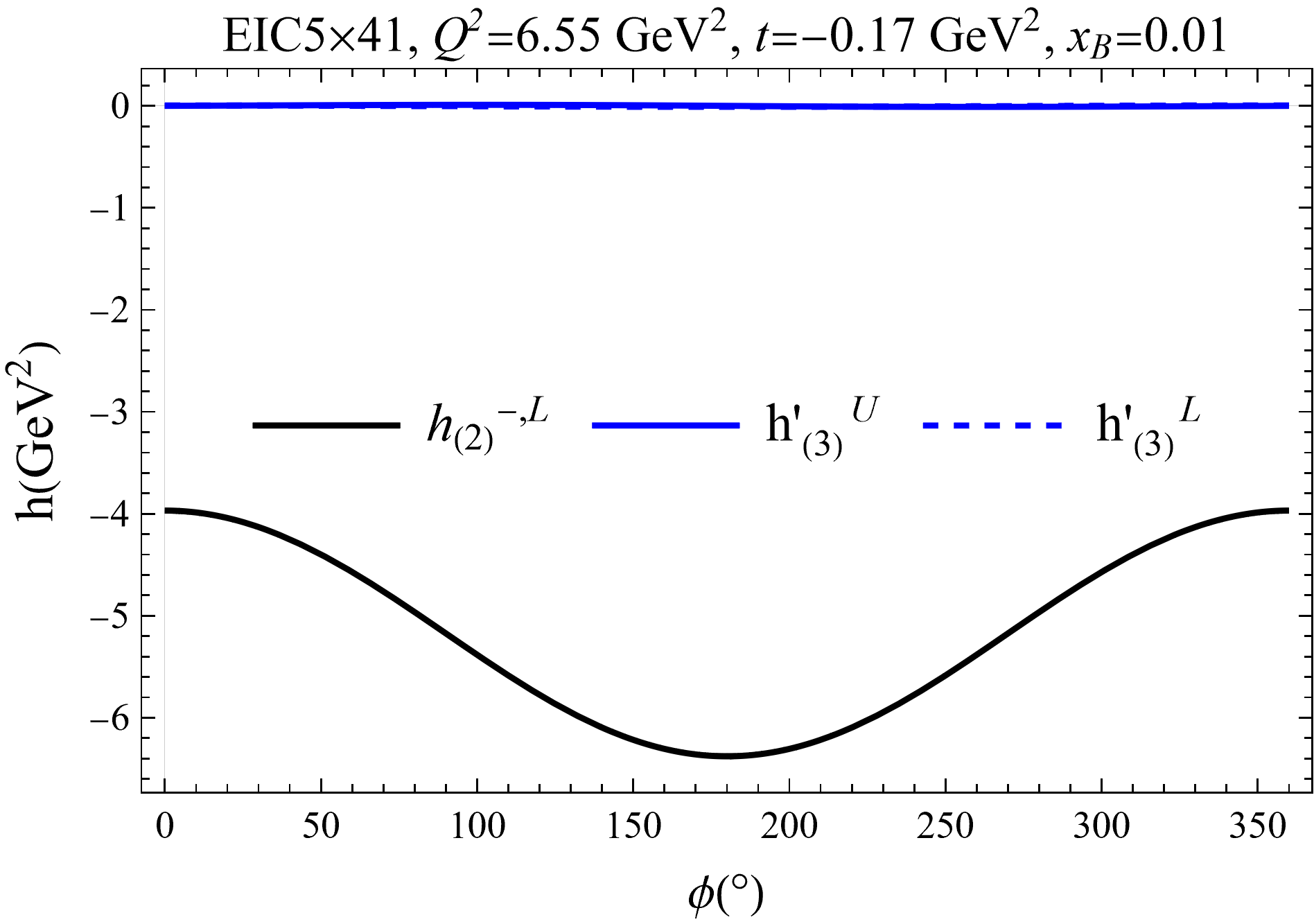}
\end{minipage}
\caption{\label{fig:dvcsscalarEIC} Comparison of the coefficients $h^{\rm{U}}_{(2)}$, $h^{\rm{U/L}}_{(3)}$ and $h^{\rm{U}}_{(4)}$, and $h^{-,\rm{L}}_{(2)}$ and $h'^{\rm{U/L}}_{(3)}$ for EIC5$\times$41 ($s = 820$ GeV$^2$). We plot the twist-two coefficients $h^{\rm{U}}_{(2)}$ and $h^{-,\rm{L}}_{(2)}$ as the background for comparison. The higher-twist scalar coefficients are strongly suppressed due to the large $Q^2$ and small $x_B$.}
\end{figure}

Besides the two kinematical points for JLab 6 GeV and 12 GeV, we also calculate and compare the scalar coefficients at the collider kinematics, for instance, the kinematical point EIC5$\times$41 with electron beam energy $5$ GeV and proton beam $41$ GeV, as shown in figure \ref{fig:dvcsscalarEIC}. The typical $x_B$ is chosen to be $x_B=0.01$ and we also set $t=-0.17$ GeV$^2$, see for instance ref. \cite{Aschenauer:2013hhw} for more details in the kinematical study of DVCS at EIC. Apparently, the twist suppression is much more evident for EIC kinematics due to the large $Q^2$ and small $x_B$.

We note that in general the pure DVCS and interference cross-sections will be more suppressed than the BH contributions for large $Q^2$, and thus it seems to be impractical to measure their contributions at extremely large $Q^2$, even though the higher-twist effects are more suppressed with large $Q^2$ and the leading twist assumption works better there. In practice, a suitable choice of $Q^2$ will be in the regions where the higher-twist effects get reasonable suppression, whereas the pure DVCS and interference contributions still have sizable effects such that one could separate them from the BH background. 
With that in mind, consider the relation between $Q^2$, center of mass energy squared $s$ (for collider experiments) and the electron beam energy $E_b$ (for fixed-target experiments) as
\begin{equation}
\label{eq:Q2fromy}
  Q^2=  x_B y \left(s-M^2\right)= x_B y 2 M E_b  \ ,
\end{equation}
one would natural focus on the region with small $y$ for extreme large center of mass energy $s$ or beam energy $E_b$. Practically, we consider the lower $y$ limit as $y\sim 0.1$ to avoid the strong $Q^2$ suppression for pure DVCS and interference cross-section. For the same reason, we focus on the kinematical points with relatively lower $Q^2$ (and lower center of mass energy correspondingly) among all the EIC kinematical points. These effects will be discussed in more details in the section \ref{sec:t3cff} for the cross-sections.

\subsection{Twist-three interference scalar coefficients}

\begin{figure}[t]
\centering
\begin{minipage}[b]{\textwidth}
\includegraphics[width=0.49\textwidth]{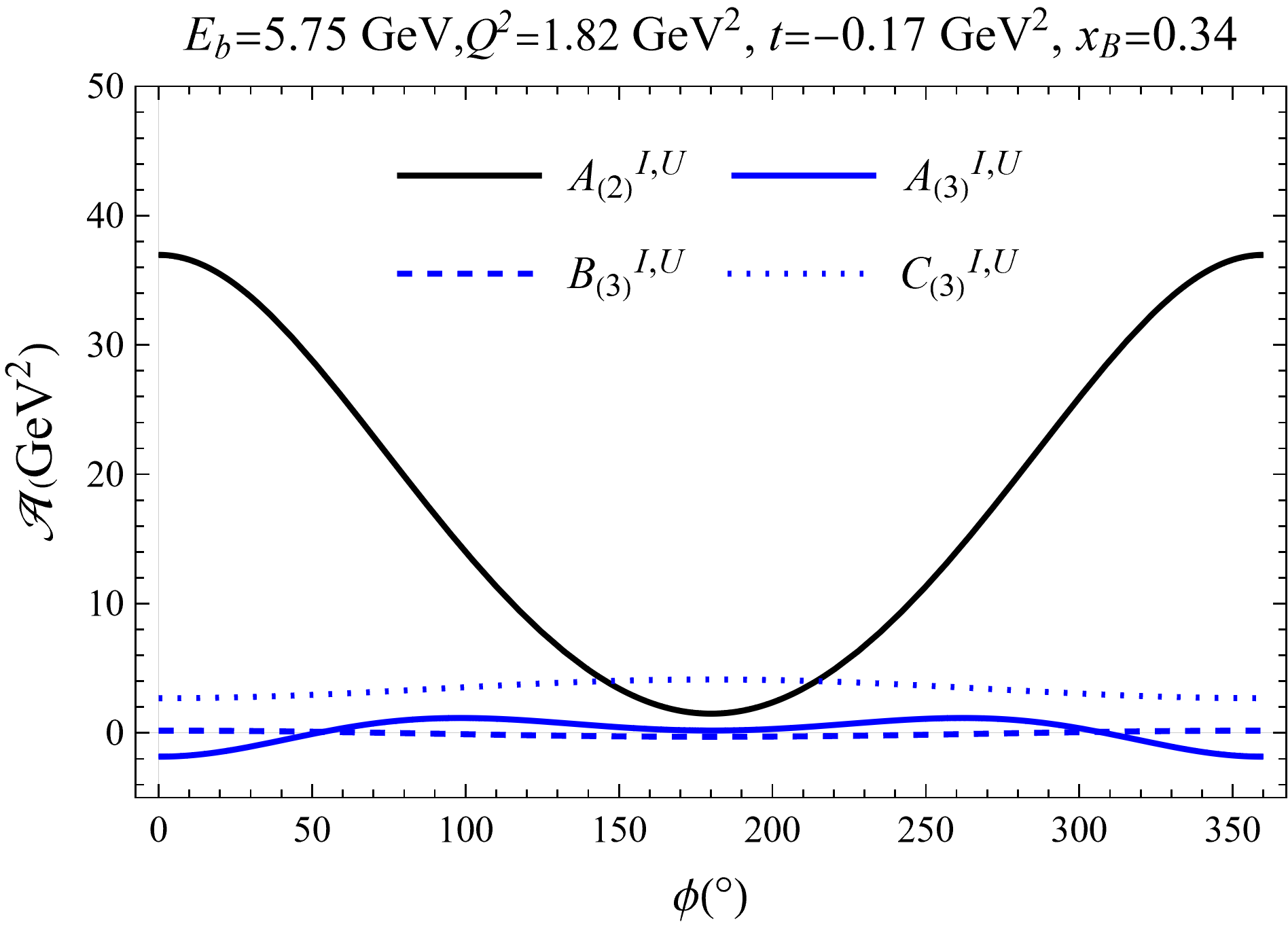}
\includegraphics[width=0.49\textwidth]{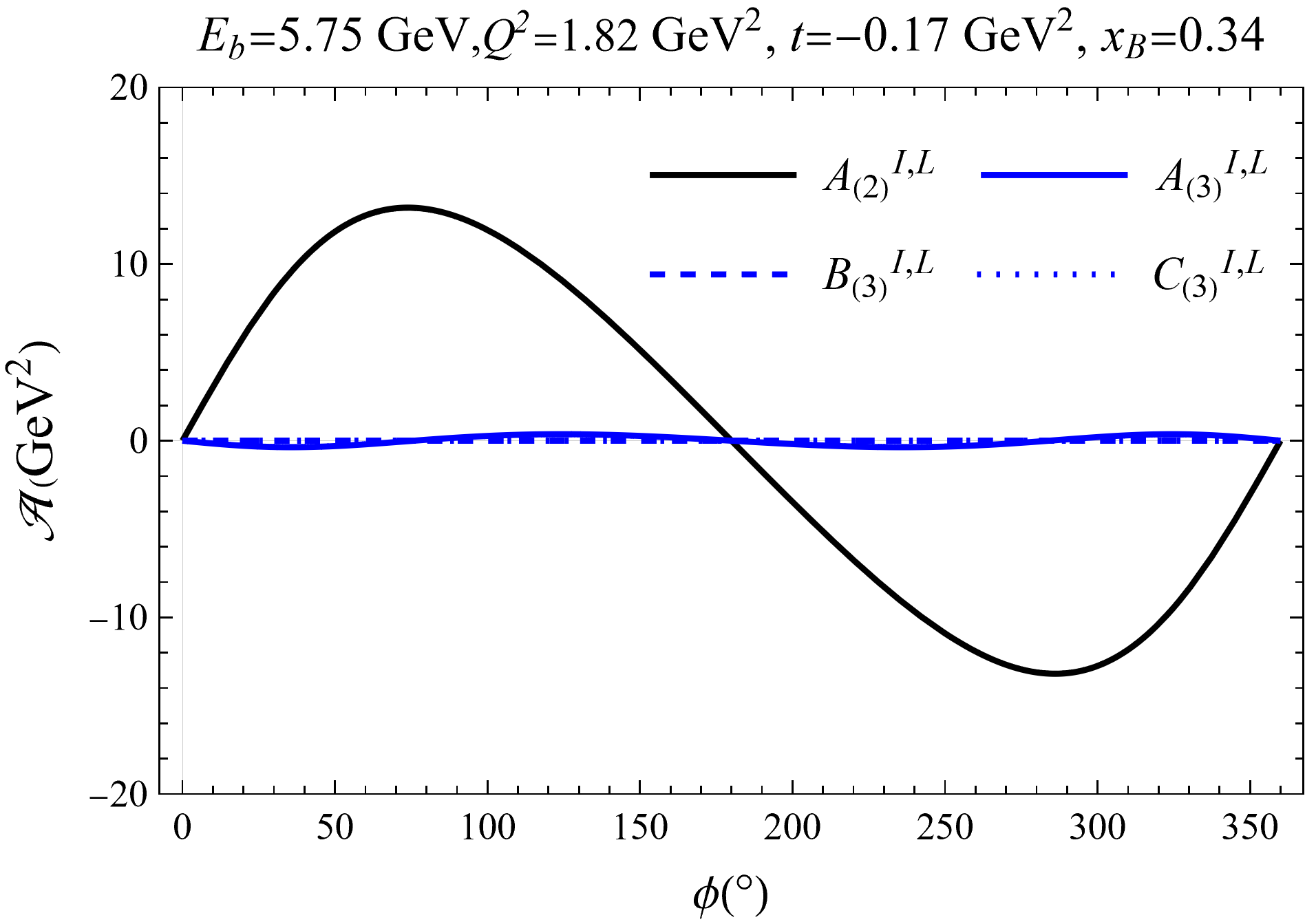}
\end{minipage}
\begin{minipage}[b]{\textwidth}
\includegraphics[width=0.49\textwidth]{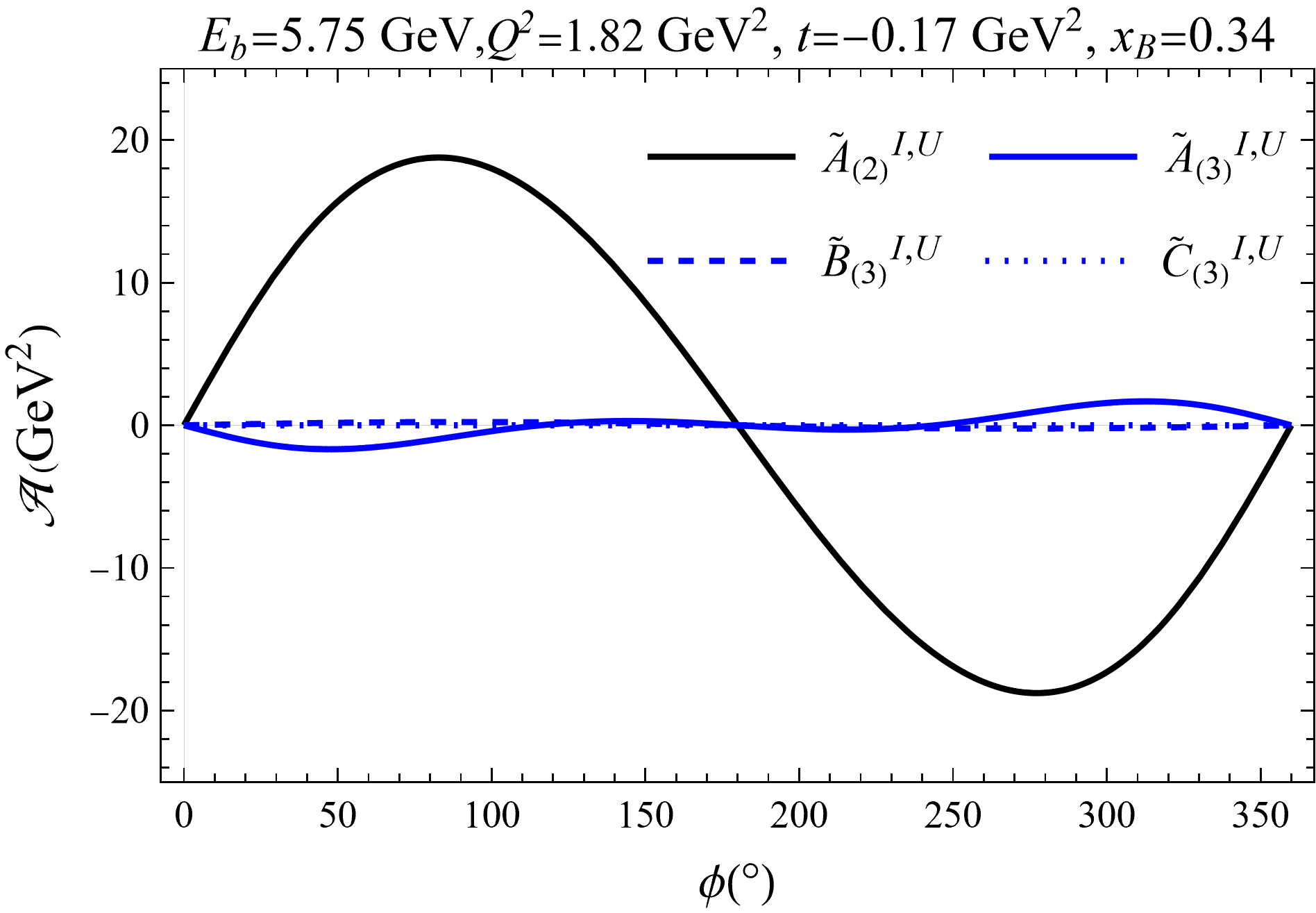}
\includegraphics[width=0.49\textwidth]{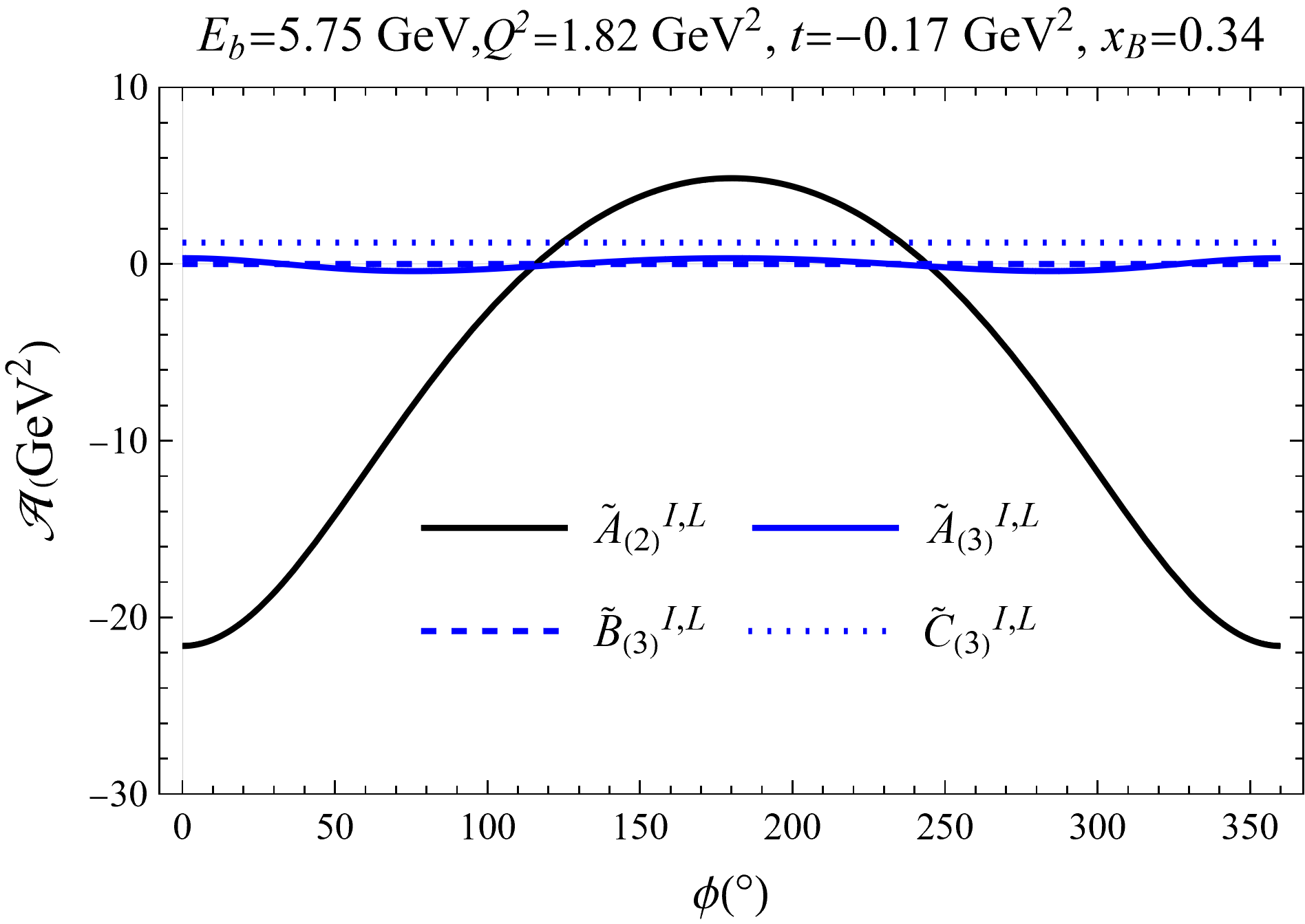}
\end{minipage}
\caption{\label{fig:intscalar} Comparison of the interference coefficients $\mathcal A^{\rm{I,U/L}}_{(3)}$ with $\mathcal{A}=\{A,B,C,\widetilde{A},\widetilde{B},\widetilde{C}\}$ for JLab 6 GeV. The four plots compare different sets of scalar coefficients that eventually enter different polarization configurations. We put in the twist-two scalar coefficients $A^{\rm{I,U/L}}_{(2)}$ and $\widetilde{A}^{\rm{I,U/L}}_{(2)}$ as the background for comparison. The twist-two coefficients (black lines) generally dominates the higher-twist ones in all plots.}
\end{figure}

\begin{figure}[t]
\centering
\begin{minipage}[b]{\textwidth}
\includegraphics[width=0.49\textwidth]{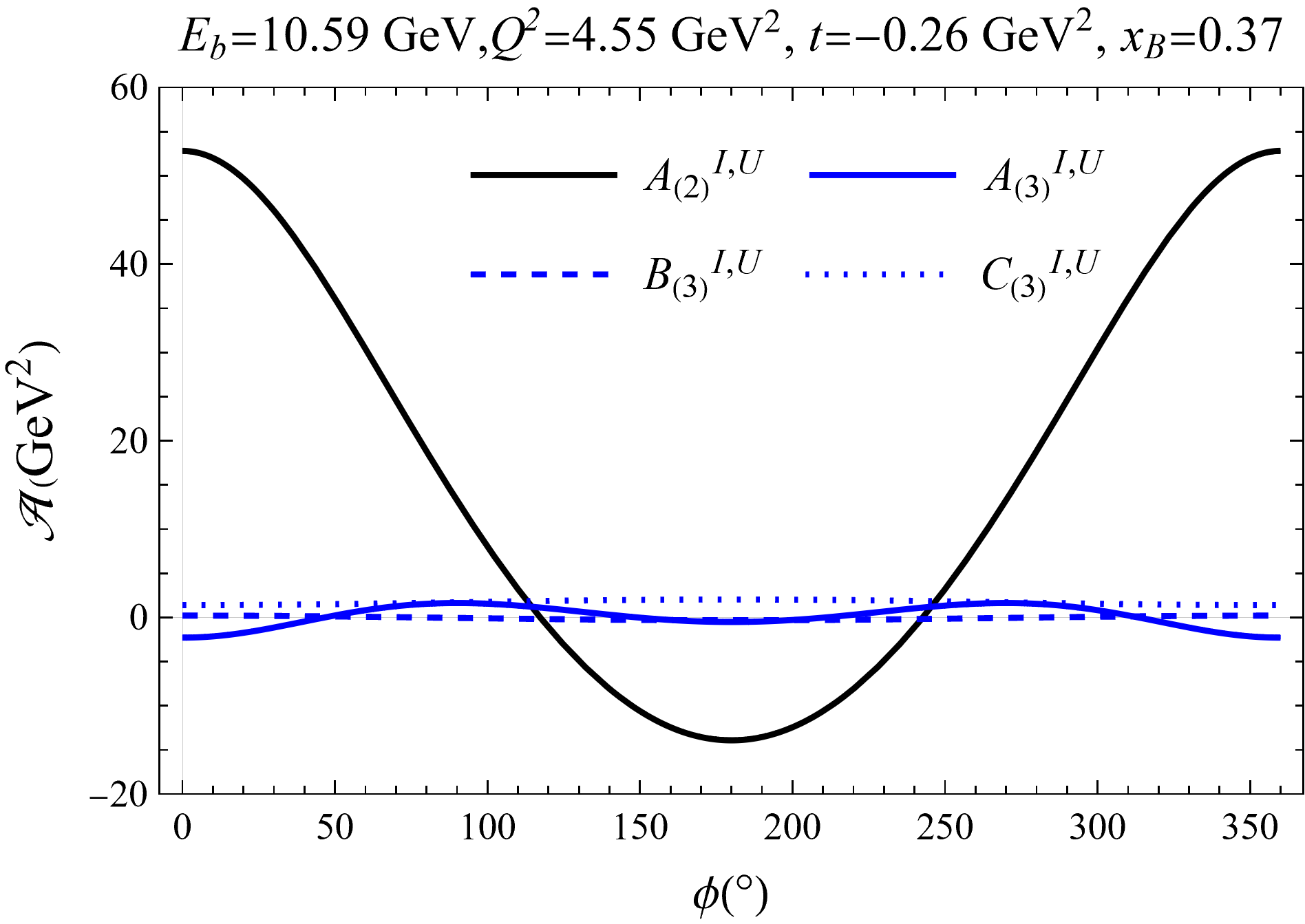}
\includegraphics[width=0.49\textwidth]{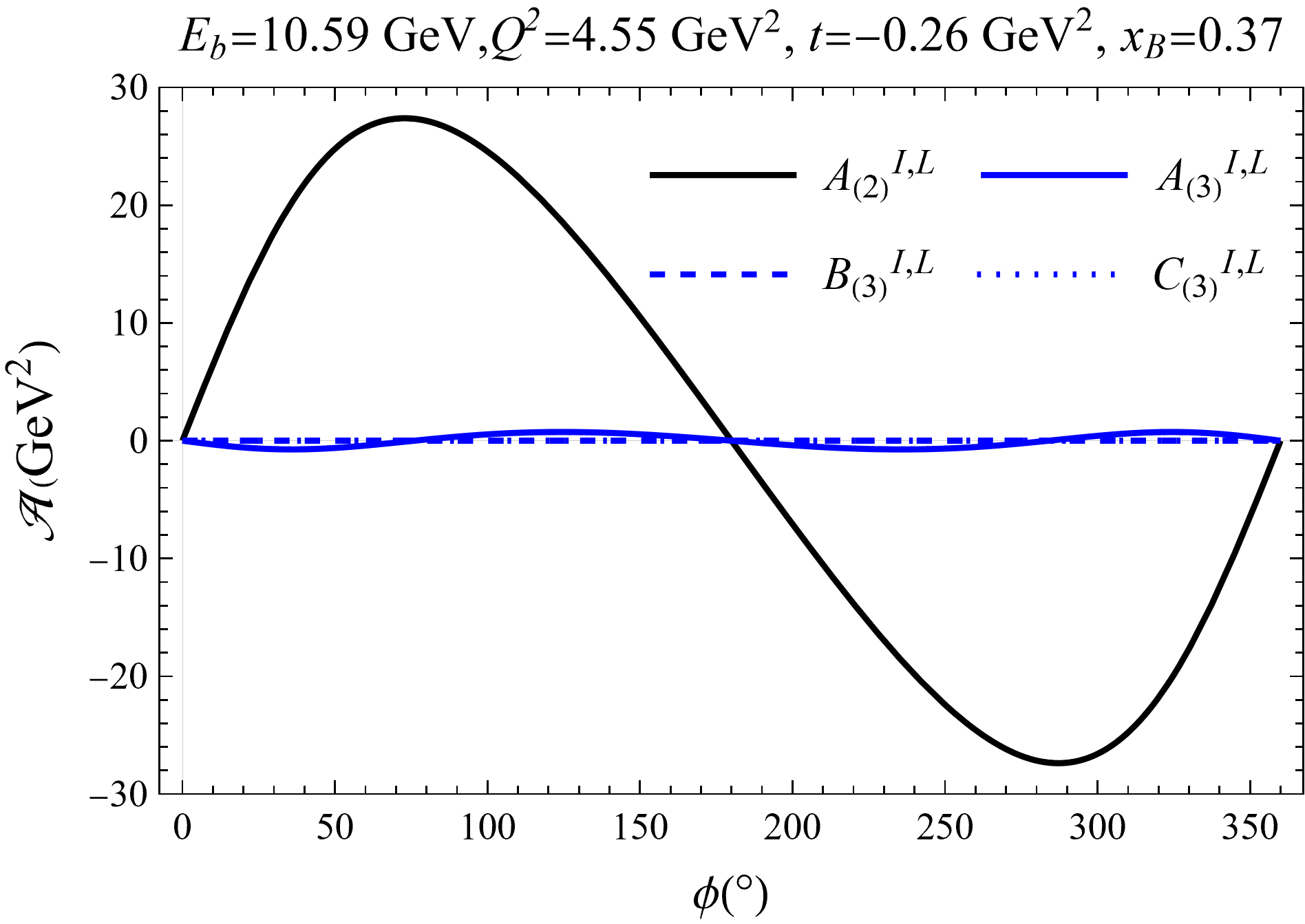}
\end{minipage}
\begin{minipage}[b]{\textwidth}
\includegraphics[width=0.49\textwidth]{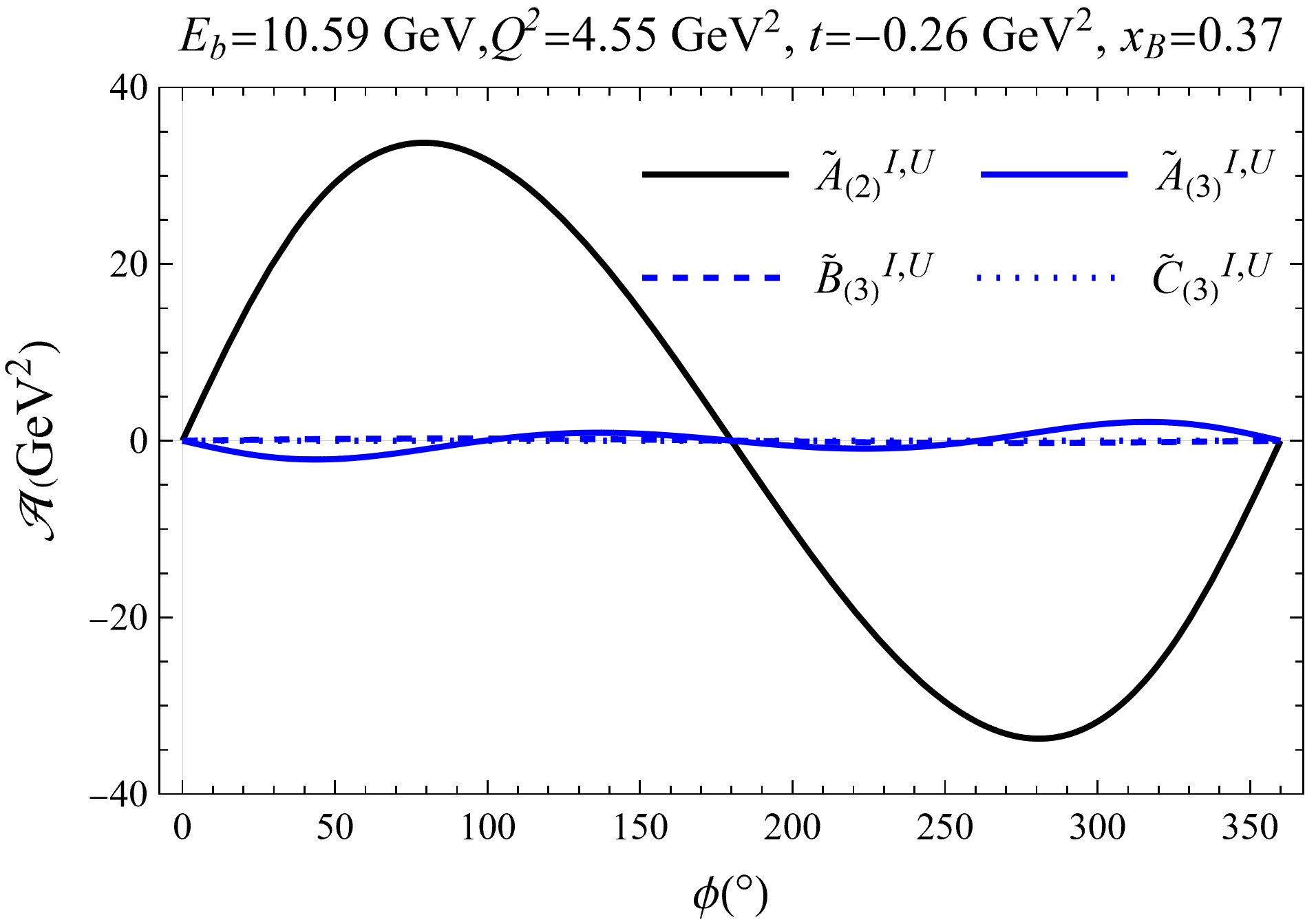}
\includegraphics[width=0.49\textwidth]{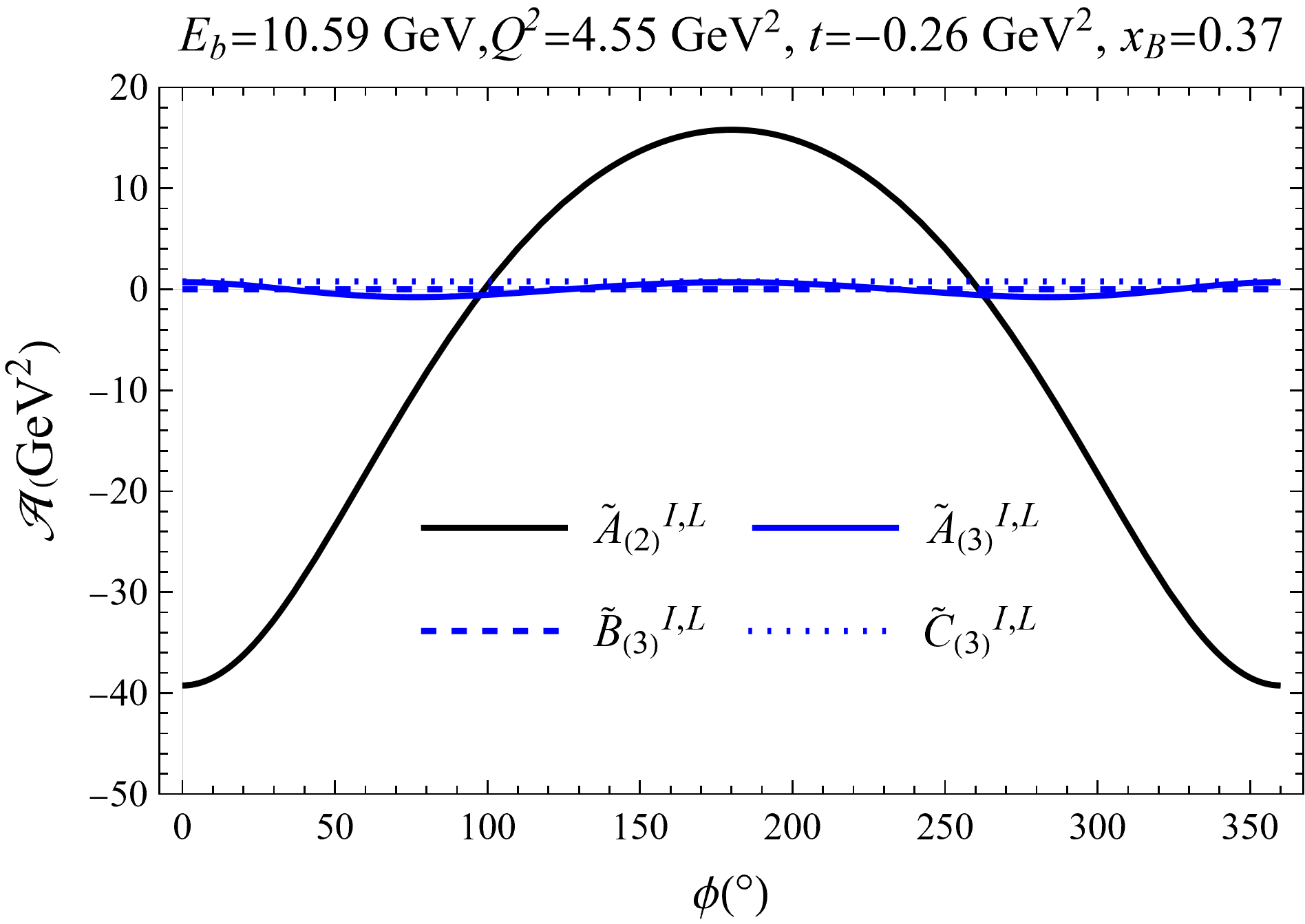}
\end{minipage}
\caption{\label{fig:intscalar2} Another comparison of the interference coefficients $\mathcal A^{\rm{I,U/L}}_{(3)}$ with $\mathcal{A}=\{A,B,C,\widetilde{A},\widetilde{B},\widetilde{C}\}$ for JLab 12 GeV. Again, the four plots compare different sets of scalar coefficients that enter different polarization configurations, and we put in the twist-two scalar coefficients $A^{\rm{I,U/L}}_{(2)}$ and $\widetilde{A}^{\rm{I,U/L}}_{(2)}$ as the background for comparison. The twist-two coefficients (black lines) dominates the higher-twist ones even more significantly than the JLab 6 GeV case in figure \ref{fig:intscalar} due to the larger $Q^2$ here.}
\end{figure}

\begin{figure}[t]
\centering
\begin{minipage}[b]{\textwidth}
\includegraphics[width=0.49\textwidth]{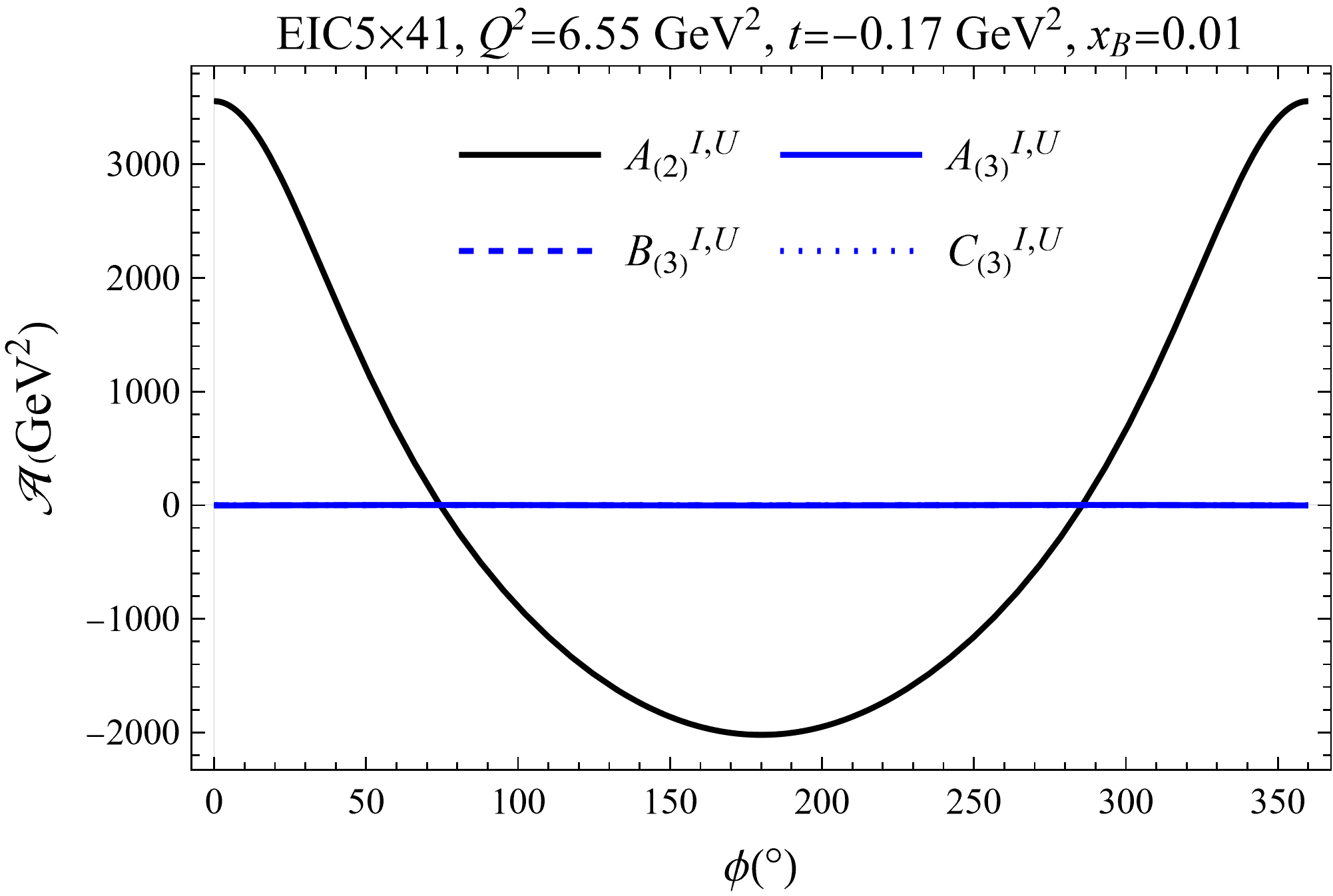}
\includegraphics[width=0.49\textwidth]{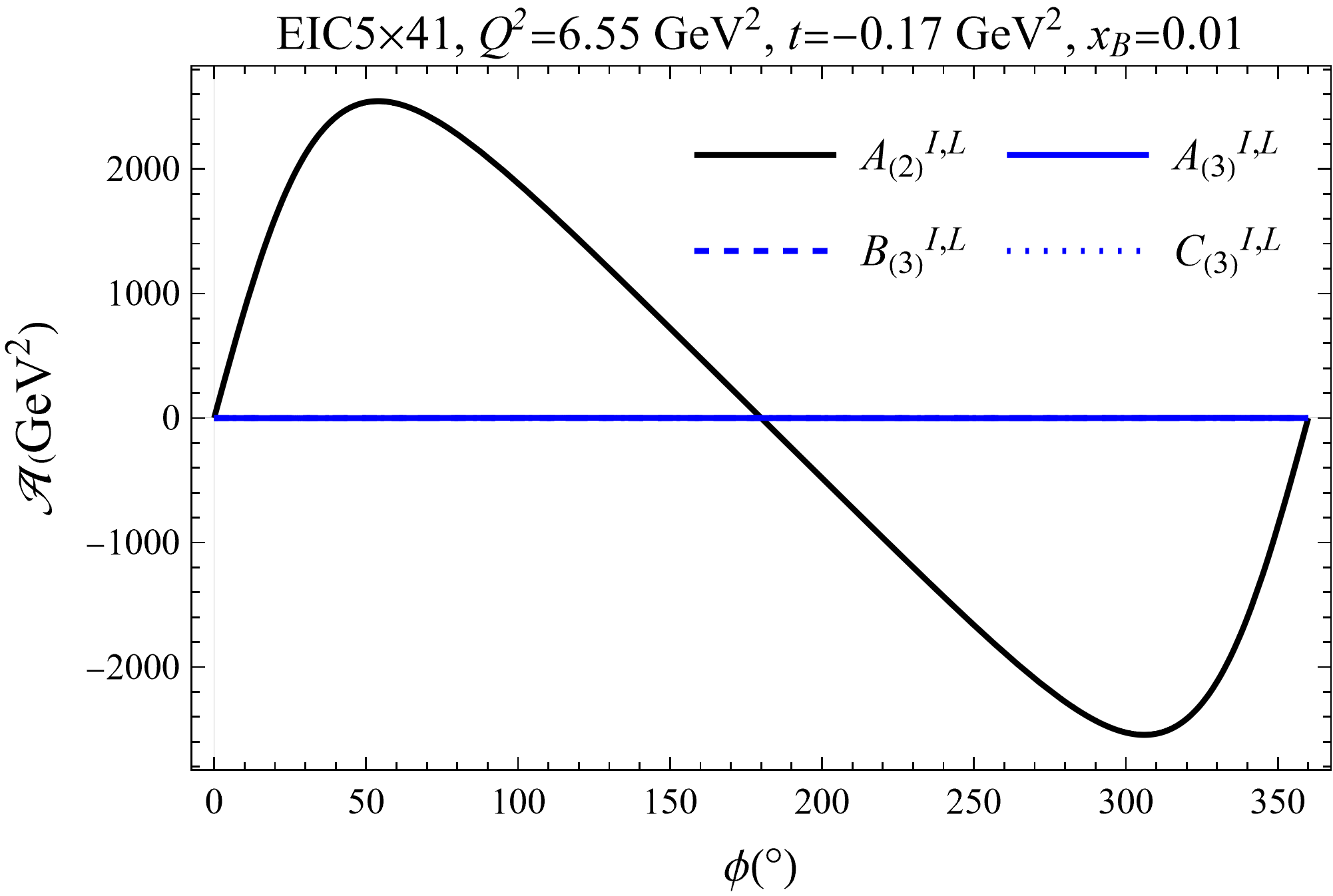}
\end{minipage}
\begin{minipage}[b]{\textwidth}
\includegraphics[width=0.49\textwidth]{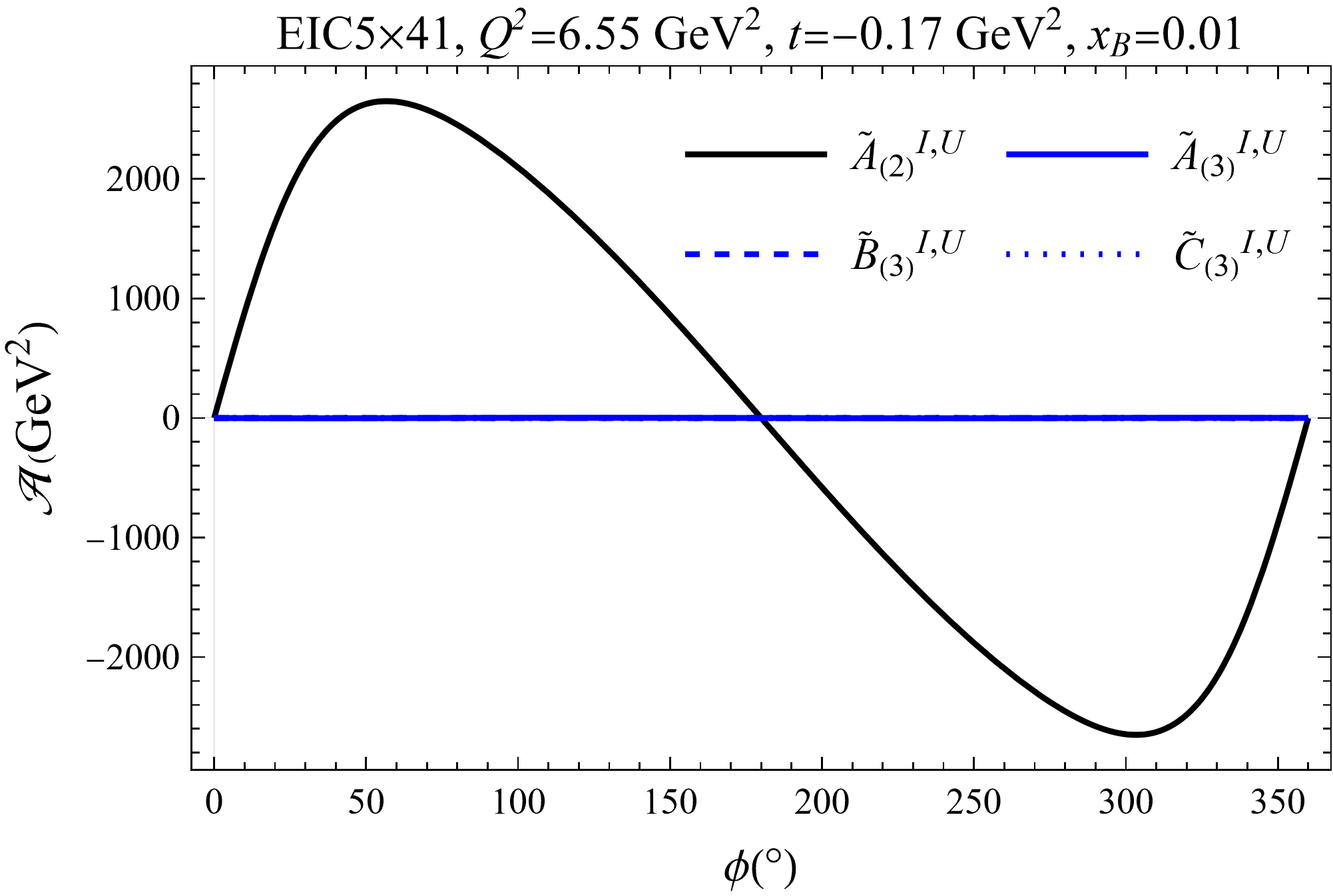}
\includegraphics[width=0.49\textwidth]{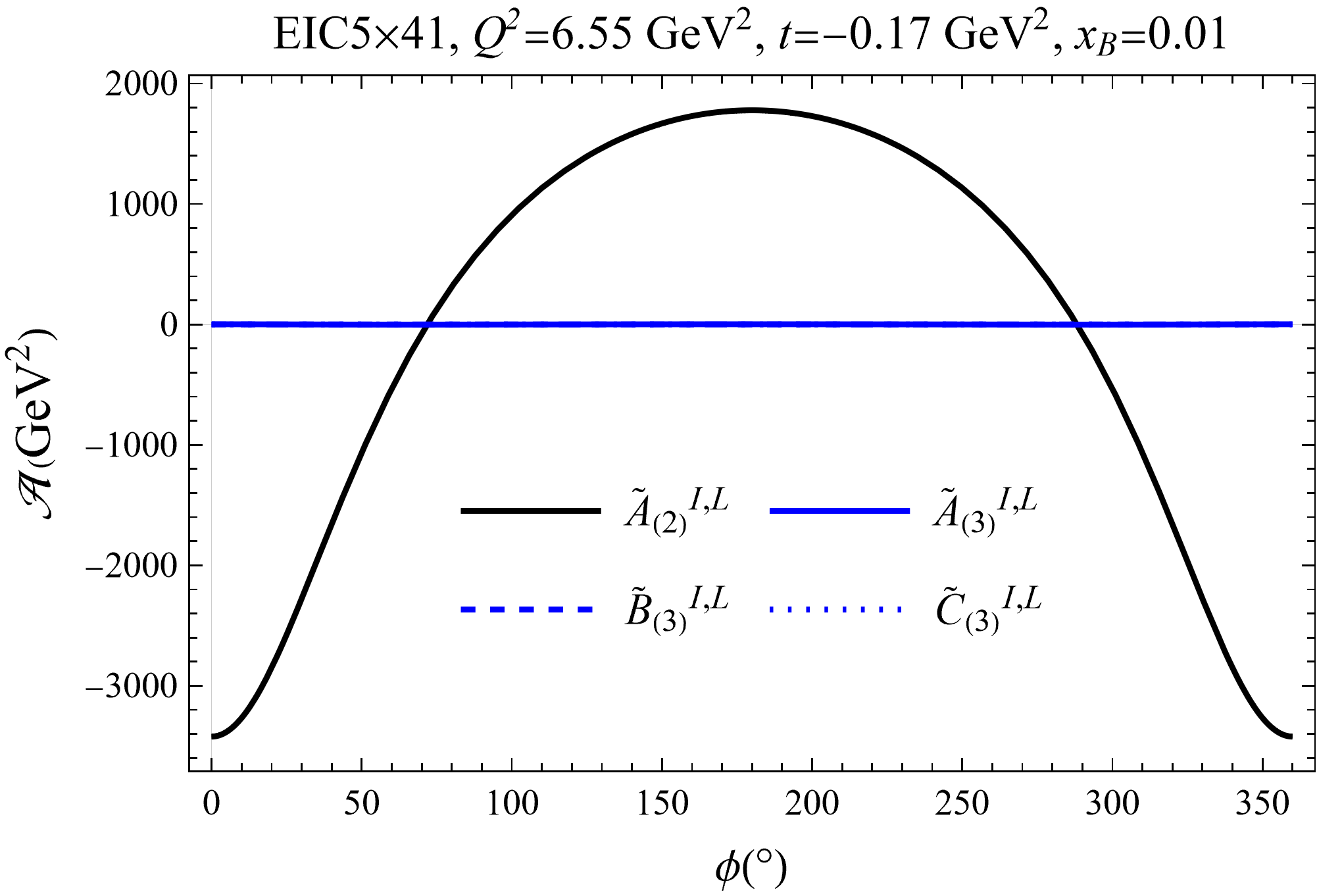}
\end{minipage}
\caption{\label{fig:intscalarEIC} Comparison of the interference coefficients $\mathcal A^{\rm{I,U/L}}_{(3)}$ with $\mathcal{A}=\{A,B,C,\widetilde{A},\widetilde{B},\widetilde{C}\}$ for EIC5$\times$41 ($s=820$ GeV$^2$). The four plots compare different sets of scalar coefficients that enter different polarization configurations, and we put in the twist-two scalar coefficients $A^{\rm{I,U/L}}_{(2)}$ and $\widetilde{A}^{\rm{I,U/L}}_{(2)}$ as the background for comparison. Apparently, the higher-twist coefficients are practically $0$ for such large $Q^2$  and small $x_B$.}
\end{figure}

The same comparisons of scalar coefficients can be done for the interference scalar coefficients, which are shown in figures \ref{fig:intscalar} -- \ref{fig:intscalarEIC}. Again, we rename those twist-two scalar coefficients $\mathcal A^{\rm{I,U/L}}$ in ref. \cite{Guo:2021gru} into $\mathcal A^{\rm{I,U/L}}_{(2)}$ with $\mathcal{A}=\{A,B,C,\widetilde{A},\widetilde{B},\widetilde{C}\}$ for consistency hereafter.

First, we notice that among the 12 twist-three scalar coefficients we compared in figures \ref{fig:intscalar} -- \ref{fig:intscalarEIC}, 4 of them, namely $A^{\rm{I,U/L}}_{(3)}$ and $\widetilde{A}^{\rm{I,U/L}}_{(3)}$, are twist-three parameters whereas the other 8 parameters are of higher order. Among the other 8 parameters, the 4 coefficients $B^{\rm{I,U/L}}_{(3)}$ and $\widetilde{B}^{\rm{I,U/L}}_{(3)}$ are numerically small at all the kinematical points we examined. Thus, we will drop them for our analysis hereafter. With another two scalar coefficients $C^{\rm{I,L}}_{(3)}$ and $\widetilde{C}^{\rm{I,U}}_{(3)}$ that are exactly zero, there remain only 2 interference scalar coefficients out of the other 8 higher-order parameters that are of interest. To conclude, practically there are 6 twist-three scalar coefficients that are relevant to our analysis, $A^{\rm{I,U/L}}_{(3)}$ and $\widetilde{A}^{\rm{I,U/L}}_{(3)}$ which are twist-three, and $C^{\rm{I,U}}_{(3)}$ and $\widetilde{C}^{\rm{I,L}}_{(3)}$ which are twist-four.

Focusing on the comparison of those relevant interference scalar coefficients, we observe the same behavior of them as the pure DVCS ones. The two coefficients $C^{\rm{I,U}}_{(3)}$ and $\widetilde{C}^{\rm{I,L}}_{(3)}$ that vanish at twist three are actually more significant than the twist-three parameters $A^{\rm{I,U/L}}_{(3)}$ and $\widetilde{A}^{\rm{I,U/L}}_{(3)}$ for low $Q^2$ such as the $Q^2=1.82 \text{ GeV}^2$ points for JLab 6 GeV in figure \ref{fig:intscalar}, whereas the twist-three parameters $A^{\rm{I,U/L}}_{(3)}$ and $\widetilde{A}^{\rm{I,U/L}}_{(3)}$ will dominate with a larger $Q^2$ of $4.55 \text{ GeV}^2$ for the JLab 12 GeV as shown in figure \ref{fig:intscalar2}, while for the EIC5$\times$41 the higher-twist parameters are all strongly suppressed by the large $Q^2$ and small $x_B$ there and none of them are relevant at this point as shown in figure \ref{fig:intscalarEIC}. Just like the case of pure DVCS coefficients, this extra suppression of those twist-three parameters $A^{\rm{I,U/L}}_{(3)}$ and $\widetilde{A}^{\rm{I,U/L}}_{(3)}$ at a relatively low $Q^2$ of $1.82$ GeV$^2$ is related to the extra factor of $\xi$ in their definitions that makes them more suppressed than the naive twist counting.

Based on the numerical results shown above, we find that the twist-three contributions are in generally suppressed in terms of the kinematical behavior of these twist-three scalar coefficients. Although there are certain channels where the pure DVCS cross-sections receive twist-three contributions only and seem to allow us to study the twist-three effects, those channels are actually contaminated by the twist-two interference contributions, and thus it will be hard to find a clean channel to measure the pure twist-three effects. Beam charge asymmetry, for which a positron beam is used instead of an electron beam, can help separate the pure DVCS and interference contributions, which will not be discussed in details in this work.

\subsection{Light-cone dependence of the scalar coefficients}

In the comparisons above, we did not emphasize how we define the light-cone vectors, which can lead to a higher-twist effect in those scalar coefficients, as discussed in ref. \cite{Guo:2021gru}. For all our twist-two quantities, we choose $\alpha=0, \beta =100$, which corresponds to defining the light cone according to the two photon momenta. The definition of $\alpha$ and $\beta$ and the light-cone dependence are discussed in details in ref. \cite{Guo:2021gru}. Whereas for the twist-three quantities, the light-cone dependence turns out to be extremely weak. A careful study of those different choices of light-cone vectors (corresponding to different choices of $\alpha$ and $\beta$ introduced in ref. \cite{Guo:2021gru}) shows that the light-cone dependence only shows up as twist-five effects. Those effects will be extremely suppressed, and they make no differences practically. Therefore, we suppress our choice of light-cone vectors for the twist-three quantities in the above plots and in the rest of the paper. This is a direct result of the covariant Compton tensor coefficients we used in eq. (\ref{eq:covCompton}).

However, it should be noted that this result should not be considered as a proof or confirmation that the light-cone dependence vanishes at twist four, as our covariant Compton tensor coefficients are the promoted coefficients originated from the light-cone Compton tensor coefficients which only have twist-three accuracy, and thus any prediction beyond twist-three is not justified with such tensor coefficients. A rigorous proof of such statement requires a comprehensive study of all twist-four effects with general light-cone vectors, which is beyond the scope of this paper. On the other hand, our results imply that such corrections can be very weak, assuming the covariant Compton tensor coefficients partially take those higher-order effects into account.

\section{Numerical studies of twist-three cross-sections}
\label{sec:t3cff}

With the scalar coefficients studied in the previous section, we are left with the twist-three CFFs. In general, the twist-three CFFs are completely independent quantities derived from twist-three GPDs, which are naturally of order $\mathcal{O}(1)$. However, there exist non-trivial relations between twist-two and twist-three GPDs due to the Lorentz invariance and QCD equations of motion, making it possible to relate the twist-three CFFs with the twist-two ones up to some extra genuine twist-three pieces, which is generally known as the WW relations. In this paper, we focus on the twist-three CFFs related to the twist-two GPDs and neglect the genuine twist-three pieces.
We note that the genuine twist-three pieces might not be negligible, but a proper estimation of their effects does not exist yet. Therefore, the WW approximation that assumes the dominance of the WW parts is the only sensible way to estimate the twist-three CFFs. Nevertheless, most of our arguments for the estimated cross-sections will not rely on the absolute values of the twist-three CFFs. They should be applicable even if the CFFs get modified by the genuine twist-three pieces.

The WW relations between twist-two and twist-three GPDs are already studied in the literature \cite{Belitsky:2000vx,Belitsky:2005qn,Kivel:2000rb,Radyushkin:2000jy,Aslan:2018zzk}. As those relations themselves do not allow one to write the twist-three CFFs directly in terms of twist-two CFFs, one is forced to employ a certain model of GPDs in order to relate the twist-two and twist-three CFFs, for which we choose a dynamical framework of GPD parameterization based on the reggeized spectator model \cite{Kriesten:2021sqc,Ahmad:2006gn,Ahmad:2009fvg,Goldstein:2010gu,Gonzalez-Hernandez:2012xap}. The twist-three cross-sections can then be studied with our scalar coefficients in the last section.

\subsection{Wandzura-Wilczek relations of twist-three GPDs}

The WW relations for twist-three GPDs have been studied in the literature, which can be written as \cite{Belitsky:2000vx,Kivel:2000rb,Radyushkin:2000jy},
\begin{align}
\label{eq:WWrel1}
\begin{split}
    W^{[\gamma^\mu]}\approx\frac{\Delta^\mu}{n\cdot \Delta} n_\nu W^{[\gamma^\nu]}+\int_{-1}^{1}\text{d}u W_{+}(x,u,\xi)G^\mu(u,\xi)+i\widetilde{\epsilon}^{\mu\nu}  \int_{-1}^{1}\text{d}u W_{-}(x,u,\xi)\widetilde{G}_\nu(u,\xi)\ ,
\end{split}\\
\label{eq:WWrel2}
\begin{split}
    W^{[\gamma^\mu\gamma^5]}\approx\frac{\Delta^\mu}{n\cdot \Delta} n_\nu W^{[\gamma^\nu\gamma^5]}+\int_{-1}^{1}\text{d}u W_{+}(x,u,\xi)\widetilde{G}^\mu(u,\xi)+i\widetilde{\epsilon}^{\mu\nu}  \int_{-1}^{1}\text{d}u W_{-}(x,u,\xi)G_\nu(u,\xi)\ ,
\end{split}
\end{align}
for which the WW kernel 
\begin{align}
W(x,u,\xi)=\frac{\theta(x-\xi)\theta(u-x)-\theta(\xi-x)\theta(x-u)}{u-\xi} \ ,
\end{align}
and two combinations
\begin{align}
W_{\pm}(x,u,\xi)=\frac{1}{2} \left[W(x,u,\xi)\pm W(x,u,-\xi)\right]\ ,
\end{align}
are defined. Note that we suppress the $t$-dependence in the GPDs in the derivation of twist-three WW CFFs, since it does not interfere with the WW kernel convolution, and it can be recovered easily. The two notations $G^\mu(u,\xi)$ and $\widetilde{G}^\mu(u,\xi)$ are defined as \cite{Belitsky:2000vx,Kivel:2000rb,Radyushkin:2000jy},
\begin{align}
\begin{split}
    G^\mu(u,\xi)=&\left<\gamma_T^{\mu}\right>(H+E)(u,\xi)-\frac{\Delta^\mu_T}{n\cdot \Delta}\left<\frac{1}{M}\right>\left(u\frac{\partial~}{\partial u}+\xi\frac{\partial~}{\partial \xi} \right) E(u,\xi)\\
    &+\frac{\Delta^\mu_T}{n\cdot \Delta}\left<n_\nu \gamma^\nu\right>\left(u\frac{\partial~}{\partial u}+\xi\frac{\partial~}{\partial \xi} \right) (H+E)(u,\xi)\ ,
\end{split}\\
\begin{split}
    \widetilde{G}^\mu(u,\xi)=&\left<\gamma_T^{\mu}\gamma^5\right>\widetilde{H}(u,\xi)+\frac{\Delta_T^{\mu}}{2}\left<\frac{\gamma^5}{M}\right>\left(1+u\frac{\partial~}{\partial u}+\xi\frac{\partial~}{\partial \xi} \right) \widetilde{E}(u,\xi)\\
    &+\frac{\Delta^\mu_T}{n\cdot \Delta}\left<n_\nu \gamma^\nu\gamma^5\right>\left(u\frac{\partial~}{\partial u}+\xi\frac{\partial~}{\partial \xi} \right) \widetilde{H}(u,\xi)\ ,
\end{split}
\end{align}
where the notation $\left<\Gamma\right> \equiv \bar u(P',S') \Gamma u(P,S)$ is introduced in the equations above. It is worth noting that here the $\Delta_T^{\mu}$ is defined with 
\begin{align}
V^{\mu}_T\equiv (V^{\mu} \bar{P}^\nu- V^{\nu} \bar{P}^\mu) n_\nu\ ,
\end{align}
which is different from the light-cone transverse projection $\Delta_\perp^{\mu}$. Their difference is non-trivial for a general choice of light-cone vectors, and thus they shall be clearly distinguished.

In the case of twist-three DVCS amplitude, we are especially interested in the combination as given in eq. (\ref{eq:t3GPDcomb}), and we have the following relation,
\begin{align}
\label{eq:WWapprox}
\begin{split}
 \widetilde{ g}^{\mu\nu}\int_{-1}^1\text{d}x & C^{q[-]}_{(0)}(x,\xi) W^{\left[\gamma_\nu\right]}(x,\xi)-i\widetilde{\epsilon}^{\mu\nu}\int_{-1}^1 \text{d}x C^{q[+]}_{(0)}(x,\xi) W^{\left[\gamma_\nu\gamma^5\right]}(x,\xi)\\
\approx \int_{-1}^1 \text{d}x\text{d}u\Bigg[ &\widetilde{ g}^{\mu\nu} \left(C^{q[-]}_{(0)}(x,\xi) W_{+}(x,u,\xi)-C^{q[+]}_{(0)}(x,\xi) W_{-}(x,u,\xi)\right)G_\nu(u,\xi)\\
&-i\widetilde{\epsilon}^{\mu\nu}\left(C^{q[+]}_{(0)}(x,\xi) W_{+}(x,u,\xi)- C^{q[-]}_{(0)}(x,\xi) W_{-}(x,u,\xi)\right) \widetilde{G}_\nu(u,\xi)\Bigg]\ .
\end{split}
\end{align}

We note in our analysis of twist-two CFFs in ref. \cite{Guo:2021gru}, we find it crucial to absorb the first terms in eqs. (\ref{eq:WWrel1}) and (\ref{eq:WWrel2}) that are proportional to the twist-two GPDs into the twist-two coefficients in order for the light-cone dependence to vanish at twist three. Therefore, for our analysis of twist-three WW CFFs, we will only consider the other two terms in eqs. (\ref{eq:WWrel1}) and (\ref{eq:WWrel2}) that are related to $G^\mu(u,\xi)$, $\widetilde{G}^\mu(u,\xi)$ and WW kernels $W_{\pm}(x,u,\xi)$, which correspond to those terms in the second and third lines of eq. (\ref{eq:WWapprox}).

Then if we define two new Wilson coefficients associated with WW kernel as \cite{Belitsky:2005qn},
\begin{align}
\begin{split}
 C^{q[-]}_{3(0)}(x,\xi) \equiv &\int_{-1}^1 \text{d}u \left(C^{q[-]}_{(0)}(u,\xi) W_{+}(u,x,\xi)-C^{q[+]}_{(0)}(u,\xi) W_{-}(u,x,\xi)\right) \ ,\\
 =& \frac{Q_q^2}{x+\xi}\ln{\frac{2\xi-i 0}{\xi-x-i 0}} + \frac{Q_q^2}{x-\xi}\ln{\frac{2\xi-i 0}{x+\xi-i 0}} \ ,
\end{split}\\
\begin{split}
 C^{q[+]}_{3(0)}(x,\xi) \equiv &\int_{-1}^1 \text{d}u \left(C^{q[+]}_{(0)}(u,\xi) W_{+}(u,x,\xi)-C^{q[-]}_{(0)}(u,\xi) W_{-}(u,x,\xi)\right)\ ,\\
 =& \frac{Q_q^2}{x+\xi}\ln{\frac{2\xi-i 0}{\xi-x-i 0}} - \frac{Q_q^2}{x-\xi}\ln{\frac{2\xi-i 0}{x+\xi-i 0}} \ ,
\end{split}
\end{align}
we can rewrite eq. (\ref{eq:WWapprox}) as
\begin{align}
\label{eq:wwDVCS}
\begin{split}
 &\widetilde{ g}^{\mu\nu}\int_{-1}^1\text{d}x C^{q[-]}_{(0)}(x,\xi) W^{\left[\gamma_\nu\right]}(x,\xi)-i\widetilde{\epsilon}^{\mu\nu}\int_{-1}^1 \text{d}x C^{q[+]}_{(0)}(x,\xi) W^{\left[\gamma_\nu\gamma^5\right]}(x,\xi)\\
\approx ~&\widetilde{ g}^{\mu\nu} \int_{-1}^1 \text{d}x C^{q[-]}_{3(0)}(x,\xi)  G_\nu(x,\xi)-i\widetilde{\epsilon}^{\mu\nu}\int_{-1}^1 \text{d}x  C^{q[+]}_{3(0)}(x,\xi) \widetilde{G}_\nu(x,\xi)\ .
\end{split}
\end{align}
Notice that the above Wilson coefficients $C^{q[\pm]}_{3(0)}(x,\xi) $ for WW kernels are regular, even though the WW kernels $W_{\pm}(x,u,\xi)$ themselves have singularities at $u\to \pm\xi$. This cancellation of discontinuities results from the specific combination of GPDs in eq. (\ref{eq:t3GPDcomb}), which makes sure the factorization theorem is not broken by the WW approximation at twist-three level, as discussed in refs. \cite{Kivel:2000cn,Kivel:2000rb,Belitsky:2005qn}. 

Then all we are left are those twist-three CFFs expressed in terms of the convolution of Wilson coefficients $ C^{q[-]}_{3(0)}(x,\xi) $ and GPDs $G^\mu(x,\xi)$ and  $\widetilde{G}^\mu(x,\xi)$. With the following definitions of twist-three CFFs from WW relations,
\begin{align}
   \mathcal G_1(\xi,t) \equiv& \int_{-1}^{1} \text{d}x C^{q[-]}_{3(0)}(x,\xi) (H+E)(x,\xi)\ ,\\
   \mathcal G_2(\xi,t) \equiv& \int_{-1}^{1} \text{d}x C^{q[-]}_{3(0)}(x,\xi) \left(x\frac{\partial~}{\partial x}+\xi\frac{\partial~}{\partial \xi}\right)E(x,\xi)\ ,\\
   \mathcal G_3(\xi,t) \equiv& \int_{-1}^{1} \text{d}x C^{q[-]}_{3(0)}(x,\xi) \left(x\frac{\partial~}{\partial x}+\xi\frac{\partial~}{\partial \xi}\right)(H+E)(x,\xi)\ ,\\
   \widetilde{\mathcal G}_1(\xi,t) \equiv& \int_{-1}^{1} \text{d}x C^{q[+]}_{3(0)}(x,\xi) \widetilde{H}(x,\xi)\ ,\\
   \widetilde{\mathcal G}_2(\xi,t) \equiv& \int_{-1}^{1} \text{d}x C^{q[+]}_{3(0)}(x,\xi) \left(1+x\frac{\partial~}{\partial x}+\xi\frac{\partial~}{\partial \xi}\right)\widetilde{E}(x,\xi)\ ,\\
   \widetilde{\mathcal G}_3(\xi,t) \equiv& \int_{-1}^{1} \text{d}x C^{q[+]}_{3(0)}(x,\xi) \left(x\frac{\partial~}{\partial x}+\xi\frac{\partial~}{\partial \xi}\right)\widetilde{H}(x,\xi)\ .
\end{align}
Together with some relations presented in appendix. \ref{app:diracalgebra} to transform different Dirac structures, we can derive the following WW approximations for the twist-three CFFs as,
\begin{align}
   \HTcff(\xi,t)=&\widetilde{\mathcal G}_1(\xi,t)+\frac{t}{4M^2} \widetilde{\mathcal G}_2(\xi,t)+\widetilde{\mathcal G}_3(\xi,t)\ , \\
   \ETcff(\xi,t)=&-\frac{{\mathcal G}_3(\xi,t)}{\xi}-\widetilde{\mathcal G}_1(\xi,t)- \widetilde{\mathcal G}_2(\xi,t)-\widetilde{\mathcal G}_3(\xi,t)\ , \\
   \HTtcff(\xi,t)=&\frac{{\mathcal G}_2(\xi,t)}{2\xi}+ \frac{\widetilde{\mathcal G}_2(\xi,t)}{2}\ ,\\
   \ETtcff(\xi,t)=&-{\mathcal G}_1(\xi,t)-{\mathcal G}_3(\xi,t)-\frac{\widetilde{\mathcal G}_3(\xi,t)}{\xi}\ .
\end{align}
Note that those CFFs with over-lines are defined from eq. (\ref{eq:CFFbar}), which involves both the vector-like twist-three CFFs from $W^{[\gamma^{j}]}$ ($\mathcal{H}_{2T}(\xi,t)$ for instance) and the axial-vector-like twist-three CFFs from $W^{[\gamma^{j}\gamma^5]}$ (${\mathcal{H}}'_{2T}(\xi,t)$ for instance). A separation of their contributions to the cross-sections formulas is totally unnecessary due to the degeneracy of twist-three CFFs as show in eq. (\ref{eq:t3GPDcomb}) and the discussions there. We also note that it is the combination in eq. (\ref{eq:t3GPDcomb}) that ensures that the singularities in the WW kernel get canceled in the those WW CFFs, which does not necessarily apply to $\mathcal{H}_{2T}(\xi,t)$ or ${\mathcal{H}}'_{2T}(\xi,t)$ respectively.

\begin{table*}[t]
\centering
\scalebox{0.99}{
\begin{tabular}{|c|c|c|c|c|c|c|c|c|c|c|}
\hline
 $x_{B}$ &  $|t|(\text{GeV}^{2})$ &  $Q^{2} (\text{GeV}^{2}) $ &Re$\mathcal H$  &  Re$\mathcal E$  &   Re$\widetilde{\mathcal H}$   &  Re$\widetilde{\mathcal E}$   & Im$\mathcal H$ &  Im$\mathcal E$  &  Im$\widetilde{\mathcal H}$  &   Im$\widetilde{\mathcal E}$   \\
 \hline
 0.34 & 0.17 & 1.82 & -4.19 & -3.49  & 1.73 & 21.0  & 2.67  & 0.785  & 4.32  & 52.0  \\
 \hline 
 0.37 & 0.26 & 4.55 &
   -4.77  & -4.31  & 1.68  & 17.2 & 1.98  & 0.525  & 3.54  & 36.0  \\
 \hline
\end{tabular}}
\centering
\scalebox{0.84}{
\begin{tabular}{|c|c|c|c|c|c|c|c|c|c|c|}
\hline
 $x_{B}$ &  $|t|(\text{GeV}^{2})$ &  $Q^{2} (\text{GeV}^{2}) $ &Re$\overline{\mathcal H}_{2T}$  &  Re$\overline{\mathcal E}_{2T}$  &   Re$\overline{\widetilde{\mathcal H}}_{2T}$   &  Re$\overline{\widetilde{\mathcal E}}_{2T}$   & Im$\overline{\mathcal H}_{2T}$ &  Im$\overline{\mathcal E}_{2T}$  &  Im$\overline{\widetilde{\mathcal H}}_{2T}$  &   Im$\overline{\widetilde{\mathcal E}}_{2T}$   \\
 \hline
 0.34 & 0.17 & 1.82 & 1.34 & -26.0  & 17.1 & -20.1  & -2.01  & -39.6  & 15.9  & -5.98  \\
 \hline 
 0.37 & 0.26 & 4.55 &
   0.258  & -10.1  & 10.9  & -11.7  & -1.69  & -27.2  & 10.7  & -4.87  \\
 \hline
\end{tabular}}
\caption{Twist-two and twist-three CFFs used for numerical comparison. The values are obtained using GPDs calculated in a spectator model and perturbatively evolved to the scale of the chosen $Q^2$ \cite{Gonzalez-Hernandez:2012xap,Kriesten:2021sqc}.} 
\label{tab:CFF}
\end{table*}

One can further separate the WW CFFs into the real and imaginary parts. For instance, we have for the real part of $\mathcal G_1(\xi,t)$,
\begin{equation}
\Re e \mathcal{G}_{1}(\xi,t) =\int_{0}^{+1}dx\Bigg[\frac{1}{x+\xi}\ln\Big| \frac{2\xi}{\xi -x}\Big|  + \frac{1}{x-\xi}\ln\Big| \frac{2\xi}{x+\xi}\Big| \Bigg]\Big(H^{+}(x,\xi,t) + E^{+}(x,\xi,t) \Big) \ ,
\end{equation}
and the imaginary part as,
\begin{eqnarray}
\Im m \mathcal{G}_{1}(\xi,t) &=& -\pi \int_{\xi}^{+1}dx \frac{1}{x+\xi}\Big(H^{+}(x,\xi,t) + E^{+}(x,\xi,t) \Big) \ ,
\end{eqnarray}
where the notation of GPDs is defined as,
\begin{eqnarray}
H^{+}(x,\xi,t)=\sum_{q} e_q^2 \left[H_q(x,\xi,t)-H_q(-x,\xi,t)\right ]\ ,
\end{eqnarray}
and similarly for $E^+(x,\xi,t)$. The the WW approximated twist-three CFFs can be calculated using GPDs from a spectator model. The quark distributions are fitted first to PDF extractions, and then to flavor separated elastic scattering form factors in the integrated limit. The GPD parametrization includes anti-quark and gluon distributions parametrized by fitting to Lattice QCD data. The GPDs are evolved perturbatively from an initial scale to the chosen $Q^2$, see GPD details in refs. \cite{Goldstein:2010gu,Gonzalez-Hernandez:2012xap,Kriesten:2021sqc}. The values of twist-three WW CFFs are presented in table  \ref{tab:CFF} together with the twist-two CFFs at the same kinematical points using the same GPD inputs.

\subsection{The $Q$ dependence of the cross-sections and the kinematical suppression}

With the estimated CFFs in table \ref{tab:CFF}, our comparisons of scalar coefficients can be turned into the cross-section comparisons. We should note that the estimated cross-sections do not correspond to any predictions of cross-sections, since those CFFs are neither fitted to experimental data nor calculated from first principal calculation like lattice QCD. However, the CFFs are from the same GPD model with the same input twist-two GPDs,  and therefore they can be considered as a self-consistent input which is necessary for cross-sections estimation. The main purpose of this subsection is to show how the cross-sections behave kinematically, assuming reasonable GPD inputs and WW approximation.

\begin{figure}[t]
\centering
\begin{minipage}[b]{\textwidth}
\includegraphics[width=0.48\textwidth]{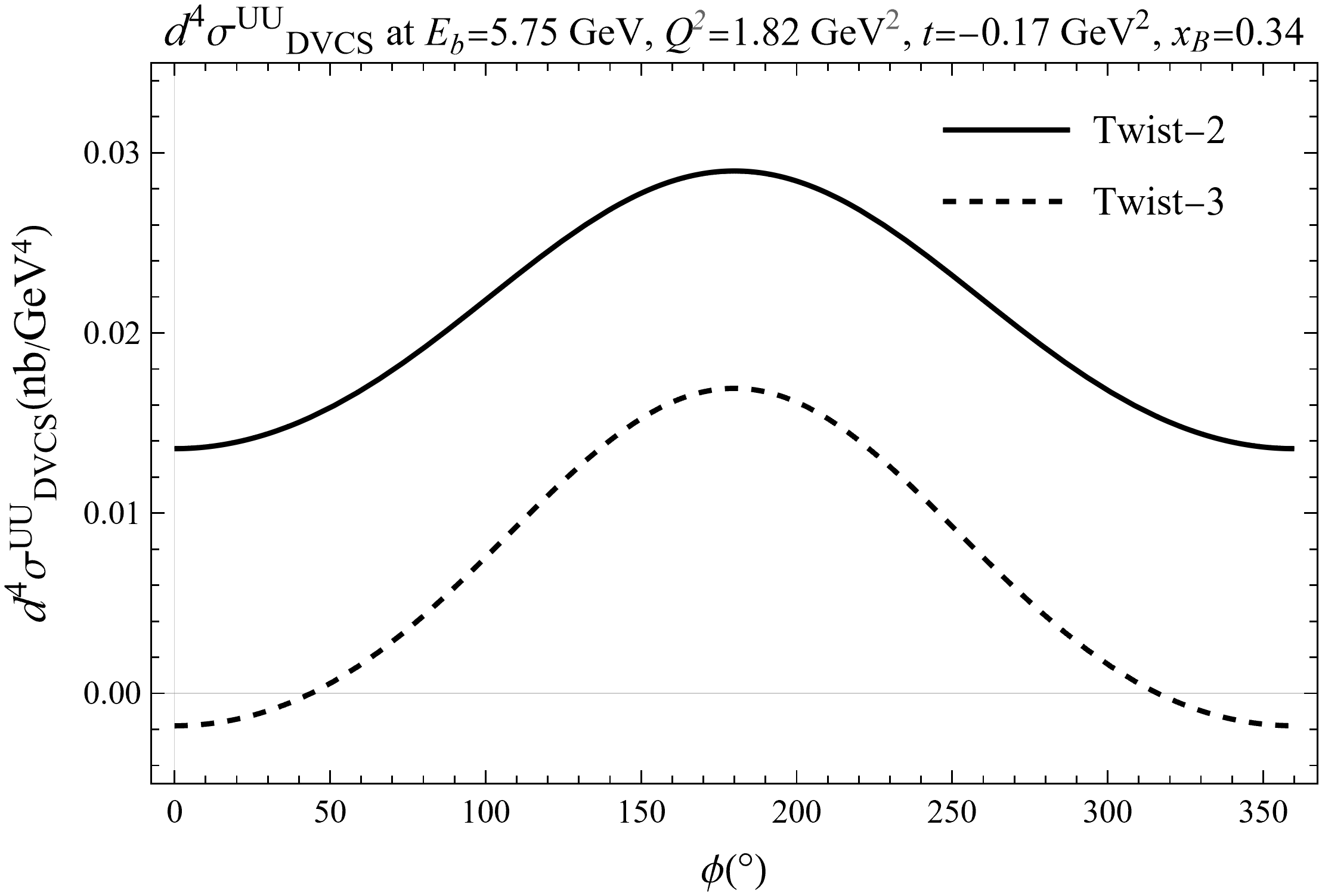}
\includegraphics[width=0.51\textwidth]{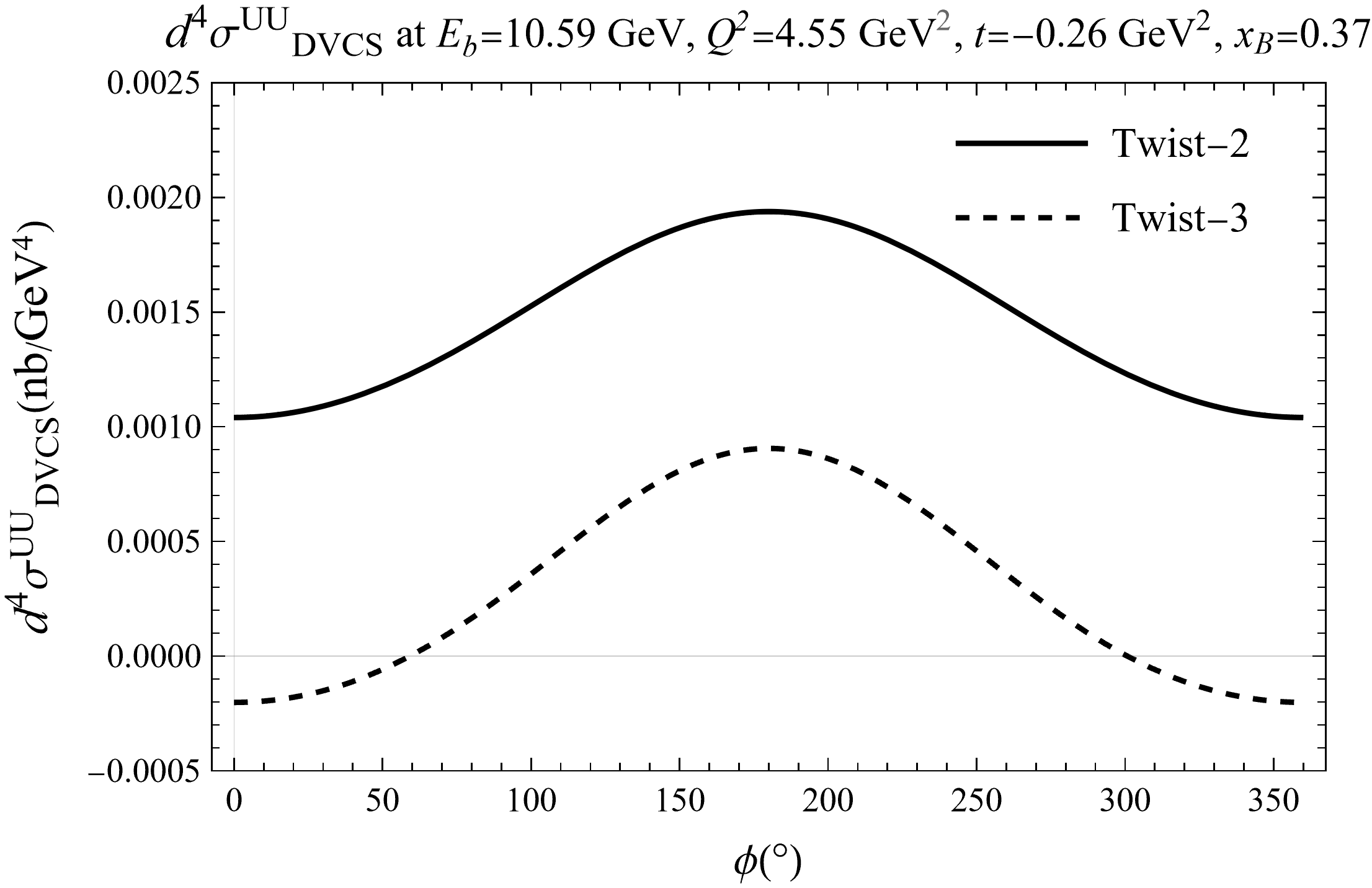}
\end{minipage}
\begin{minipage}[b]{\textwidth}
\includegraphics[width=0.48\textwidth]{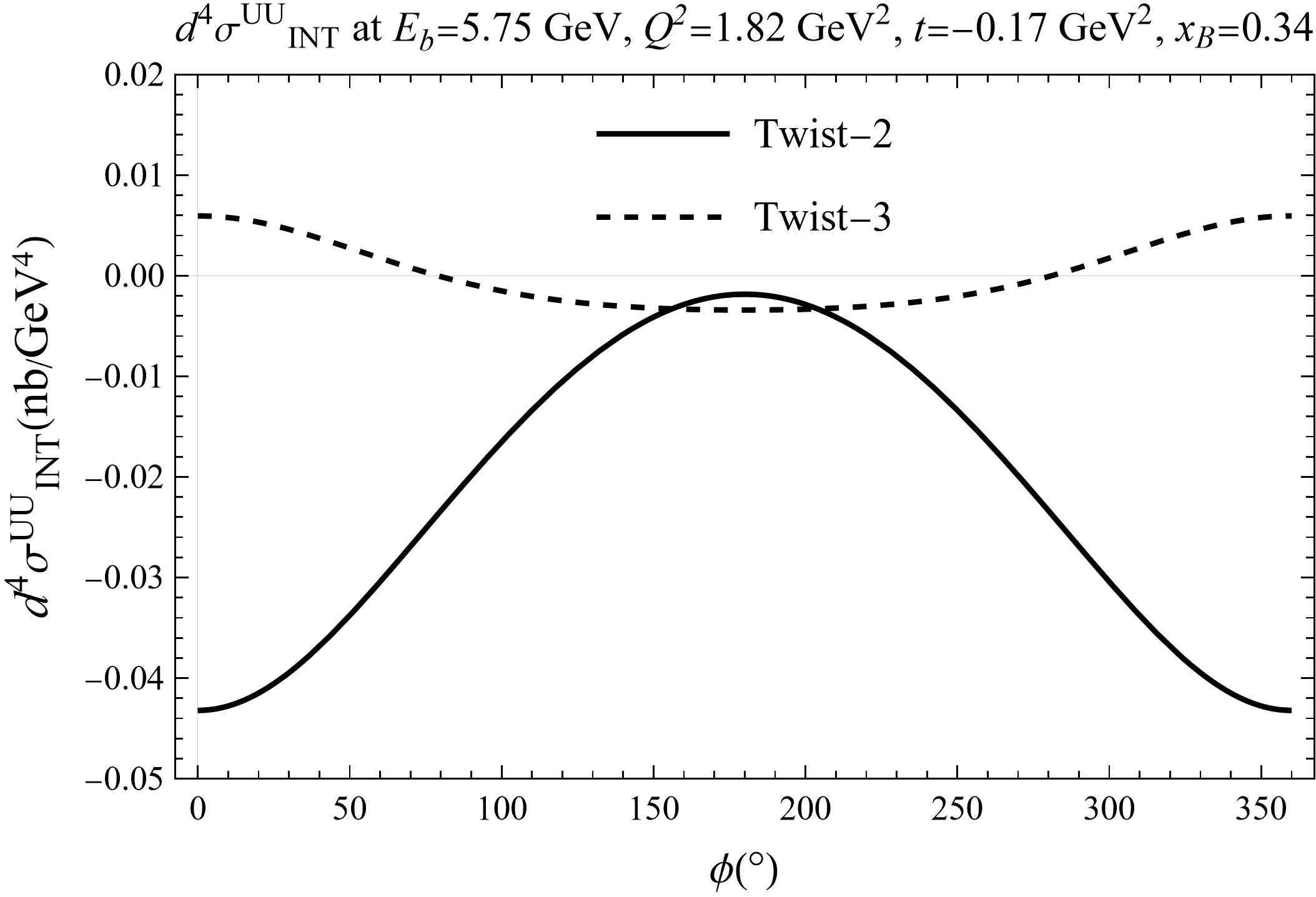}
\includegraphics[width=0.5\textwidth]{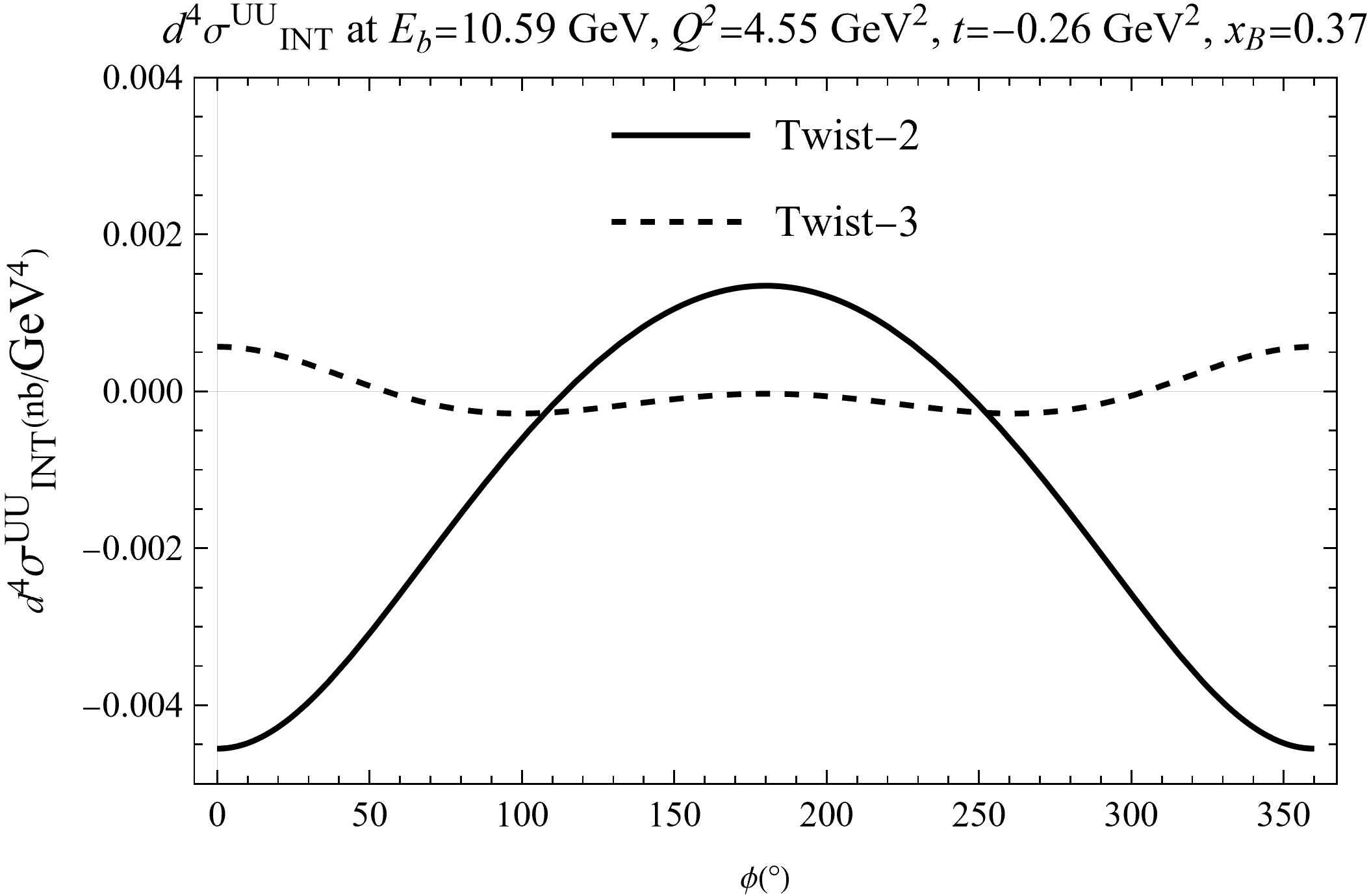}
\end{minipage}
\caption{\label{fig:xsection} A comparison of pure DVCS cross-sections (first row) and interference cross-sections (second row) at JLab 6 GeV (left) and 12 GeV (right) kinematics. The twist-two (solid lines) and twist-three (dashed lines) cross-sections are associated with the twist-two and twist-three CFFs, respectively. The twist-three pure DVCS cross-sections can be quite comparable to the twist-two ones, while the twist-three interference contributions are more suppressed.}
\end{figure}

As an example, we consider the four-fold unpolarized cross-sections,
\begin{equation}
    \frac{\text{d}^4 \sigma^{\rm{UU}}}{\text{d}x_B\text{d}Q^2\text{d}|t| \text{d}\phi}=\int_0^{2\pi} \text{d}\phi_S\frac{\text{d}^5 \sigma^{\rm{UU}}}{\text{d}x_B\text{d}Q^2\text{d}|t| \text{d}\phi\text{d}\phi_S}\ ,
\end{equation}
which is shown in figure \ref{fig:xsection}. In the plots, especially the top left one for pure DVCS cross-section at JLab $6$ GeV, we can see that the twist-three cross-sections are clearly comparable to the twist-two cross-sections for the relatively low $Q^2$, even though the scalar coefficients are shown to be suppressed at this point in figure \ref{fig:dvcsscalar}. The reason behind this observation is that although we assume CFFs to be $\mathcal{O}(1)$ quantities, and consequently their combinations in the cross-sections formulas should be as well, the actual values of those combinations could be greater than one. Then for a relatively weak kinematical suppression, the final twist-three cross-sections can be comparable when those scalar coefficients are combined with larger CFFs which are still $\mathcal{O}(1)$. Therefore, for someone who is interested in the leading-twist effects, it is necessary to go to the region where twist-three are strongly suppressed to avoid such problem. However, as we mentioned before, the pure DVCS and interference cross-sections are suppressed relative to the BH contributions for large $Q^2$. Then we should be extremely careful of what region of kinematical space to go to study the leading-twist effects. 

\begin{figure}[t]
\centering
\begin{minipage}[b]{\textwidth}
\includegraphics[width=0.49\textwidth]{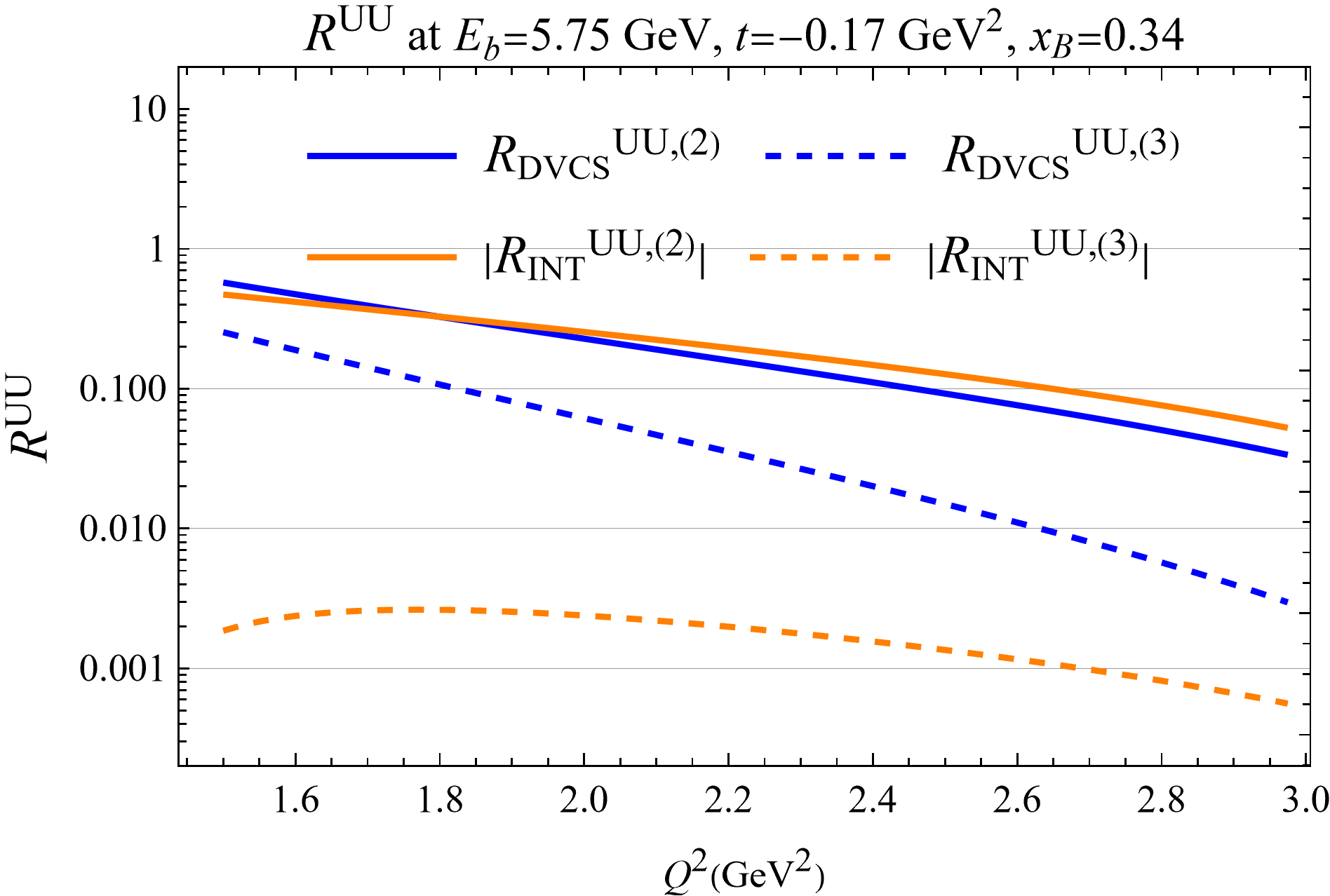}
\includegraphics[width=0.49\textwidth]{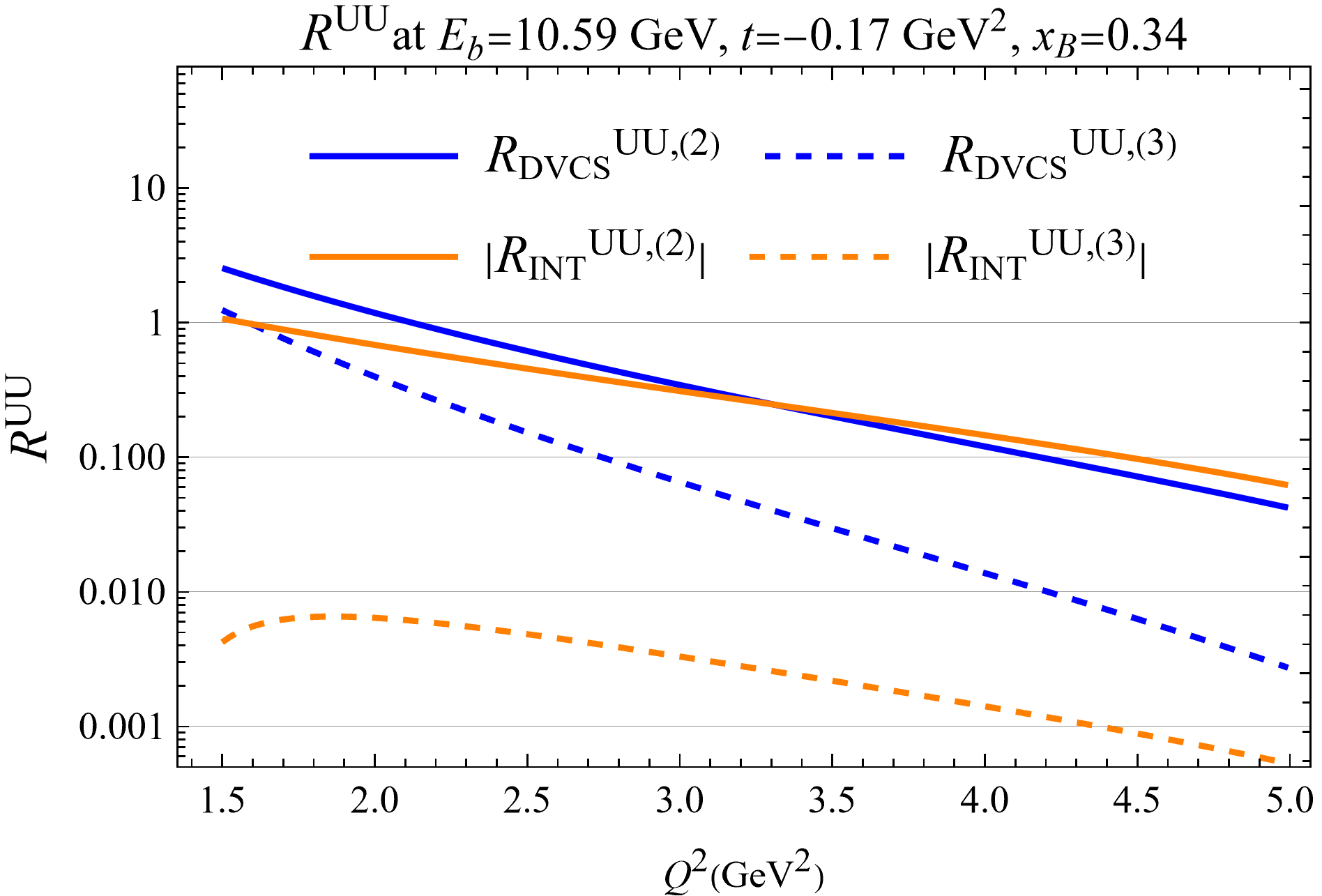}
\end{minipage}
\caption{\label{fig:ratiojlab6and12} A comparison of cross-section ratios at JLab 6 GeV (left) and 12 GeV (right) kinematics. The absolute values of them are taken when negative values are encountered. The CFFs in table \ref{tab:CFF} are used, assuming weak $Q^2$-dependence of them. The plots show clearly the suppression of the twist-two (solid lines) and twist-three (dashed lines) contributions relative to the BH background as the $Q^2$ increases. In addition, it is also apparent that the twist-three effects get suppressed faster, as indicated by their high-twist nature.}
\end{figure}

In order to push the above analysis further, we first integrate out the azimuthal variable $\phi$, which is not necessary for the kinematical analysis here (though we note that the azimuthal dependence is crucial for separating the pure DVCS and interference contributions, see ref. \cite{Belitsky:2001ns,Shiells:2021xqo}), and we get the three-fold cross-sections
\begin{equation}
\label{eq:3foldxsection}
    \frac{\text{d}^3 \sigma^{\rm{UU}}}{\text{d}x_B\text{d}Q^2\text{d}|t| } =\int_0^{2\pi}\text{d}\phi \text{d}\phi_S\frac{\text{d}^5 \sigma^{\rm{UU}}}{\text{d}x_B\text{d}Q^2\text{d}|t| \text{d}\phi\text{d}\phi_S}\ .
\end{equation}
Since we are interested in their ratios instead of the absolute cross-sections, we normalize the pure DVCS and interference cross-sections by the BH ones and define the ratios
\begin{equation}
   R_i^{\rm{UU}}(x_B,Q^2,t)=\frac{\text{d}^3 \sigma^{\rm{UU}}_i}{\text{d}x_B\text{d}Q^2\text{d}|t| } \left(\frac{\text{d}^3 \sigma^{\rm{UU}}_{\text{BH}}}{\text{d}x_B\text{d}Q^2\text{d}|t| }\right)^{-1}\ ,
\end{equation}
with $i\in\left\{\text{DVCS},\text{INT}\right\}$, and take the absolute value of the ratios if necessary. At last, we perform a twist separation for those ratios as 
\begin{equation}
   R_i^{\rm{UU}}=R_i^{\rm{UU},(2)}+R_i^{\rm{UU},(3)}+\cdots\ .
\end{equation}
Here the $R_i^{\rm{UU},(2)}$s are associated with twist-two CFFs, and the $R_i^{\rm{UU},(3)}$s are associated with twist-three CFFs, whereas higher-order CFFs are neglected. Ideally, one would look for the region where $R_i^{\rm{UU},(3)} \ll R_i^{\rm{UU},(2)}$ such that twist-three effects are negligible, while $R_i^{\rm{UU},(2)}$ are still sizable such that one could separate the twist-two effects from BH contribution. To find the region numerically, we calculated those ratios for the JLab $6$ GeV and $12$ GeV kinematics, as shown in figure \ref{fig:ratiojlab6and12}. 
The plot of JLab 6 GeV shows explicitly the conflict we just stated above --- in the region where the twist-three effects (dashed lines) are suppressed compared to the twist-two ones (solid lines), the twist-two effects themselves are also suppressed compared to the BH contribution.

However, this conflict seems to be relieved for the JLab 12 GeV plots on the right. As for $Q^2$ of about $4$ GeV$^2$, the twist-three effects are reasonably suppressed (less than $2\%$ of the BH background), while the twist-two effects are about $20\%$ of the BH background which are still sizable. This enhancement at large beam energy is known as the small-$y$ enhancement --- the ratios of pure DVCS and interference cross-sections to the BH background are inversely proportional to certain powers of $y$, so their contributions get enhanced in the small $y$ regions, which correspond to the regions with higher beam energy $E_b$ or higher center of mass energy $s$ for fixed $Q^2$ according to eq. (\ref{eq:Q2fromy}). More specifically, in terms of the cross-sections, the BH contributions get suppressed for increasing $E_b$ or $s$, while the pure DVCS and interference contributions stay at around the same order. Thus, the ratios of the pure DVCS and interference contributions to the BH background will be enhanced for increasing $E_b$ or $s$.

\begin{figure}[t]
\centering
\begin{minipage}[b]{\textwidth}
\includegraphics[width=0.49\textwidth]{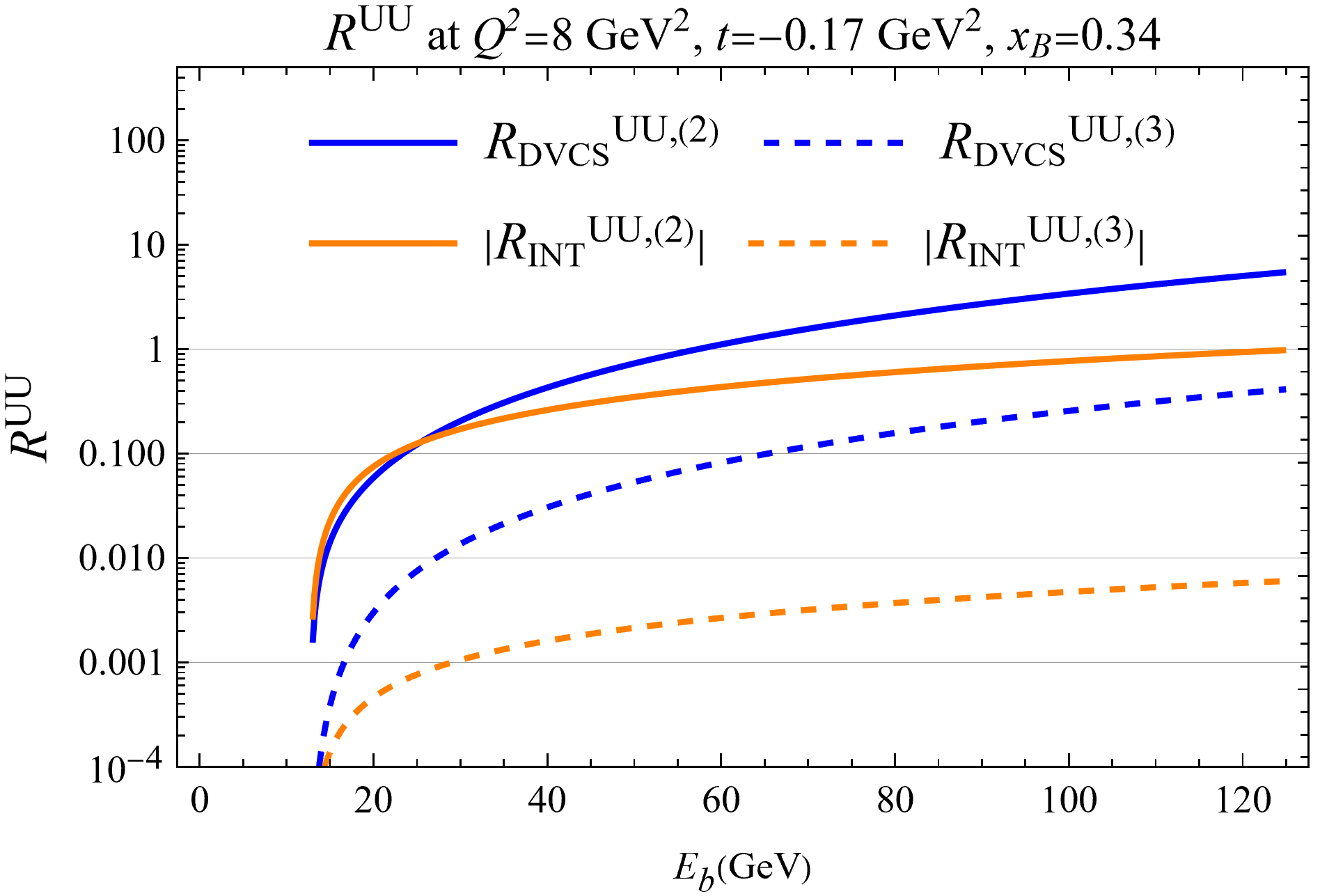}
\includegraphics[width=0.49\textwidth]{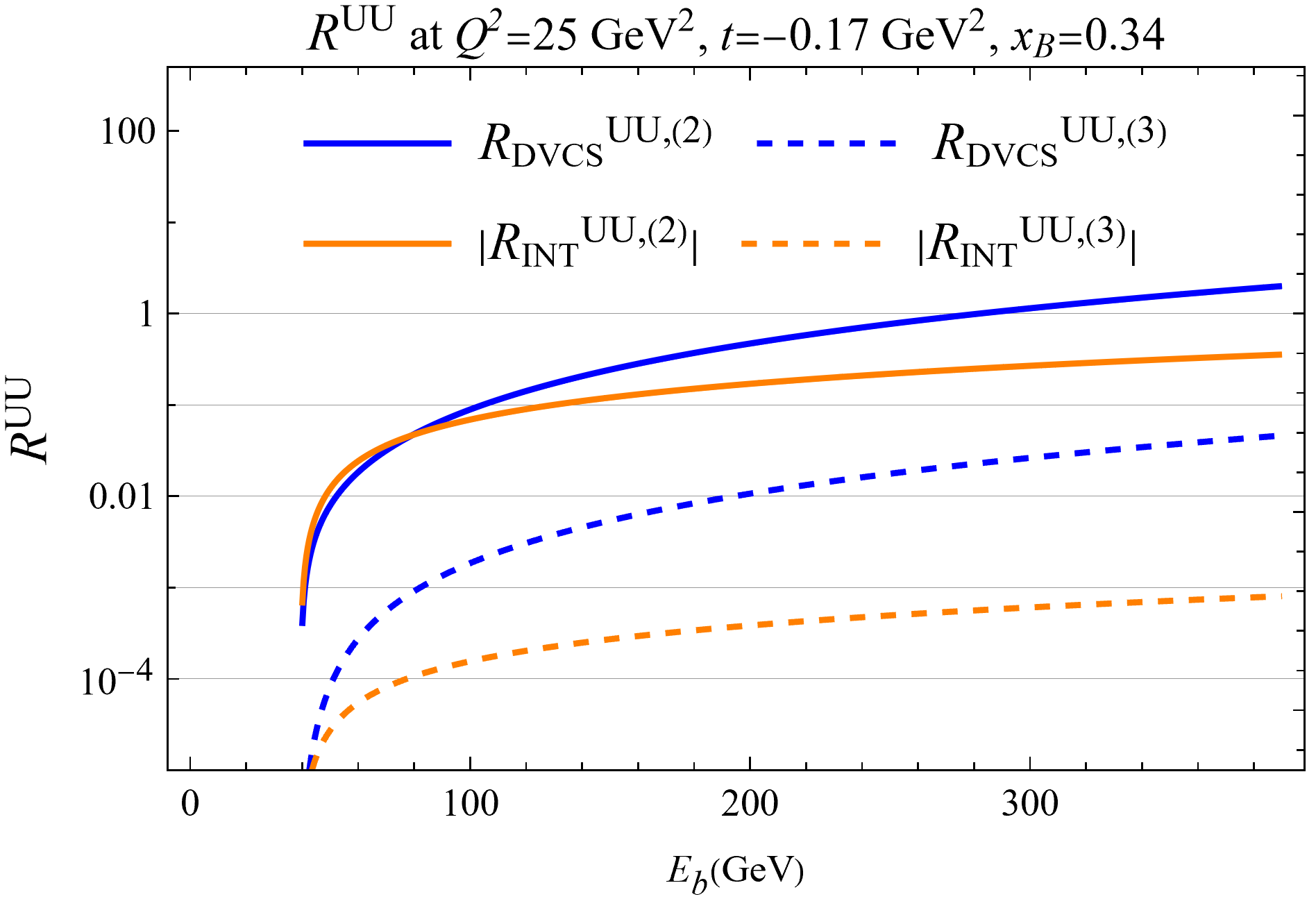}
\end{minipage}
\caption{\label{fig:ratiofixedQ2} A comparison of cross-section ratios of $Q^2 = 8 \text{ GeV}^2$ (left) and $Q^2 = 25 \text{ GeV}^2$ (right) kinematics. The absolute values of them are taken when negative values are encountered, and the CFFs in table \ref{tab:CFF} are used assuming weak $Q^2$ dependence of them. While the twist-three ratios (dashed lines) are always suppressed relative to the twist-two ratios (solid lines) due to the fixed large $Q^2$, both ratios increase as $E_b$ increases, indicating the enhancement of pure DVCS and interference contributions compared with the BH background. Note that the $x$-axis $E_b$ can be converted into $s$ using $s=M^2+2 M E_b$, which applies to collider kinematics as such the EIC5$\times$41 as well.}
\end{figure}

In figure \ref{fig:ratiofixedQ2}, we show this enhancement explicitly with numerical calculation. We note that the typical $x_B$ of EIC kinematics is much smaller than the $x_B=0.34$ here. However, the $E_b$ dependence of cross-sections ratios will be independent of that. As shown in the plots, while the higher-twist effects are suppressed by the size of the fixed $Q^2$, the cross-section ratios grow as we increase the beam energy $E_b$ (or equivalently center of mass energy $s$) and eventually go above $1$. 
This makes it possible to extract the twist-two effects from the BH background with large $E_b$ or $s$, as the extraction will not be reliable if the ratios of those twist-two effects $R^{\rm{UU},(2)}_i$ are below the relative uncertainties of the cross-section measurements or the relative uncertainties of the theoretically calculated BH contributions. Therefore, we conclude that measurements at large $Q^2$ with large beam energy $E_b$ are most suitable for the extraction of leading-twist effects.

\subsection{The $x_B$ and $t$ dependence of the cross-section coefficients}

With the explicit calculations of the $Q^2$ and $y$ dependence of twist-three cross-sections in the last subsection, one would naturally consider the $x_B$ and $t$ dependence of them as well. However, the situation gets complicated since the CFFs can not be considered as constants for different $x_B$ and $t$, and their kinematic dependence will be involved inevitably. Unfortunately, due to the lack of experimental constraints, the $x_B$ and $t$ dependence of CFFs will depend on the model of extrapolation, and consequently the predictions of cross-sections will too.

One possible way to study the cross-sections without the knowledge of the CFFs is to separate the CFFs dependence from the cross-sections. For instance, consider the unpolarized three-fold cross-section in eq. (\ref{eq:3foldxsection}). The three-fold cross-sections can be split into contributions of different twist as,
\begin{align}
\text{d}^3 \sigma^{\rm{UU}}_{\rm{DVCS}}=\text{d}^3 \sigma^{\rm{UU},(2)}_{\rm{DVCS}}+ \text{d}^3 \sigma^{\rm{UU},(3)}_{\rm{DVCS}}+\text{d}^3 \sigma^{\rm{UU},(4)}_{\rm{DVCS}}\ ,
\end{align}
for the pure DVCS cross-sections according to eq. (\ref{eq:twistdecdvcsstruct}), and 
\begin{align}
\text{d}^3 \sigma^{\rm{UU}}_{\rm{INT}}=\text{d}^3 \sigma^{\rm{UU},(2)}_{\rm{INT}}+ \text{d}^3 \sigma^{\rm{UU},(3)}_{\rm{INT}}\ ,
\end{align}
for the interference cross-sections according to eq. (\ref{eq:inttwistsep}), where the other higher-twist terms are dropped. Each term on the right-hand side depends on both the kinematical variables $(x_B,t,Q^2,\cdots)$ as well the CFFs. Nevertheless, the CFF dependence for each term is factorizable. For example, the $\text{d}^3 \sigma^{\rm{UU},(3)}_{\rm{DVCS}}$ can be written as
\begin{align}
\label{eq:rexsection}
\text{d}^3 \sigma^{\rm{UU},(3)}_{\rm{DVCS}} (x_B,t,Q^2,\cdots)= \text{d}^3 \widetilde{\sigma}^{\rm{UU},(3)}_{\rm{DVCS}}(x_B,t,Q^2,\cdots)  \times \mathcal F_{\rm{UU}}^{(3)}(x_B,t)\ ,
\end{align}
according to eq. (\ref{eq:fuu3}) where we suppress the $Q^2$ dependence of CFFs due to radiative corrections. The $\mathcal F_{\rm{UU}}^{(3)}$ contains all the CFF dependence which is given by the terms in the bracket of eq. (\ref{eq:fuu3}),
\begin{align}
\begin{split}
\mathcal F_{\rm{UU}}^{(3)}=&\text{Re}\Bigg[-\Ecff^*\HTcff+\Hcff^* \ETcff+2\left(\Hcff+\frac{t}{4M^2}\Ecff\right)^*\HTtcff-\xi (\Hcff+\Ecff)^* \ETtcff\\
  &\qquad+\xi\Etcff^* \HTcff-\xi\Htcff^*\ETcff+\Htcff^*\ETtcff\Bigg]\ .
\end{split}
\end{align}
The remaining coefficient $ \text{d}^3 \widetilde{\sigma}^{\rm{UU},(3)}_{\rm{DVCS}}$ will be independent of CFFs and has the same unit as the three-fold cross-section, which can be written in terms of the scalar coefficients as,
\begin{align}
\begin{split}
 \text{d}^3 \widetilde{\sigma}^{\rm{UU},(3)}_{\rm{DVCS}}(x_B,t,Q^2,\cdots) =\frac{\Gamma }{Q^4} \int\text{d}\phi ~4 h^{\rm{U}}_{(3)}(\phi,x_B,t,Q^2,\cdots) \ , 
\end{split}
\end{align}
where we define the constant
\begin{equation}
\Gamma \equiv \frac{\alpha_{\rm{EM}}^3 x_B y^2}{8\pi Q^4\sqrt{1+\gamma^2}} \ ,
\end{equation}
and the $h^{\rm{U}}_{(3)}$ is the scalar coefficient in eq. (\ref{eq:fuu3}). 

We note that the cross-section $\text{d}^3 \sigma^{\rm{UU},(3)}_{\rm{DVCS}}$ can have different $x_B$  and $t$ dependence from the coefficient $\text{d}^3 \widetilde{\sigma}^{\rm{UU},(3)}_{\rm{DVCS}}$, due to the $x_B$ and $t$ dependence of $\mathcal F_{\rm{UU}}^{(3)}$, which is associated with the unknown non-perturbative physics. Using the coefficients $\text{d}^3 \widetilde{\sigma}^{\rm{UU}}_{\rm{DVCS}}$ to compare the $x_B$ and $t$ dependence of cross-sections assumes the universal $x_B$ and $t$ dependence of $\mathcal F_{\rm{UU}}$. The test of this assumption is crucial to our analysis, which is, however, beyond the scope of this work or any perturbative frameworks.

With the cross-section coefficients introduced in eq. (\ref{eq:rexsection}), we define the following quantities in the same manner, 
\begin{align}
\begin{split}
 \text{d}^3 \widetilde{\sigma}^{\rm{UU},(2)}_{\rm{DVCS}}(x_B,t,Q^2,\cdots) =  \frac{\Gamma}{Q^4} \int\text{d}\phi ~4 h^{\rm{U}}_{(2)}(\phi,x_B,t,Q^2,\cdots) \ , 
\end{split}\\
\begin{split}
 \text{d}^3 \widetilde{\sigma}^{\rm{UU},(4)}_{\rm{DVCS}}(x_B,t,Q^2,\cdots) =\frac{\Gamma}{Q^4} \int\text{d}\phi ~4 h^{\rm{U}}_{(4)}(\phi,x_B,t,Q^2,\cdots) \ , 
\end{split}
\end{align}
for the pure DVCS contributions and 
\begin{align}
\begin{split}
 \text{d}^3 \widetilde{\sigma}^{\rm{UU},(2)}_{\rm{INT,A}}(x_B,t,Q^2,\cdots) =\frac{\Gamma }{Q^2 |t|} \int\text{d}\phi ~A^{\rm{I,U}}_{(2)}(\phi,x_B,t,Q^2,\cdots) \ , 
\end{split}\\
\begin{split}
 \text{d}^3 \widetilde{\sigma}^{\rm{UU},(3)}_{\rm{INT,A}}(x_B,t,Q^2,\cdots) =\frac{\Gamma }{Q^2 |t|} \int\text{d}\phi ~A^{\rm{I,U}}_{(3)}(\phi,x_B,t,Q^2,\cdots) \ , 
\end{split}\\
\begin{split}
 \text{d}^3 \widetilde{\sigma}^{\rm{UU},(2)}_{\rm{INT,C}}(x_B,t,Q^2,\cdots) =\frac{\Gamma }{Q^2 |t|} \int\text{d}\phi ~C^{\rm{I,U}}_{(2)}(\phi,x_B,t,Q^2,\cdots) \ , 
\end{split}\\
\begin{split}
 \text{d}^3 \widetilde{\sigma}^{\rm{UU},(3)}_{\rm{INT,C}}(x_B,t,Q^2,\cdots) =\frac{\Gamma }{Q^2 |t|} \int\text{d}\phi ~C^{\rm{I,U}}_{(3)}(\phi,x_B,t,Q^2,\cdots) \ , 
\end{split}
\end{align}
for the interference contributions. The unpolarized interference cross-sections actually involve three scalar coefficients $A^{\rm{I,U}}$, $B^{\rm{I,U}}$ and $C^{\rm{I,U}}$. Since the $B^{\rm{I,U}}$ is kinematically suppressed by large $Q^2$ and practically negligible compared with the others, we only focus on the other two coefficients for which the above quantities are defined. Different from our analysis in section \ref{sec:t3scalar}, here we integrate out the azimuthal $\phi$ dependence and focus on the $x_B$  and $t$ dependence.

\begin{figure}[t]
\centering
\begin{minipage}[b]{\textwidth}
\includegraphics[width=\textwidth]{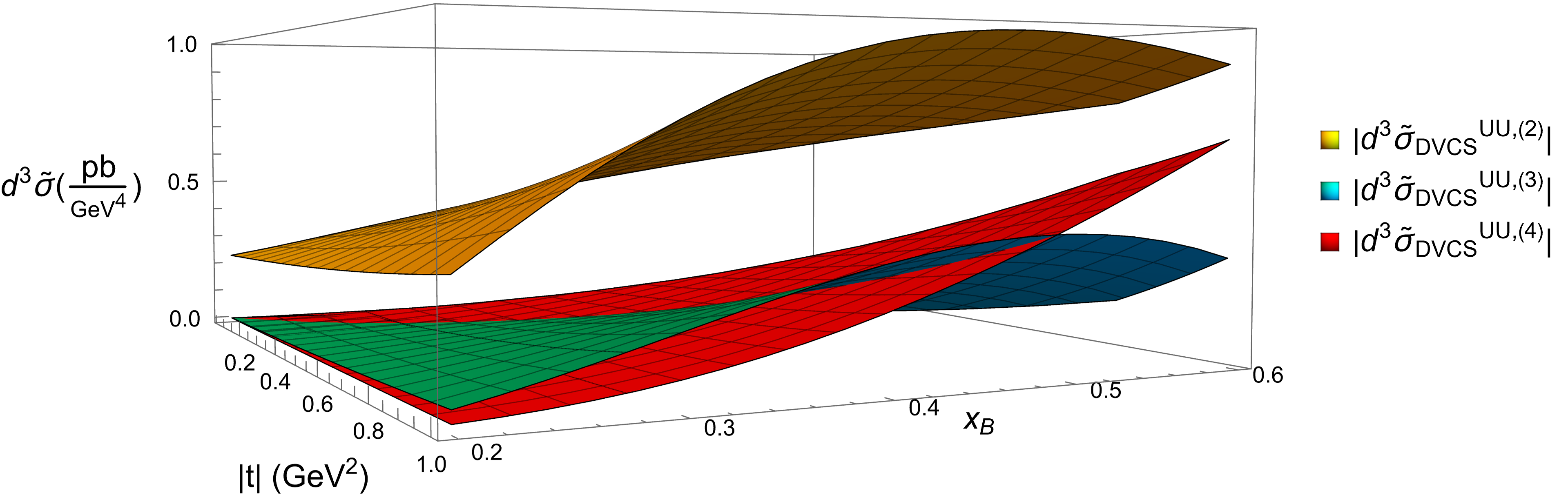}
\end{minipage}
\caption{\label{fig:DVCSxsection3D} A comparison of twist-two and twist-three pure DVCS cross-section coefficients at JLab 12 GeV kinematics with $E_b=10.59 \text{ GeV}$ and $Q^2=4 \text{ GeV}^2$. The absolute values are taken for comparison of the size. The leading-twist (the top yellow surface) effect is clearly dominating the other higher-twist effects in the small $x_B$ regions, while the higher-twist effects (the lower two surfaces) get more relevant as $x_B$ increases.}
\end{figure}

\begin{figure}[t]
\centering
\begin{minipage}[b]{\textwidth}
\includegraphics[width=\textwidth]{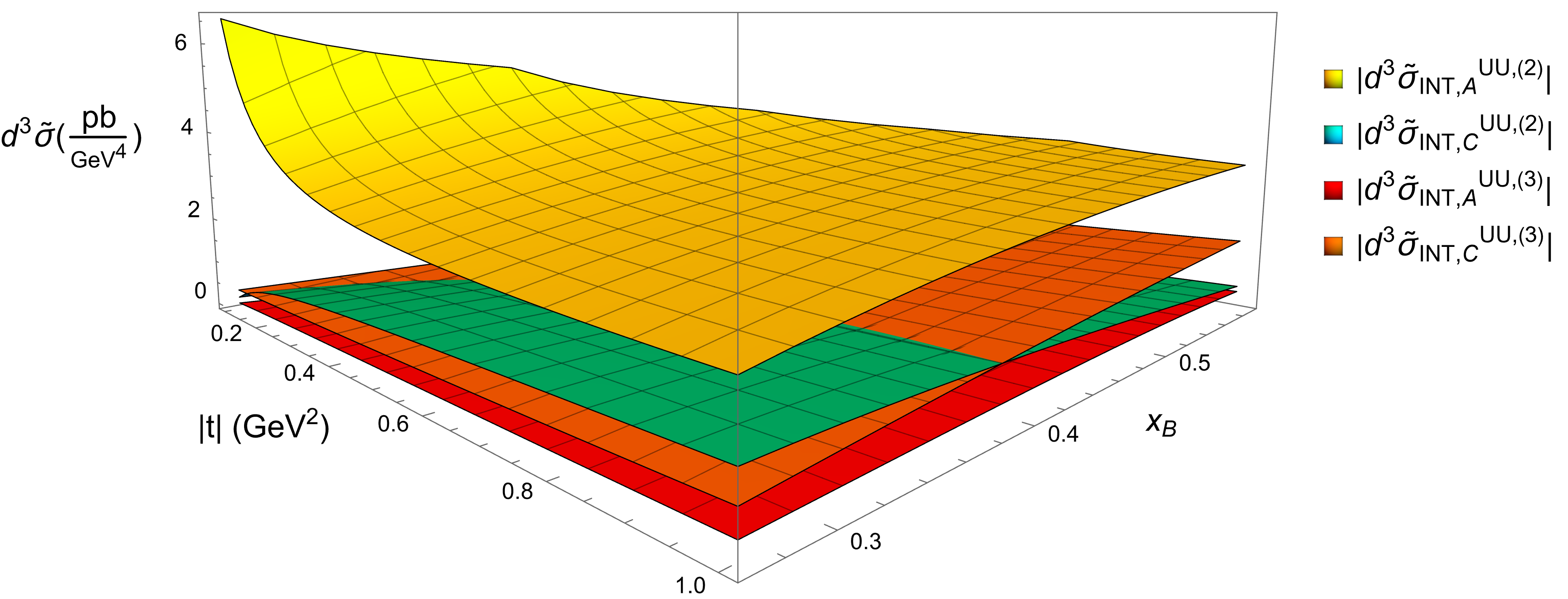}
\end{minipage}
\caption{\label{fig:INTxsection3D} A comparison of twist-two and twist-three interference cross-section coefficients at JLab 12 GeV kinematics with $E_b=10.59 \text{ GeV}$ and $Q^2=4 \text{ GeV}^2$. The absolute values are taken for comparison of the size.  The twist suppression for the interference cross-sections works well too, as the leading-twist effect (specifically the top yellow surface) is dominating. Unlike the pure DVCS case, the twist suppression is not affected as $x_B$ increases.  }
\end{figure}

We consider the typical JLab 12 GeV kinematics with $E_b= 10.59 \text{ GeV}$ with a fixed $Q^2=4 \text{ GeV}^2$. At this choice of $Q^2$, we expect the leading-twist contributions to dominate, while the twist-three effects are not strongly suppressed and might have sizable effects, which shall be tested numerically. In order to do so, the above cross-section coefficients are calculated for a wide range of $x_B$ and $t$, for which we vary $x_B$ and $t$ in the range
\begin{align}
    0.25<x_B<0.6 \qquad \text{and}\qquad |t|_{\rm{min}}<|t|<1 \text{ GeV}^2 \ ,
\end{align}
based on the simulations in ref. \cite{CLAS:2015uuo}. Note that the absolute value of momentum transfer $|t|$ has a kinematic lower bound $|t|_{\rm{min}}$  given by
\begin{align}
 -4M^2\xi^2-t(1-\xi^2) >0 \qquad \to \qquad |t|_{\rm{min}}= \frac{4M^2 \xi^2}{1-\xi^2}\approx \frac{M^2 x_B^2}{1-x_B} \ ,
\end{align}
so large $x_B$ will be associated with large momentum transfer $|t|$.
The comparisons of the cross-sections coefficients are shown in figure \ref{fig:DVCSxsection3D}
for pure DVCS contributions and figure \ref{fig:INTxsection3D} for interference contributions.

\begin{figure}[t]
\centering
\includegraphics[width=\textwidth]{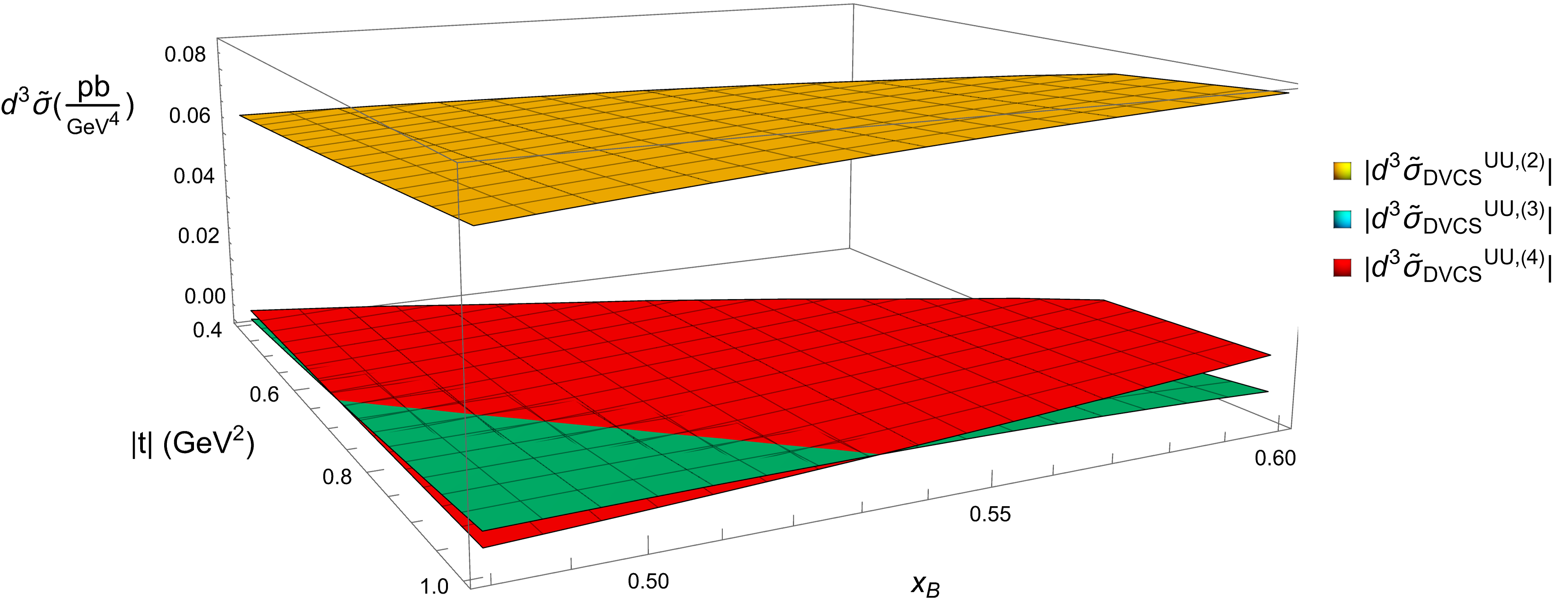}
\caption{\label{fig:xsection3Dp2} A comparison of twist-two and twist-three pure DVCS cross-section coefficients at JLab 12 GeV kinematics with $E_b=10.59 \text{ GeV}$ and $Q^2=8 \text{ GeV}^2$ in the high $x_B$ regions. The absolute values are taken for comparison of the size. The higher-twist effects are indeed suppressed for such large $Q^2$, with the leading-twist top yellow surface dominating. The same twist suppression can be shown for the interference contributions, which it is already the case for the lower $Q^2=4 \text{ GeV}^2$ as shown in figure \ref{fig:INTxsection3D}.}
\end{figure}

According to the numerical results, while the twist suppression generally works well for the interference contribution, it seems the higher-twist pure DVCS contributions get more important for large $x_B$ which is associated with large $|t|$. This requires larger $Q^2$ in order to suppress the higher-twist effects, since they are generally suppressed by the factor $t/Q^2$. Indeed, as shown in the recent JLab Hall A DVCS measurements at high $x_B$ \cite{JeffersonLabHallA:2022pnx}, a large $Q^2$ of $8.40$ GeV$^2$ associated with beam energy $E_b=10.59$ GeV is set, when $x_B$ reaches $0.6$. By repeating the same calculations above, we show that the higher-twist effects are indeed kinematically suppressed, as shown in figure \ref{fig:xsection3Dp2}. On the other hand, as discussed in the last subsection, increasing $Q^2$ with fixed beam energy $E_b$ leads to the dominance of BH contributions. For instance, the three-fold BH cross-section is about $1.2 \text{ pb}/\text{GeV}^4$ at $Q^2=8.40 \text{ GeV}^2, x_B=0.6, E_b=10.59 \text{ GeV}$ and $t=-0.91\text{ GeV}^{2}$. Comparing this with the pure DVCS cross-section coefficient in figure \ref{fig:xsection3Dp2} which is about $0.08 \text{ pb}/\text{GeV}^4$ at this point, it indicates that the pure DVCS contributions are typical one order of magnitude lower than the BH background here, assuming the combination of twist-two CFFs $\mathcal F_{\rm{UU}}^{(2)}$ is of order 1. According to the measurements and fitting in ref. \cite{JeffersonLabHallA:2022pnx}, the BH cross-section is indeed about $3$--$4$ times the pure DVCS cross-section with fitted CFFs at this kinematical point.

It is also worth noting that the analysis will be more complicated if the azimuthal $\phi$ dependence is kept rather than integrated out, since the coefficients of different harmonics functions could have different kinematical behaviors. An analysis with azimuthal dependence of all polarization configurations can be done in the same manner as our analysis here, together with the general set-up as described in ref. \cite{Shiells:2021xqo}.

\section{Conclusion}
\label{sec:conc}
We present the DVCS cross-section formulas with both twist-two and twist-three CFFs for all polarization configurations. We show that the twist-three cross-sections formulas can be expressed in terms of the twist-three scalar coefficients and twist-three CFFs, similar to the twist-two case. We present our cross-section formulas in a frame-independent manner, which apply to both fixed-target and collider coordinates. We also study the transformation relation between the two coordinates explicitly.

With the cross-section formulas, we compare the twist-two and twist-three contributions to the cross-sections at typical JLab 6 GeV, JLab 12 GeV and EIC kinematics. We show that the twist-three scalar coefficients are indeed kinematically suppressed even for a relatively low $Q^2$ of $1.82$ GeV$^2$ for the typical JLab 6 GeV kinematical points. Our results justify the kinematical suppression of higher-twist effects and allow us to extract leading-twist CFFs from cross-sections measurements with large $Q^2$. On the other hand, those twist-three effects can in principle be measured with relatively low $Q^2$ once the leading-twist effects are determined with precision, allowing us to access the higher-twist CFFs from experiments.

We estimate the twist-three CFFs with the WW approximation and a dynamical framework of GPD parameterization based on the reggeized spectator model. Those estimated twist-three CFFs are the same order of the twist-two ones for the given inputs. With those WW twist-three CFFs as well as the twist-two CFFs calculated from the same twist-two GPDs, we compare the twist-two and twist-three cross-sections. We show that the twist-three cross-sections can have a sizable effect for relatively low $Q^2$, and one will have to go to the large $Q^2$ regions in order for the dominance of leading-twist effects. We also argue that the suppression of leading-twist effects relative to the BH background at large $Q^2$ can be compensated by going to the small-$y$ regions, which corresponds to the high beam energy or higher center of mass energy regions. We also study the $x_B$ and $t$ dependence of the cross-section coefficients with JLab 12 GeV kinematics, which indicates that the higher-twist effects could be sizable for large $x_B$ and correspondingly large $|t|$, which require higher $Q^2$.

\section*{Acknowledgments}
We thank V. Braun and F. Yuan for discussions related to the subject of this paper. This research is supported by the U.S. Department of Energy, Office of Science, Office of Nuclear Physics, under contract number DE-SC0020682, and the Center for Nuclear Femtography, Southeastern Universities Research Association, Washington D.C. Y. Guo is partially supported by a graduate fellowship from Center for Nuclear Femtography, SURA, Washington DC.
\newpage
\appendix
\section{Relations between fixed-target and collider coordinates}
\label{app:eicframe}
In this appendix, we show how the vectors can be transformed between the two different coordinates for fixed-target and collider experiments. In general, the vectors are related through,
\begin{equation}
    V^{(c)}= \Lambda(K^z,\eta) \Lambda(J^y,\theta) V\ ,
\end{equation}
where the vector $V$ in the fixed-target coordinates is first rotated about the $y$-axis, such that the electron beam will be oriented in the $-z$ direction, then the system will be boosted to the given electron/proton beam energy configuration. The electron beam four-momentum can be written as
\begin{equation}
    k^{\mu}=|k_0|(1,\sin\theta_l,0,\cos\theta_l)\ ,
\end{equation}
where we write the four-vectors $r^\mu$ in terms of its components as $(r^0,r^1,r^2,r^3)$ and we define~\cite{Kriesten:2019jep},
\begin{equation}
    \sin \theta_l=\frac{\gamma}{\sqrt{1+\gamma^2}}\sqrt{1-y-\frac{y^2\gamma^2}{4}}\ ,
\end{equation}
Then the rotation matrix can be written as,
\begin{equation}\Lambda(J^z,\theta=\pi-\theta_l)=
   \begin{pmatrix}
    1&0&0&0\\
    0&\cos(\pi-\theta_l)&0&\sin(\pi-\theta_l)\\
    0&0&1&0\\
    0&-\sin(\pi-\theta_l)&0&\cos(\pi-\theta_l)
   \end{pmatrix}\ ,
\end{equation}
such that $\Lambda(J^z,\theta=\pi-\theta_l)  k^{\mu}$ is in the $-z$ direction. Then for the final photon momentum $q'^\mu=|q'_0|(1,\sin\theta \cos\phi,\sin\theta \cos\phi,\cos\theta)$ where $\theta$ are given by 
\begin{equation}
    \cos\theta =-\frac{1+\frac{\gamma^2}{2}\frac{Q^2+t}{Q^2+x_B t}}{\sqrt{1+\gamma^2}}\ ,
\end{equation}
and satisfies $\sin(\theta)>0$. Then one can solve $\phi^{(c)}$ with the equation,
\begin{equation}
    \frac{\sin\phi^{(c)}}{\cos\phi^{(c)}}=  \frac{\sin\phi\sin\theta}{-\cos\theta_l \cos\phi\sin\theta+\cos\theta \sin\theta_l}\ ,
\end{equation}
for which one can easily write down a formal solution of $\phi^{(c)}$ using the $\arctan$ function, though it should be carefully treated as $\phi^{(c)}\in (0,2\pi)$ whereas $\arctan$ takes the value from $(-\pi/2,\pi/2)$. Another transverse quantity that needs to be transformed is the transverse polarization vectors $S_T^{\mu}$. However, the transverse direction depends on the coordinates one choose, so we should take both transverse and longitudinal polarization vectors into account, for which we write
\begin{align}
    S^\mu&=2\Lambda_L(0,0,0,1) +2\Lambda_T \left(0,\cos\phi_S,\sin\phi_S,0\right)\ , \\
    S^{(c)\;\mu}&=2\Lambda_{L}^{(c)}(0,0,0,1) +2\Lambda_{T}^{(c)} \left(0,\cos\phi_{S}^{(c)},\sin\phi_{S}^{(c)},0\right)\ ,
\end{align}
In order to solve the transformation, we define two angles representing the polarization configuration by
\begin{align}
    \theta_S=\frac{\Lambda_T}{\sqrt{\Lambda_T^2+\Lambda_L^2}}\ , \qquad  \theta_{S}^{(c)}=\frac{\Lambda_{T}^{(c)}}{\sqrt{\left(\Lambda_{T}^{(c)}\right)^2+\left(\Lambda_{L}^{(c)}\right)^2}}\ ,
\end{align}
and the equations reads
\begin{align}
    \frac{\sin\phi^{(c)}_S}{\cos\phi^{(c)}_S}&=  \frac{\sin\phi_S\sin\theta_S}{-\cos\theta_l \cos\phi_S\sin\theta_S+\cos\theta_S \sin\theta_l}\ ,\\
   \cos\theta_S^{(c)}&=-\cos\theta_S\cos\theta_l-\cos\phi_S\sin\theta_S\sin \theta_l \ ,
\end{align}
for which again, one can write down a formal solution of $\phi^{(c)}_S$ and $\theta^{(c)}_S$ easily using inverse trigonometric function while taking care of the restricted domain of those functions. We also note that the $\phi^{(c)}_S$ actually satisfies a very similar equation as $\phi^{(c)}$, since they are derived with the same rotation matrix. There are two comments regarding this transformation. First, in the leading twist limit $\theta_l \to \pi$ and those equations reduce to $\phi^{(c)}=\phi$, $\phi^{(c)}_S=\phi_S$ and $\theta^{(c)}_S=\theta_S$, which shows that their differences are indeed higher-twist effects. Another comment is that for non-zero $\theta_l$, $\theta^{(c)}_S\not = 0$ even if $\theta_S = 0$, which shows that indeed the transverse and longitudinal polarization will get mixed when transform between these two coordinates.

\section{Twist-three pure DVCS structure functions}
\label{app:dvcsstructurefunc}

In this appendix, we present all the hadronic tensors as well as structure functions for pure DVCS contributions that are related to twist-three CFFs. Those expressions are calculated with the help of the  \textsc{FeynCalc}~\cite{Mertig:1990an,Shtabovenko:2016sxi,Shtabovenko:2020gxv} package, and the code is integrated to our previous twist-two scalar coefficients code public at ref. \cite{Guo:2021git}. We introduce the normalization constant,
\begin{equation}
N\equiv \frac{\sqrt{-4M^2\xi^2-t(1-\xi^2)}}{M}\ ,
\end{equation}
which will be used for presenting those hadronic tensor. The four twist-three hadronic tensor $H_{\rm{DVCS,U},(3)}^{\rho\sigma}$, $H_{\rm{DVCS,L},(3)}^{\rho\sigma}$, $H_{\rm{DVCS,T,in},(3)}^{\rho\sigma}$ and $H_{\rm{DVCS,T,out},(3)}^{\rho\sigma}$ can be written with the tensor structures defined in eq. (\ref{eq:h3ampdef}) as
\begin{align}
\label{eq:dvcsut3}
\begin{split}
  H_{\rm{DVCS,U},(3)}^{\rho\sigma}=&4\mathscr{H}^{\left(\rho\sigma\right)}_{(3)} \text{Re}\Bigg[-\Ecff^*\HTcff+\Hcff^* \ETcff+2\left(\Hcff+\frac{t}{4M^2}\Ecff\right)^*\HTtcff-\xi (\Hcff+\Ecff)^* \ETtcff\Bigg]\\
  +&4i \mathscr{H}^{\left[\rho\sigma\right]}_{(3)} \text{Im}\Bigg[-\Ecff^*\HTcff+\Hcff^* \ETcff+2\left(\Hcff+\frac{t}{4M^2}\Ecff\right)^*\HTtcff-\xi (\Hcff+\Ecff)^* \ETtcff\Bigg]\\
  +&4\widetilde{\mathscr{H}}^{\left(\rho\sigma\right)}_{(3)}\text{Re}\Bigg[\xi\Etcff^* \HTcff-\xi\Htcff^*\ETcff+\Htcff^*\ETtcff\Bigg]\\
  +&4i\widetilde{\mathscr{H}}^{\left[\rho\sigma\right]}_{(3)}\text{Im}\Bigg[\xi\Etcff^* \HTcff-\xi\Htcff^*\ETcff+\Htcff^*\ETtcff\Bigg]\ ,
\end{split}
\end{align}
and for $H_{\rm{DVCS, L}}^{\rho\sigma}$
\begin{align}
\begin{split}
  H_{\rm{DVCS,L},(3)}^{\rho\sigma}=&-4\mathscr{H}'^{\left(\rho\sigma\right)}_{(3)} \text{Im}\Bigg[\xi\Etcff^* \HTcff-\left(\Htcff-\frac{\xi^2}{1+\xi}\Etcff\right)^*\ETcff\\
  &\qquad\qquad\qquad-2\left(\Htcff+\xi\left(\frac{t}{4M^2}-\frac{\xi}{1+\xi}\right)\Etcff\right)^* \HTtcff+\xi\left(\Htcff+\frac{\xi}{1+\xi} \Etcff\right)^*\ETtcff\Bigg]\\
  +&4i\mathscr{H}'^{\left[\rho\sigma\right]}_{(3)} \text{Re}\Bigg[\xi\Etcff^* \HTcff-\left(\Htcff-\frac{\xi^2}{1+\xi}\Etcff\right)^*\ETcff\\
  &\qquad\qquad\qquad-2\left(\Htcff+\xi\left(\frac{t}{4M^2}-\frac{\xi}{1+\xi}\right)\Etcff\right)^* \HTtcff+\xi\left(\Htcff+\frac{\xi}{1+\xi} \Etcff\right)^*\ETtcff\Bigg]\\
  +&4\widetilde{\mathscr{H}}'^{\left(\rho\sigma\right)}_{(3)}\text{Im}\Bigg[\Ecff^*\HTcff-\xi\left( \Hcff+\frac{\xi}{1+\xi}\Ecff\right)^*\ETcff+\left( \Hcff+\frac{\xi}{1+\xi}\Ecff\right)^*\ETtcff\Bigg]\\
  -&4i\widetilde{\mathscr{H}}'^{\left[\rho\sigma\right]}_{(3)}\text{Re}\Bigg[\Ecff^*\HTcff-\xi\left( \Hcff+\frac{\xi}{1+\xi}\Ecff\right)^*\ETcff+\left( \Hcff+\frac{\xi}{1+\xi}\Ecff\right)^*\ETtcff\Bigg]\ ,
\end{split}
\end{align}
\begin{align}
\begin{split}
  H_{\rm{DVCS,L},(4)}^{\rho\sigma}=0\ ,
\end{split}
\end{align}
and for $H_{\rm{DVCS,T,in}}^{\rho\sigma}$
\begin{align}
\begin{split}
  &H_{\rm{DVCS,T,in},(3)}^{\rho\sigma}\\
  =&\frac{4}{N}\mathscr{H}'^{(\rho\sigma)}_{(3)} \text{Im}\Bigg[2\left((\xi^2-1)\Htcff+\xi^2 \Etcff\right)^* \HTcff+2\xi\left(\xi\Htcff+\left(\frac{\xi^2}{1+\xi}+\frac{t}{4M^2}\right)\Etcff\right)^*\ETcff\\
  &\qquad\qquad\qquad-N^2\left(\Htcff+\frac{\xi}{1+\xi}\Etcff\right)^*\HTtcff-2\xi\left(\Htcff-\xi\left(\frac{\xi}{1+\xi}-\frac{t}{4M^2}\right)\Etcff\right)^*\ETtcff\Bigg]\\
  &-\frac{4i}{N}\mathscr{H}'^{[\rho\sigma]}_{(3)} \text{Re}\Bigg[2\left((\xi^2-1)\Htcff+\xi^2 \Etcff\right)^* \HTcff+2\xi\left(\xi\Htcff+\left(\frac{\xi^2}{1+\xi}+\frac{t}{4M^2}\right)\Etcff\right)^*\ETcff\\
  &\qquad\qquad\qquad-N^2\left(\Htcff+\frac{\xi}{1+\xi}\Etcff\right)^*\HTtcff-2\xi\left(\Htcff-\xi\left(\frac{\xi}{1+\xi}-\frac{t}{4M^2}\right)\Etcff\right)^*\ETtcff\Bigg]\\
  &+\frac{4}{N}\widetilde{\mathscr{H}}'^{(\rho\sigma)}_{(3)}\text{Im}\Bigg[2\left((\xi^2-1)\Hcff+\xi^2 \Ecff\right)^* \HTcff+2\xi\left(\xi\Hcff+\left(\frac{\xi^2}{1+\xi}+\frac{t}{4M^2}\right)\Ecff\right)^*\ETcff\\
  &\qquad\qquad\qquad-2\left(\xi\Hcff+\left(\frac{\xi^2}{1+\xi}+\frac{t}{4M^2}\right)\Ecff\right)^*\ETtcff
  \Bigg]\\
  &-\frac{4i}{N}\widetilde{\mathscr{H}}'^{[\rho\sigma]}_{(3)}\text{Re}\Bigg[2\left((\xi^2-1)\Hcff+\xi^2 \Ecff\right)^* \HTcff+2\xi\left(\xi\Hcff+\left(\frac{\xi^2}{1+\xi}+\frac{t}{4M^2}\right)\Ecff\right)^*\ETcff\\
  &\qquad\qquad\qquad-2\left(\xi\Hcff+\left(\frac{\xi^2}{1+\xi}+\frac{t}{4M^2}\right)\Ecff\right)^*\ETtcff\Bigg]\ ,
\end{split}
\end{align}
and for $H_{\rm{DVCS,T,out}}^{\rho\sigma}$
\begin{align}
\begin{split}
  &H_{\rm{DVCS, T,out},(3)}^{\rho\sigma}\\
  =&\frac{4}{N}\mathscr{H}^{(\rho\sigma)}_{(3)} \text{Im}\Bigg[2\left((\xi^2-1)\Hcff+\xi^2 \Ecff\right)^* \HTcff+2\left(\xi^2 \Hcff+\left(\xi^2+\frac{t}{4M^2}\right)\Ecff\right)^*\ETcff\\
  &\qquad\qquad\qquad-N^2(\Hcff+\Ecff)^*\HTtcff-2\xi\left(\Hcff+\frac{t}{4M^2}\Ecff\right)^*\ETtcff\Bigg]\\
  &-\frac{4i}{N}\mathscr{H}^{[\rho\sigma]}_{(3)} \text{Re}\Bigg[2\left((\xi^2-1)\Hcff+\xi^2 \Ecff\right)^* \HTcff+2\left(\xi^2 \Hcff+\left(\xi^2+\frac{t}{4M^2}\right)\Ecff\right)^*\ETcff\\
  &\qquad\qquad\qquad-N^2(\Hcff+\Ecff)^*\HTtcff-2\xi\left(\Hcff+\frac{t}{4M^2}\Ecff\right)^*\ETtcff\Bigg]\\
  &+\frac{4}{N}\widetilde{\mathscr{H}}^{(\rho\sigma)}_{(3)}\text{Im}\Bigg[2\left((\xi^2-1)\Htcff+\xi^2 \Etcff\right)^* \HTcff+2\xi^2\left(\Htcff+\frac{t}{4M^2}\Etcff\right)^*\ETcff\\
  &\qquad\qquad\qquad-2\xi\left(\Htcff+\frac{t}{4M^2}\Etcff\right)^*\ETtcff\Bigg]\\
  &-\frac{4i}{N}\widetilde{\mathscr{H}}^{[\rho\sigma]}_{(3)}\text{Re}\Bigg[2\left((\xi^2-1)\Htcff+\xi^2 \Etcff\right)^* \HTcff+2\xi^2\left(\Htcff+\frac{t}{4M^2}\Etcff\right)^*\ETcff\\
  &\qquad\qquad\qquad-2\xi\left(\Htcff+\frac{t}{4M^2}\Etcff\right)^*\ETtcff\Bigg]\ ,
  \end{split}
\end{align}
Besides, there are two kinematically twist-four hadronic tensors that come with the square of the twist-three CFFs, which are $H_{\rm{DVCS,U},(4)}^{\rho\sigma}$
\begin{align}
\label{eq:dvcsut4}
\begin{split}
  H_{\rm{DVCS,U},(4)}^{\rho\sigma}=&8\mathscr{H}^{\rho\sigma}_{(4)} \text{Re}\Bigg\{(1-\xi^2) \left|\bar{\mathcal H}_{2T}\right|^2-2\xi \bar{\mathcal H}^*_{2T} \left(\xi\ETcff-\ETtcff\right)-\frac{t}{4M^2} \left|\xi\ETcff-\ETtcff\right|^2\\
  &+\frac{N^2}{2}\left[\left(\bar{\mathcal H}_{2T}+\bar{\mathcal E}_{2T}\right)^*\bar{\widetilde{\mathcal H}}_{2T}+\frac{1}{4}\left|\bar{\mathcal E}_{2T}\right|^2-\frac{1}{4}\left|\bar{\widetilde{\mathcal E}}_{2T}\right|^2+\left(1-\frac{t}{4M^2}\right)\left|\bar{\widetilde{\mathcal H}}_{2T}\right|^2\right]\Bigg\}\ ,
\end{split}
\end{align}
and $H_{\rm{DVCS, T,out},(4)}^{\rho\sigma}$
\begin{align}
\label{eq:dvcsTout4}
\begin{split}
  H_{\rm{DVCS, T,out},(4)}^{\rho\sigma}=&4N \mathscr{H}^{\rho\sigma}_{(4)} \text{Im}\Bigg[\Big((\xi-1)\left(\ETcff+\ETtcff\right)-2\HTtcff\Big)^*\HTcff\\
  &+\Big(\xi\ETtcff-2\frac{t}{4M^2} \HTcff\Big)^*\ETcff+2\xi\left(1-\frac{t}{4 M^2}\right)\ETtcff^*\HTtcff\Bigg]\ ,
\end{split}
\end{align}
respectively, whereas the other two hadronic tensors have no kinematically twist-four contribution 
\begin{align}
\begin{split}
  H_{\rm{DVCS,L},(4)}^{\rho\sigma}=H_{\rm{DVCS,T,in},(4)}^{\rho\sigma}=0\ .
\end{split}
\end{align}
Then we can derive all the pure DVCS structure functions using these hadronic tensor. With their definition in eq. (\ref{eq:dvcsstruct}), we can perform the contraction and express them in terms of the scalar amplitude defined in eq. (\ref{eq:dvcsscalaramp}) as,
\begin{align}
\label{eq:fuu3}
\begin{split}
  F_{\rm{UU}}^{(3)}=&4h^{\rm{U}}_{(3)} \text{Re}\Bigg[-\Ecff^*\HTcff+\Hcff^* \ETcff+2\left(\Hcff+\frac{t}{4M^2}\Ecff\right)^*\HTtcff-\xi (\Hcff+\Ecff)^* \ETtcff\\
  &\qquad\qquad+\xi\Etcff^* \HTcff-\xi\Htcff^*\ETcff+\Htcff^*\ETtcff\Bigg]\ ,\\
\end{split}
\end{align}
\begin{align}
\begin{split}
  F_{\rm{LU}}^{(3)}=&-4h^{\rm{L}}_{(3)} \text{Im}\Bigg[-\Ecff^*\HTcff+\Hcff^* \ETcff+\left(\Hcff+\frac{t}{4M^2}\Ecff\right)^*2\HTtcff-\xi (\Hcff+\Ecff)^* \ETtcff\\
  &\qquad\qquad+\xi\Etcff^* \HTcff-\xi\Htcff^*\ETcff+\Htcff^*\ETtcff\Bigg]\ ,\\
\end{split}
\end{align}
\begin{align}
\begin{split}
  F_{\rm{UL}}^{(3)}=&4h'^{\rm{U}}_{(3)} \text{Im}\Bigg[-\xi\Etcff^* \HTcff+\left(\Htcff-\frac{\xi^2}{1+\xi}\Etcff\right)^*\ETcff+2\left(\Htcff+\xi\left(\frac{t}{4M^2}-\frac{\xi}{1+\xi}\right)\Etcff\right)^* \HTtcff\\
  &-\xi\left(\Htcff+\frac{\xi}{1+\xi} \Etcff\right)^*\ETtcff+\Ecff^*\HTcff-\xi\left( \Hcff+\frac{\xi}{1+\xi}\Ecff\right)^*\ETcff+\left( \Hcff+\frac{\xi}{1+\xi}\Ecff\right)^*\ETtcff\Bigg]\ ,
\end{split}
\end{align}
\begin{align}
\begin{split}
  F_{\rm{LL}}^{(3)}=&4h'^{\rm{L}}_{(3)} \text{Re}\Bigg[-\xi\Etcff^* \HTcff+\left(\Htcff-\frac{\xi^2}{1+\xi}\Etcff\right)^*\ETcff+2\left(\Htcff+\xi\left(\frac{t}{4M^2}-\frac{\xi}{1+\xi}\right)\Etcff\right)^* \HTtcff\\
  &-\xi\left(\Htcff+\frac{\xi}{1+\xi} \Etcff\right)^*\ETtcff+\Ecff^*\HTcff-\xi\left( \Hcff+\frac{\xi}{1+\xi}\Ecff\right)^*\ETcff+\left( \Hcff+\frac{\xi}{1+\xi}\Ecff\right)^*\ETtcff\Bigg]\ ,
\end{split}
\end{align}
\begin{align}
\begin{split}
  F_{\rm{UT,in}}^{(3)}=&\frac{4}{N}h'^{\rm{U}}_{(3)} \text{Im}\Bigg[2\left((\xi^2-1)\Htcff+\xi^2 \Etcff\right)^* \HTcff+2\xi\left(\xi\Htcff+\left(\frac{\xi^2}{1+\xi}+\frac{t}{4M^2}\right)\Etcff\right)^*\ETcff\\
  &\qquad\qquad-N^2\left(\Htcff+\frac{\xi}{1+\xi}\Etcff\right)^*\HTtcff-2\xi\left(\Htcff-\xi\left(\frac{\xi}{1+\xi}-\frac{t}{4M^2}\right)\Etcff\right)^*\ETtcff\\
  &\qquad\qquad+2\left((\xi^2-1)\Hcff+\xi^2 \Ecff\right)^* \HTcff+2\xi\left(\xi\Hcff+\left(\frac{\xi^2}{1+\xi}+\frac{t}{4M^2}\right)\Ecff\right)^*\ETcff\\
  &\qquad\qquad-2\left(\xi\Hcff+\left(\frac{\xi^2}{1+\xi}+\frac{t}{4M^2}\right)\Ecff\right)^*\ETtcff
  \Bigg]\ ,
\end{split}
\end{align}
\begin{align}
\begin{split}
  F_{\rm{LT,in}}^{(3)}=&\frac{4}{N}h'^{\rm{L}}_{(3)} \text{Re}\Bigg[2\left((\xi^2-1)\Htcff+\xi^2 \Etcff\right)^* \HTcff+2\xi\left(\xi\Htcff+\left(\frac{\xi^2}{1+\xi}+\frac{t}{4M^2}\right)\Etcff\right)^*\ETcff\\
  &\qquad\qquad-N^2\left(\Htcff+\frac{\xi}{1+\xi}\Etcff\right)^*\HTtcff-2\xi\left(\Htcff-\xi\left(\frac{\xi}{1+\xi}-\frac{t}{4M^2}\right)\Etcff\right)^*\ETtcff\\
  &\qquad\qquad+2\left((\xi^2-1)\Hcff+\xi^2 \Ecff\right)^* \HTcff+2\xi\left(\xi\Hcff+\left(\frac{\xi^2}{1+\xi}+\frac{t}{4M^2}\right)\Ecff\right)^*\ETcff\\
  &\qquad\qquad-2\left(\xi\Hcff+\left(\frac{\xi^2}{1+\xi}+\frac{t}{4M^2}\right)\Ecff\right)^*\ETtcff
  \Bigg]\ ,
\end{split}
\end{align}
\begin{align}
\begin{split}
  F_{\rm{UT,out}}^{(3)}=&\frac{4}{N}h^{\rm{U}}_{(3)} \text{Im}\Bigg[2\left((\xi^2-1)\Hcff+\xi^2 \Ecff\right)^* \HTcff+2\left(\xi^2 \Hcff+\left(\xi^2+\frac{t}{4M^2}\right)\Ecff\right)^*\ETcff\\
  &\qquad\qquad-N^2(\Hcff+\Ecff)^*\HTtcff-2\xi\left(\Hcff+\frac{t}{4M^2}\Ecff\right)^*\ETtcff\\
  &\qquad\qquad+2\left((\xi^2-1)\Htcff+\xi^2 \Etcff\right)^* \HTcff+2\xi^2\left(\Htcff+\frac{t}{4M^2}\Etcff\right)^*\ETcff\\
  &\qquad\qquad-2\xi\left(\Htcff+\frac{t}{4M^2}\Etcff\right)^*\ETtcff\Bigg]\ ,
  \end{split}
\end{align}
\begin{align}
\begin{split}
  F_{\rm{LT,out}}^{(3)}=&\frac{4}{N}h^{\rm{L}}_{(3)} \text{Re}\Bigg[2\left((\xi^2-1)\Hcff+\xi^2 \Ecff\right)^* \HTcff+2\left(\xi^2 \Hcff+\left(\xi^2+\frac{t}{4M^2}\right)\Ecff\right)^*\ETcff\\
  &\qquad\qquad-N^2(\Hcff+\Ecff)^*\HTtcff-2\xi\left(\Hcff+\frac{t}{4M^2}\Ecff\right)^*\ETtcff\\
  &\qquad\qquad+2\left((\xi^2-1)\Htcff+\xi^2 \Etcff\right)^* \HTcff+2\xi^2\left(\Htcff+\frac{t}{4M^2}\Etcff\right)^*\ETcff\\
  &\qquad\qquad-2\xi\left(\Htcff+\frac{t}{4M^2}\Etcff\right)^*\ETtcff\Bigg]\ ,
  \end{split}
\end{align}
\begin{align}
\begin{split}
   F_{\rm{UU}}^{(4)}=&8 h^{\rm{U}}_{(4)} \text{Re}\Bigg\{(1-\xi^2) \left|\bar{\mathcal H}_{2T}\right|^2-2\xi \bar{\mathcal H}^*_{2T} \left(\xi\ETcff-\ETtcff\right)-\frac{t}{4M^2} \left|\xi\ETcff-\ETtcff\right|^2\\
  &+\frac{ N^2}{2} \Bigg[\left(\bar{\mathcal H}_{2T}+\bar{\mathcal E}_{2T}\right)^*\bar{\widetilde{\mathcal H}}_{2T}+\frac{1}{4}\left|\bar{\mathcal E}_{2T}\right|^2-\frac{1}{4}\left|\bar{\widetilde{\mathcal E}}_{2T}\right|^2+\left(1-\frac{t}{4M^2}\right)\left|\bar{\widetilde{\mathcal H}}_{2T}\right|^2\Bigg]\Bigg\}\ ,
\end{split}
\end{align}
\begin{align}
\begin{split}
  F_{\rm{UT,out}}^{(4)}=&4N h^{\rm{U}}_{(4)} \text{Im}\Bigg\{\Big[(\xi-1)\left(\ETcff+\ETtcff\right)-2\HTtcff\Big]^*\HTcff\\
 &\qquad\qquad+\Big(\xi\ETtcff-2\frac{t}{4M^2} \HTcff\Big)^*\ETcff+2\xi\left(1-\frac{t}{4 M^2}\right)\ETtcff^*\HTtcff\Bigg\}\ ,
\end{split}
\end{align}
\begin{align}
\label{eq:flu4}
\begin{split}
  F_{\rm{LU}}^{(4)}=F_{\rm{LT,out}}^{(4)}=F_{\rm{UL}}^{(4)}= F_{\rm{LL}}^{(4)}=F_{\rm{UT,in}}^{(4)}= F_{\rm{LT,in}}^{(4)}=0\ ,
\end{split}
\end{align}
Therefore, it will be sufficient to have those five coefficients $h^{\rm{L/U}}_{(3)}$, $h'^{\rm{L/U}}_{(3)}$ and $h^{\rm{U}}_{(4)}$ in order to get the pure DVCS cross-section related to the twist-three CFFs, besides the CFFs themselves. And we also present those scalar coefficients in terms of the scalar products of all the four-vectors as,
\begin{align}
\begin{split}
  h^{\rm{U}}_{(3)}=\frac{\xi q^2(2k-q)\cdot q'}{2(\bar P\cdot q)(q\cdot q')^3}\Bigg\{&(q\cdot q')\Big[2(k\cdot \bar P)(q\cdot q')-q^2(\bar P\cdot q')\Big]\\
  &+2(k\cdot q')\Big[q^2(\bar P\cdot q')-(\bar P\cdot q)(q\cdot q')\Big]\Bigg\}\ ,
\end{split}
\end{align}
\begin{align}
\begin{split}
  h^{\rm{L}}_{(3)}=\frac{\xi q^2\epsilon^{k\bar P q q'}}{(\bar P\cdot q)(q\cdot q')}
\end{split}
\end{align}
\begin{align}
\begin{split}
  h'^{\rm{U}}_{(3)}=\frac{\xi q^2(2k-q)\cdot q'\epsilon^{k\bar P q q'}}{(\bar P\cdot q)(q\cdot q')^2}
\end{split}
\end{align}
\begin{align}
\begin{split}
  h'^{\rm{L}}_{(3)}=\frac{-\xi q^2}{2(\bar P\cdot q)(q\cdot q')^2}\Bigg\{&(q\cdot q')\Big[2(\bar P\cdot k)(q\cdot q')-q^2(\bar P\cdot q')\Big]\\
  &+2(k\cdot q')\Big[q^2(\bar P\cdot q')-(\bar P\cdot q)(q\cdot q')\Big]\Bigg\}\ ,
\end{split}
\end{align}
\begin{align}
\begin{split}
  h^{\rm{U}}_{(4)}=\frac{M^2 q^4(k\cdot q')(k-q)\cdot q'}{(\bar P\cdot q)^2(q\cdot q')^2}
\end{split}
\end{align}
With those expressions, all the cross-sections can be calculated by combining all the four-vectors and plugging in the twist-three CFFs at given kinematics.
\section{Twist-three interference structure functions}
\label{app:structurefunc}
In this appendix, we present all the hadronic tensors as well as structure functions for interference contributions that are related to twist-three CFFs. The four hadronic tensor $H_{\rm{INT,U},(3)}^{\rho\sigma}$, $H_{\rm{INT,U,L},(3)}^{\rho\sigma}$, $H_{\rm{INT,U,T,in},(3)}^{\rho\sigma}$ and $H_{\rm{INT,U,T,out},(3)}^{\rho\sigma}$ can be written as,
\begin{align}
\begin{split}
       H_{\rm{INT,U},(3)}^{\mu\rho\sigma }= &-4\xi \bar P^\sigma \bar P_\gamma \mathscr{T}^{\mu\rho,\gamma}_{(3)}\left[F_1 (\ETcff+2\HTtcff)^*-F_2\left(\HTcff-\frac{t}{4M^2}2\HTtcff\right)^*\right]\\
      &-t n^\sigma \bar P_\gamma \mathscr{T}^{\mu\rho,\gamma}_{(3)} (F_1+F_2) \ETtcff^*\\
      &-4 M^2  \mathscr{T}^{\mu\rho,\sigma}_{(3)}\left(F_1+F_2\right)\left[\xi \HTcff+\frac{t}{4M^2}\left(\xi \ETcff-\ETtcff\right)\right]^* \ ,
\end{split}
\end{align}
and for $H_{\rm{INT,L}}^{\mu\rho\sigma}$,
\begin{align}
    \begin{split}
      &H_{\rm{INT,L},(3)}^{\mu\rho\sigma}\\
      = &4\xi \bar P^\sigma  \bar P_\gamma \widetilde{\mathscr{T}}^{\mu\rho,\gamma}_{(3)} \Bigg[F_1 \left(\frac{2\xi}{1+\xi}\HTtcff+\ETtcff\right)^*+F_2\left(\HTcff+\frac{\xi}{1+\xi}\left(\ETcff+\ETtcff+2\HTtcff\right)\right)^*\Bigg]\\
      &+ t n^\sigma \bar P_\gamma \widetilde{\mathscr{T}}^{\mu\rho,\gamma}_{(3)} (F_1+F_2) \Bigg[\ETcff^*+\frac{2}{1+\xi}\HTtcff^*\Bigg]\\
      &-4  M^2 \widetilde{\mathscr{T}}^{\mu\rho,\sigma}_{(3)} \left(F_1+F_2\right)\Bigg[\xi\HTcff^*+\left(\frac{\xi^2}{1+\xi}+\frac{t}{4M^2}\right)\ETcff^*\\
      &\qquad\qquad\qquad\qquad\qquad+2\left(\frac{\xi^2}{1+\xi}+(1-\xi)\frac{t}{4M^2}\right)\HTtcff^*+\xi\left(\frac{\xi}{1+\xi}-\frac{t}{4M^2}\right)\ETtcff^*\Bigg] \ ,
\end{split}
\end{align}
and for $H_{\rm{INT,T,in}}^{\mu\rho\sigma}$,
\begin{align}
    \begin{split}
      &H_{\rm{INT,T,in},(3)}^{\mu\rho\sigma}\\
      = &-\frac{8\xi}{N} \bar P^\sigma  \bar P_\gamma \widetilde{\mathscr{T}}^{\mu\rho,\gamma}_{(3)}\Bigg[F_1\left((1+\xi)\HTcff-2\xi\left(\frac{t}{4M^2}-\frac{\xi}{1+\xi}\right) \HTtcff+\xi \ETtcff\right)^*\\
      &+F_2\left(\xi\HTcff-\xi\left(\frac{t}{4M^2}-\frac{\xi}{1+\xi}\right)\left(
      \ETcff+2\HTtcff\right)+\left(\frac{t}{4M^2}+\frac{\xi^2}{1+\xi}\right)\ETtcff\right)^*\Bigg]\\
      &-\frac{2}{N} t n^\sigma \bar P_\gamma \widetilde{\mathscr{T}}^{\mu\rho,\gamma}_{(3)} (F_1+F_2) \Bigg[(1+\xi)\HTcff+\xi\ETcff-\left(\frac{t}{4M^2}-\frac{\xi}{1+\xi}\right)2\HTtcff\Bigg]^*\\
      &-2 N M^2  \widetilde{\mathscr{T}}^{\mu\rho,\sigma}_{(3)}\left(F_1+F_2\right)\Bigg[\HTcff+\frac{\xi}{1+\xi}\left( \ETcff+\ETtcff\right)-\left(\frac{t}{4M^2}-\frac{\xi}{1+\xi}\right)2 \HTtcff\Bigg]^* \ ,
\end{split}
\end{align}
and for $H_{\rm{INT,T,out}}^{\mu\rho\sigma}$,
\begin{align}
    \begin{split}
      H_{\rm{INT,T,out},(3)}^{\mu\rho\sigma}= &\frac{8i}{N}\xi \bar P^\sigma \bar P_\gamma \mathscr{T}^{\mu\rho,\gamma}_{(3)}\Bigg[F_1 \left((1+\xi)\HTcff+\xi \ETtcff-\frac{t}{4M^2} 2\HTtcff\right)^*\\
      &\qquad\qquad\qquad\quad+F_2\left(\xi \HTcff-\frac{t}{4M^2}\left(\ETcff-\xi\ETtcff+2\HTtcff\right)\right)^*\Bigg]\\
      &+\frac{2 i}{N} t n^\sigma \bar P_\gamma \mathscr{T}^{\mu\rho,\gamma}_{(3)}(F_1+F_2) \Bigg[(1+\xi)\HTcff+\xi \ETcff+2\xi\left(1-\frac{t}{4M^2}\right)\HTtcff\Bigg]^*\\
      &+2 N i M^2 \mathscr{T}^{\mu\rho,\sigma}_{(3)}\left(F_1+F_2\right)\HTcff^* \ ,
\end{split}
\end{align}

Then we can derive all the interference structure functions using these hadronic tensor. By simply contracting according to the definitions in eq. (\ref{eq:intxsecdef}), we have the following expressions for all the $F^{\rm{I}}$s in terms of the coefficients defined in eq. (\ref{eq:intstructfunc}),
\begin{align}
\begin{split}
    F^{\rm{I}}_{\rm{UU},(3)}=&-\text{Re}\Bigg\{A^{\rm{I,U}}_{(3)}\left[F_1 (\ETcff+2\HTtcff)^*-F_2\left(\HTcff-\frac{t}{4M^2}2\HTtcff\right)^*\right]\\
      &+B^{\rm{I,U}}_{(3)} (F_1+F_2) \ETtcff^*+C^{\rm{I,U}}_{(3)}\left(F_1+F_2\right)\left[\xi \HTcff+\frac{t}{4M^2}\left(\xi \ETcff-\ETtcff\right)\right]^* \Bigg\}\ ,
\end{split}
\end{align}
\begin{align}
\begin{split}
    F^{\rm {I}}_{\rm{LU},(3)}=&\text{Im}\Bigg\{A^{\rm{I,L}}_{(3)}\left[F_1 (\ETcff+2\HTtcff)^*-F_2\left(\HTcff-\frac{t}{4M^2}2\HTtcff\right)^*\right]\\
      &+B^{\rm{I,L}}_{(3)} (F_1+F_2) \ETtcff^*+C^{\rm{I,L}}_{(3)}\left(F_1+F_2\right)\left[\xi \HTcff+\frac{t}{4M^2}\left(\xi \ETcff-\ETtcff\right)\right]^* \Bigg\}\ ,
\end{split}
\end{align}
\begin{align}
\begin{split}
    F^{\rm {I}}_{\rm{UL},(3)} = &\text{Im}\Bigg\{\widetilde{A}^{\rm{I,U}}_{(3)} \left[F_1 \left(\frac{2\xi}{1+\xi}\HTtcff+\ETtcff\right)^*+F_2\left(\HTcff+\frac{\xi}{1+\xi}\left(\ETcff+\ETtcff+2\HTtcff\right)\right)^*\right]\\
      &+\widetilde{B}^{\rm{I,U}}_{(3)} (F_1+F_2) \Bigg(\ETcff^*+\frac{2}{1+\xi}\HTtcff^*\Bigg)\\
      &-\widetilde{C}^{\rm{I,U}}_{(3)} \left(F_1+F_2\right)\Bigg[\xi\HTcff^*+\left(\frac{\xi^2}{1+\xi}+\frac{t}{4M^2}\right)\ETcff^*+2\left(\frac{\xi^2}{1+\xi}+(1-\xi)\frac{t}{4M^2}\right)\HTtcff^*\\
      &\qquad\qquad\qquad\qquad+\xi\left(\frac{\xi}{1+\xi}-\frac{t}{4M^2}\right)\ETtcff^*\Bigg]\Bigg\} \ ,
\end{split}
\end{align}
\begin{align}
\begin{split}
    F^{\rm {I}}_{\rm{LL},(3)} = &\text{Re}\Bigg\{\widetilde{A}^{\rm{I,L}}_{(3)} \left(F_1 \left[\frac{2\xi}{1+\xi}\HTtcff+\ETtcff\right)^*+F_2\left(\HTcff+\frac{\xi}{1+\xi}\left(\ETcff+\ETtcff+2\HTtcff\right)\right)^*\right]\\
      &+\widetilde{B}^{\rm{I,L}}_{(3)} (F_1+F_2) \Bigg(\ETcff^*+\frac{2}{1+\xi}\HTtcff^*\Bigg)\\
      &-\widetilde{C}^{\rm{I,L}}_{(3)} \left(F_1+F_2\right)\Bigg[\xi\HTcff^*+\left(\frac{\xi^2}{1+\xi}+\frac{t}{4M^2}\right)\ETcff^*+2\left(\frac{\xi^2}{1+\xi}+(1-\xi)\frac{t}{4M^2}\right)\HTtcff^*\\
      &\qquad\qquad\qquad\qquad+\xi\left(\frac{\xi}{1+\xi}-\frac{t}{4M^2}\right)\ETtcff^*\Bigg]\Bigg\} \ ,
\end{split}
\end{align}
\begin{align}
\begin{split}
    F^{\rm {I}}_{\rm{UT,in},(3)} =& -\frac{2}{N}\text{Im}\Bigg\{\widetilde{A}^{\rm{I,U}}_{(3)} \Bigg[F_1\left((1+\xi)\HTcff-2\xi\left(\frac{t}{4M^2}-\frac{\xi}{1+\xi}\right) \HTtcff+\xi \ETtcff\right)^*\\
      &+F_2\left(\xi\HTcff-\xi\left(\frac{t}{4M^2}-\frac{\xi}{1+\xi}\right)\left(
      \ETcff+2\HTtcff\right)+\left(\frac{t}{4M^2}+\frac{\xi^2}{1+\xi}\right)\ETtcff\right)^*\Bigg]\\
      &+\widetilde{B}^{\rm{I,U}}_{(3)} (F_1+F_2) \Bigg((1+\xi)\HTcff+\xi\ETcff-\left(\frac{t}{4M^2}-\frac{\xi}{1+\xi}\right)2\HTtcff\Bigg)^*\\
      &+\frac{N^2}{4}\widetilde{C}^{\rm{I,U}}_{(3)} \left(F_1+F_2\right)\Bigg[\HTcff+\frac{\xi}{1+\xi}\left( \ETcff+\ETtcff\right)-\left(\frac{t}{4M^2}-\frac{\xi}{1+\xi}\right)2 \HTtcff\Bigg]^*\Bigg\} \ ,
\end{split}
\end{align}
\begin{align}
\begin{split}
    F^{\rm {I}}_{\rm{LT,in},(3)} = &-\frac{2}{N}\text{Re}\Bigg\{\widetilde{A}^{\rm{I,L}}_{(3)} \Bigg[F_1\left((1+\xi)\HTcff-2\xi\left(\frac{t}{4M^2}-\frac{\xi}{1+\xi}\right) \HTtcff+\xi \ETtcff\right)^*\\
      &+F_2\left(\xi\HTcff-\xi\left(\frac{t}{4M^2}-\frac{\xi}{1+\xi}\right)\left(
      \ETcff+2\HTtcff\right)+\left(\frac{t}{4M^2}+\frac{\xi^2}{1+\xi}\right)\ETtcff\right)^*\Bigg]\\
      &+\widetilde{B}^{\rm{I,L}}_{(3)} (F_1+F_2) \Bigg((1+\xi)\HTcff+\xi\ETcff-\left(\frac{t}{4M^2}-\frac{\xi}{1+\xi}\right)2\HTtcff\Bigg)^*\\
      &+\frac{N^2}{4}\widetilde{C}^{\rm{I,L}}_{(3)} \left(F_1+F_2\right)\Bigg[\HTcff+\frac{\xi}{1+\xi}\left( \ETcff+\ETtcff\right)-\left(\frac{t}{4M^2}-\frac{\xi}{1+\xi}\right)2 \HTtcff\Bigg]^*\Bigg\} \ ,
\end{split}
\end{align}
\begin{align}
\begin{split}
    F^{\rm {I}}_{\rm{UT,out},(3)} = -\frac{2}{N}\text{Im}\Bigg\{&{A}^{\rm{I,U}}_{(3)} \Bigg[F_1 \left((1+\xi)\HTcff+\xi \ETtcff-\frac{t}{4M^2} 2\HTtcff\right)^*\\
      &\qquad\quad+F_2\left(\xi \HTcff-\frac{t}{4M^2}\left(\ETcff-\xi\ETtcff+2\HTtcff\right)\right)^*\Bigg]\\
      &+{B}^{\rm{I,U}}_{(3)} (F_1+F_2) \Bigg[(1+\xi)\HTcff+\xi \ETcff+2\xi\left(1-\frac{t}{4M^2}\right)\HTtcff\Bigg]^*\\
      &+\frac{N^2}{4}{C}^{\rm{I,U}}_{(3)} \left(F_1+F_2\right)\HTcff^*\Bigg\} \ ,
\end{split}
\end{align}
\begin{align}
\begin{split}
    F^{\rm {I}}_{\rm{LT,out},(3)} = -\frac{2}{N}\text{Re}\Bigg\{&{A}^{\rm{I,L}}_{(3)} \Bigg[F_1 \left((1+\xi)\HTcff+\xi \ETtcff-\frac{t}{4M^2} 2\HTtcff\right)^*\\
      &\qquad\quad+F_2\left(\xi \HTcff-\frac{t}{4M^2}\left(\ETcff-\xi\ETtcff+2\HTtcff\right)\right)^*\Bigg]\\
      &+{B}^{\rm{I,L}}_{(3)} (F_1+F_2) \Bigg[(1+\xi)\HTcff+\xi \ETcff+2\xi\left(1-\frac{t}{4M^2}\right)\HTtcff\Bigg]^*\\
      &+\frac{N^2}{4}{C}^{\rm{I,L}}_{(3)} \left(F_1+F_2\right)\HTcff^*\Bigg\} \ ,
\end{split}
\end{align}
Therefore, it will be sufficient to have those coefficients ${\mathcal A}^{\rm{I,L/U}}_{(3)}$ with $\mathcal A=\{A,B,C,\widetilde{A},\widetilde{B},\widetilde{C}\}$ in order to get the interference cross-sections related to the twist-three CFFs, besides the CFFs themselves. Those scalar coefficients are expressible in terms of the contractions of the four-vectors as,
\begin{align}
\begin{split}
A^{\rm{I,U}}_{(3)}=&\frac{4\xi q^2}{
 2(\bar P\cdot q)(k\cdot q')(q\cdot q')^2 (k-q)\cdot q' }\Bigg\{ 8(k\cdot q')^3\Big[(P\cdot q)(P\cdot q') - M^2(q\cdot q')\Big]\\
 &+ 
   4(k\cdot q')^2\Big[3M^2(q\cdot q')^2+(q\cdot q')(P\cdot q)^2 - (P\cdot q)(P\cdot q')[q^2 + 4(q\cdot q')] \\
   &\qquad\qquad\qquad+ 2(P\cdot k)\Big((P\cdot q')\left(q^2 + (q\cdot q')\right)-(P\cdot q)(q\cdot q')\Big)   \Big]\\
   &+ 
   2(k\cdot q')\Big[q^4(P\cdot q')^2 - 2(q\cdot q')^2\Big(M^2(q\cdot q')-2(P\cdot k)^2 \Big)\\
   &\qquad\qquad\qquad- 4(P\cdot q')(q\cdot q')\Big((P\cdot k)\left(q^2 + (q\cdot q')\right)-(P\cdot q)(q\cdot q')\Big) \Big]\\
   &-(q\cdot q')\Big[(P\cdot q')q^2 - 2(P\cdot k)(q\cdot q')\Big]^2\Bigg\}
\end{split}
\end{align}
\begin{align}
\begin{split}
B^{\rm{I,U}}_{(3)}\approx 0\ , 
\end{split}
\end{align}
\begin{align}
\begin{split}
C^{\rm{I,U}}_{(3)}=-\frac{16 t M^2 q^2 \left[(q\cdot q')-2(k\cdot q')\right]}{2(\bar P\cdot q)(q\cdot q')^2}\ ,
\end{split}
\end{align}
\begin{align}
\begin{split}
\widetilde{A}^{\rm{I,U}}_{(3)}=&\frac{8\xi q^2\epsilon^{k P q q'}}{2(\bar P\cdot q)(k\cdot q')(q\cdot q')^2(k-q)\cdot q'}\Bigg\{4(P\cdot q')(k\cdot q')^2\\
&+(k\cdot q')(q\cdot q')(2k-q-2q')\cdot P+(q\cdot q')\Big[q^2(P\cdot q')-2(P\cdot k)(q\cdot q')\Big]\Bigg\}
\end{split}
\end{align}
\begin{align}
\begin{split}
\widetilde{B}^{\rm{I,U}}_{(3)}\approx 0\ , \qquad \widetilde{C}^{\rm{I,U}}_{(3)}=0\ ,
\end{split}
\end{align}
\begin{align}
\begin{split}
{A}^{\rm{I,L}}_{(3)}=\frac{8\xi q^2\epsilon^{k P q q'}}{2(\bar P\cdot q)(k\cdot q')(q\cdot q')^2(k-q)\cdot q'}\Bigg\{&(q\cdot q')\Big[q^2(P\cdot q')-2(P\cdot k)(q\cdot q')\Big]\\
&+(k\cdot q')\Big[2(P\cdot q)(q\cdot q')-2q^2(P\cdot q')\Big]\Bigg\}
\end{split}
\end{align}
\begin{align}
\begin{split}
{B}^{\rm{I,L}}_{(3)}\approx 0\ , \qquad  {C}^{\rm{I,L}}_{(3)}=0\ ,
\end{split}
\end{align}
\begin{align}
\begin{split}
\widetilde{A}^{\rm{I,L}}_{(3)}=&\frac{4\xi q^2}{2(\bar P\cdot q)(k\cdot q')(q\cdot q')^2(k-q)\cdot q'}\Bigg\{(q\cdot q')\Big[q^2(P\cdot q')-2(P\cdot k)(q\cdot q')\Big]^2\\
&+4(k\cdot q')^2\Bigg[q^2(P\cdot q')^2+(q\cdot q')\Big((P\cdot q)^2+P^2(q'-q)\cdot q\Big)\\
&\qquad\qquad\qquad-2(P\cdot q)(P\cdot q')(q\cdot q')\Bigg]\\
&-4(k\cdot q')(q\cdot q')\Bigg[q^2(P\cdot q')^2+P^2(q\cdot q')^2+2(P\cdot k)\Big((P\cdot q)(q\cdot q')-q^2(P\cdot q')\Big)\\
&\qquad\qquad\qquad\qquad+(P\cdot q)(P\cdot q')(q-2q')\cdot q-P^2 q^2(q\cdot q')\Bigg]\Bigg\}\ ,
\end{split}
\end{align}
\begin{align}
\begin{split}
\widetilde{B}^{\rm{I,L}}_{(3)}\approx 0\ ,
\end{split}
\end{align}
\begin{align}
\begin{split}
 \widetilde{C}^{\rm{I,L}}_{(3)}= -\frac{16 t M^2 q^2 \left[(q\cdot q')-2(k\cdot q')\right]}{2(\bar P\cdot q)(q\cdot q')}\ .
\end{split}
\end{align}
Note that we approximate all the coefficients that almost vanish numerically to zero.

\section{Transformations of Dirac matrices}
\label{app:diracalgebra}
In eq. (\ref{eq:wwDVCS}), we relate the twist-three CFFs with the twist-two GPDs with WW relations. However, the right-hand side of this equation involves a different set of Dirac structures involving $\gamma^5$, while the twist-three GPDs are parameterized in eq. (\ref{eq:Meiztw3a}) without those Dirac structures. Therefore, we need to transform those Dirac matrices with $\gamma^5$ into those without $\gamma^5$. To do so, we simply need to use the definition of $\gamma^5\equiv -\frac{i}{24}\epsilon^{\mu\nu\rho\sigma} \gamma_\mu\gamma_\nu\gamma_\rho\gamma_\sigma$ and simplify all the Dirac structures using equation of motions.
Notice that there are three different Dirac structures for $-i\widetilde{\epsilon}^{\mu\nu}\widetilde G_{\nu}(x,\xi)$, namely $-i\widetilde{\epsilon}^{\mu\nu} \gamma_{T,\nu}\gamma^5$, $-i\widetilde{\epsilon}^{\mu\nu} \Delta_{T,\nu}\gamma^5$ and $-i\widetilde{\epsilon}^{\mu\nu} \Delta_{T,\nu} n\cdot \gamma \gamma^5$. With the help of the approximate relation (which becomes exact if we define the light-cone vectors using the two photon momenta) in eq. (\ref{eq:epsapprox}), we have the following identities,
\begin{align}
\label{eq:diraciden1}
\begin{split}
\left<-i\widetilde{\epsilon}^{\mu\nu} \gamma_{T,\nu}\gamma^5\right>=\left<-iM\sigma^{\mu\nu}n_\nu-\frac{1}{2}\Big[\Delta^\mu (n\cdot \gamma)-\gamma^\mu (n\cdot \Delta)\Big]+\cdots\right>\ ,
\end{split}
\end{align}
\begin{align}
\label{eq:diraciden2}
\begin{split}
\left< \frac{-i\widetilde{\epsilon}^{\mu\nu} \Delta_{T,\nu}\gamma^5}{M}\right>
=\Big<&-2\frac{t}{4M^2} iM\sigma^{\mu\nu}n_\nu-\Big[\Delta^\mu(n\cdot \gamma)-\gamma^\mu(n\cdot \Delta)\Big]\\
&+\frac{1}{M}\Big[\Delta^\mu(n\cdot \bar P)-\bar P^\mu (n\cdot \Delta)\Big]\Big>\ ,
\end{split}
\end{align}
\begin{align}
\label{eq:diraciden3}
\begin{split}
\left<-i\widetilde{\epsilon}^{\mu\nu} \Delta_{T,\nu} n\cdot \gamma \gamma^5\right>=\Big<&2\xi  iM\sigma^{\mu\nu}n_\nu+\xi\Big[\Delta^\mu(n\cdot \gamma)-\gamma^\mu(n\cdot \Delta)\Big]\\
&+2\Big[\bar P^\mu(n\cdot \gamma)-\gamma^\mu(n\cdot \bar P)\Big]+\cdots\Big>\ ,
\end{split}
\end{align}
where again we introduce the notation $\left<\Gamma\right>$ for $\bar u(P',S')\Gamma u(P,S)$ and $\cdots$ stands for structures $\propto n^\mu$ which are twist-four. Since there are exactly the same set of Dirac structures on the right-hand sides of eqs. (\ref{eq:diraciden1}) -- (\ref{eq:diraciden3}) as the ones in eq. (\ref{eq:Meiztw3a}), they allow us to convert the Dirac structures on the right-hand side of eq. (\ref{eq:wwDVCS}) into the structures in eq. (\ref{eq:Meiztw3a}) and match the CFFs.

\bibliographystyle{jhep}
\bibliography{refs.bib}
\end{document}